\documentclass[twocolumn]{aastex61}

\usepackage{amsmath}

\PassOptionsToPackage{breaklinks=true}{hyperref}
\def\hitomi     {\emph{Hitomi}}
\def\chandra    {\emph{Chandra}}

\def\planck    {\emph{Planck}}
\def\lax{\lesssim}
\def\gax{\gtrsim}
\def\msun       {$M_{\odot}$}

\received{December 20, 2016}
\revised{April 6, 2017}
\accepted{\today}
\submitjournal{ApJ}

\shorttitle{}
\shortauthors{Giacintucci et al.}

\begin{document}

\title{Occurrence of radio minihalos in a mass-limited sample of galaxy
  clusters}

\correspondingauthor{Simona Giacintucci}
\email{simona.giacintucci@nrl.navy.mil}

\author{Simona Giacintucci}
\affiliation{Naval Research Laboratory, 
4555 Overlook Avenue SW, Code 7213, 
Washington, DC 20375, USA}

\author{Maxim Markevitch}
\affiliation{NASA/Goddard Space Flight Center,
Greenbelt, MD 20771, USA}

\author{Rossella Cassano}
\affiliation{INAF - Istituto di Radioastronomia,
via Gobetti 101, I-40129 Bologna, Italy}

\author{Tiziana Venturi}
\affiliation{INAF - Istituto di Radioastronomia,
via Gobetti 101, I-40129 Bologna, Italy}

\author{Tracy E. Clarke}
\affiliation{Naval Research Laboratory,
4555 Overlook Avenue SW, Code 7213,
Washington, DC 20375, USA}

\author{Gianfranco Brunetti}
\affiliation{INAF - Istituto di Radioastronomia,
via Gobetti 101, I-40129 Bologna, Italy}

\begin{abstract}

We investigate the occurrence of radio minihalos --- diffuse radio sources
of unknown origin observed in the cores of some galaxy clusters --- in a
statistical sample of 58 clusters drawn from the \planck\ Sunyaev-Zel'dovich
cluster catalog using a mass cut ($M_{500}>6\times 10^{14}$\msun). We
supplement our statistical sample with a similarly-sized non-statistical
sample mostly consisting of clusters in the ACCEPT X-ray catalog with
suitable X-ray and radio data, which includes lower-mass clusters. 
Where necessary (for 9 clusters), we reanalyzed the {\em Very
  Large Array}\/ archival radio data to determine if a mihinalo is present.
Our total sample includes all 28 currently known and recently discovered
radio minihalos, including 6 candidates. We classify clusters as cool-core
or non-cool core according to the value of the specific entropy floor in the
cluster center, rederived or newly derived from the \chandra\ X-ray density
and temperature profiles where necessary (for 27 clusters). Contrary
to the common wisdom that minihalos are rare, we find that almost all cool
cores --- at least 12 out of 15 (80\%) --- in our complete sample of massive
clusters exhibit minihalos. The supplementary sample shows that the occurrence 
of minihalos may be lower in lower-mass cool-core clusters. No minihalos are 
found in non-cool-cores or ``warm cores''.  These findings will help test theories 
of the origin of minihalos and provide information on the physical processes and
energetics of the cluster cores.
\end{abstract}

\keywords{catalogs --- galaxies: clusters: general --- surveys --- X-rays:
galaxies: clusters --- radio continuum: galaxies: clusters}


\section{Introduction}

A striking feature of a number of galaxy clusters is the presence of diffuse
synchrotron radiation in the form of large peripheral radio relics and two
classes of centrally-located radio sources --- Mpc-size {\em giant radio
  halos} and smaller-scale {\em minihalos} \citep[see][for a
review]{2012A&ARv..20...54F}.
The importance of these extended, steep spectrum%
\footnote{$\alpha_{\rm radio} > 1$, for $S_{\nu} \propto \nu^{-\alpha_{\rm
      radio}}$, where $S_{\nu}$ is the flux density at the frequency $\nu$
  and $\alpha_{\rm radio}$ is the radio spectral index.}
radio sources is nowadays widely recognized as their existence requires
magnetic fields and ultra-relativistic electrons to be distributed
throughout a large fraction of the cluster volume
\citep[e.g.,][]{2014IJMPD..2330007B}.

Giant radio halos are associated with unrelaxed clusters
\citep[e.g.,][]{2010ApJ...721L..82C,2013ApJ...777..141C,2015A&A...579A..92K,2015A&A...575A.127P,2015A&A...580A..97C,2015ApJ...813...77Y}
without a central cool core \citep[]{2013A&A...556A..44R}, with a few
possible outliers \citep[Sommer et al.
2017]{2014MNRAS.444L..44B,2016MNRAS.459.2940K}.  Minihalos, instead, are
typically found in globally relaxed, cool-core clusters
\citep[e.g.,][]{2002A&A...386..456G,2009A&A...499..371G,2013ApJ...777..163H,2014ApJ...781....9G,
  2014ApJ...786L..17V, 2015A&A...579A..92K,2015ApJ...813...77Y}.  Their
emission envelops the central active radio galaxy, nearly always found at
the center of a cool-core cluster (e.g., Mittal et al.  2009), and extends
quite far from it ($r\sim$50--300 kpc), typically filling the cluster
cooling region. Minihalos are faint and usually quite amorphous in shape,
thus very different from typical extended radio galaxies with lobes and
jets.  Minihalos also differ from the {\em dying} radio galaxies that are
sometimes found at the cluster centers, whose extended emission typically
has an ultra-steep radio spectrum ($\alpha_{\rm radio}\gax 2$) and a
morphology that somehow preserves the original lobed structure of the active
phase, when the source was still fed by the central nucleus (e.g., Saikia \&
Jamrozy 2009, Murgia et al. 2011, 2012).

Minihalos often appear bounded by one or two X-ray cold fronts
\citep{2008ApJ...675L...9M,2014ApJ...781....9G,2014ApJ...795...73G} that
result from sloshing of the cool gas in the central core
\citep[e.g.,][]{2006ApJ...650..102A}.  Based on this coincidence, it has
been proposed that minihalos arise from the reacceleration of seed
relativistic electrons in the magnetized cool core (Gitti et al. 2002, 2004)
by sloshing-induced turbulence
\citep{2008ApJ...675L...9M,2013ApJ...762...78Z,2014ApJ...795...73G}.
Numerical simulations show that sloshing motions can amplify magnetic fields
and develop turbulence in the area enclosed by the cold fronts, which may
lead to the generation of diffuse radio emission confined to the sloshing
region \citep{2013ApJ...762...78Z}. A recent direct measurement of the gas
velocities in the Perseus cool core with the \hitomi\ X-ray satellite
revealed the presence of turbulence sufficient for the above scenario, and
possibly for balancing the radiative cooling in the core
\citep{2016Natur.535..117H}. An alternative explanation of minihalos is
hadronic models (e.g., Pfrommer \& Ensslin 2004; Fujita et al. 2007;
Zandanel et al. 2014), where the radio-emitting electrons are generated
through the interaction of cosmic ray (CR) protons with the thermal protons
in the intracluster medium. Both turbulence and CR protons should also
contribute to balancing the radiative cooling in the cluster cores
\citep[e.g.,][]{2014Natur.515...85Z,2011ApJ...738..182F,
  2008MNRAS.384..251G,2016arXiv160906321J}. Thus the relation between radio
minihalos and cool cores may exist at a fundamental level (e.g., Bravi et
al. 2016, Fujita \& Ohira 2013), making these radio phenomena important for
understanding the physics of cool cores.

\bigskip The study of minihalos has been limited by their small number
\citep[e.g.,][]{2004A&A...417....1G,2008A&A...486L..31C,2009A&A...499..371G},
with only about 10 confirmed detections as of 2011
\citep{2012A&ARv..20...54F}. The number of known minihalos has then rapidly
increased, with 22 confirmed detections%
\footnote{The former minihalo in A\,2390 (Bacchi et al. 2003) is excluded
  because it was recently found to be larger (Sommer et al. 2017),
  borderline between giant halos (as defined in Cassano \& Brunetti 2005)
  and minihalos. We note that this cluster has an unusually large cool core
  (Vikhlinin et al. 2005), comparable in size to the diffuse radio source.}
and 6 candidates up to date \citep[Pandey-Pommier et al.
2016]{2013ApJ...777..163H,2014ApJ...781....9G,2014ApJ...786L..17V,2015A&A...579A..92K},
including two new minihalos and one candidate detection that will be
presented in a forthcoming paper (Giacintucci et al. 2017, hereafter G17).

The new detections have allowed exploratory investigations of the
statistical properties of minihalos and of their host clusters
\citep{2014ApJ...781....9G,2015ApJ...813...77Y,2015A&A...579A..92K,2015aska.confE..76G,2016MNRAS.455L..41B,2016arXiv160906321J,2016arXiv160906322J},
that support the association of minihalos with relaxed, cool-core clusters.
Furthermore, it has been noted that minihalos tend to be observed in the
most massive and hottest clusters with cool cores
\citep{2014ApJ...781....9G}.  If true, this observation will provide a
constraint to discriminate between different models for the minihalo origin.
In a broader context, it can provide information on the energy budget and
physical processes in the cluster cool cores, where the radiative cooling is
believed to be balanced by some non-gravitational heating source(s)
\citep[e.g.,][]{2007ARA&A..45..117M}.  The aim of the present work is to
quantify this observation in a statistical way.  For this purpose, we have
selected a complete, mass-limited sample of 75 clusters from the {\em
  Planck}\/ Sunyaev-Zel'dovich (SZ) cluster catalog
\citep{2014A&A...571A..29P} and performed an X-ray and radio analysis of the
majority of the sample members (77\%) that have {\em Chandra}\/ data. We
complemented this {\em Chandra} statistical sample with
a large additional sample of 48 clusters in the ACCEPT%
\footnote{Archive of Chandra Cluster Entropy Profile Tables.  }
X-ray catalog \citep[][hereafter C09]{2009ApJS..182...12C}, with suitable
radio data, plus the Phoenix cluster \citep[SPT-CL
J2344-4243,][]{2015ApJ...811..111M}, for a total of 106 clusters.  Where
necessary (for 27 clusters), we used the X-ray {\em Chandra} data to
derive the profiles of gas density, temperature and specific entropy.  For
those clusters in both samples without published radio observations on the
minihalo angular scales, we used archival {\em Very Large Array} ({\em VLA})
data to investigate the presence of a minihalo and present the results here
(no minihalos were detected).  We also include in our analysis the 3 new
minihalo detections with the {\em Giant Metrewave Radio Telescope} that that will be 
presented in G17.  We then compared the cluster mass 
and X-ray properties with the presence of a minihalo for our mass-limited as
well as combined samples.

We adopt a $\Lambda$CDM cosmology with H$_0$=70 km s$^{-1}$ Mpc$^{-1}$, 
$\Omega_m=0.3$ and $\Omega_{\Lambda}=0.7$. All errors are quoted at
the $68\%$ confidence level.

\section{Definition of a minihalo}
\label{sec:def}

To distinguish minihalos from other diffuse radio phenomena in
clusters, such as radio galaxies (in active or 
dying phase), large radio halos, and relics, we use the following
physically-motivated definition. A minihalo is a diffuse radio source
at the cluster center with the following properties:

\begin{itemize}

\item[1.] The emission does not consist of radio lobes or tails, nor does it
show, at any angular resolution, any morphological connection (jets) to
the central AGN (such sources would be radio galaxies).

\item[2.] The minimum radius of approximately 50 kpc. At smaller radii,
diffusion and other transport mechanisms, such as sloshing motions,
can plausibly spread the relativistic electrons from the central AGN 
within their synchrotron radiative cooling time (e.g., see \S6.4 in Giacintucci 
et al. 2014a) without the
need for additional physics.

\item[3.] The maximum radius of $0.2R_{500}$.%
\footnote{$R_{500}$ is the radius that encloses a mean overdensity of 500
  with respect to the critical density at the cluster redshift.}
This radius separates two physically distinct cluster regions. Based
on X-ray and SZ observations, density, temperature and pressure
profiles of the thermal intracluster medium (ICM) outside this radius 
are self-similar, whereas a large scatter of the ICM profiles is
observed at $r\lax 0.2R_{500}$ (e.g., McDonald et al. 2017 and
references therein). This is caused by the increased importance of
non-gravitational processes such as cooling, AGN and stellar feedback
in the cores. Thus, diffuse radio emission that is confined within this
radius can have a different origin, possibly related to processes in
the core, from the emission on a larger scale (halos and peripheral
relics). The bulk of the emission of large radio halos originates
well outside $r=0.2R_{500}$ (e.g., Cassano et al.\ 2007).

\end{itemize}

To determine whether a diffuse source fits the above definition we need
radio data of a certain minimum quality. For item 1, we need sufficiently high
resolution to image the range of scales from few tens of kpc to few kpc in order to
determine the morphology of the central radio galaxy and rule out the
possibility of the diffuse emission being part of it. It is also needed to
discriminate between genuine diffuse emission and a blend of individual
radio galaxies in (or projected onto) the core and remove their contribution
as well as that of the central radio galaxy. At the same time, for items 2
and 3, sensitive, lower-resolution radio observations with a good sampling
of the $uv$\/ plane, particularly at short antenna spacings (that
correspond to larger angular scales in the sky), are crucial to detect
diffuse emission on a larger scale, including the scales
beyond our adopted maximum size for the minihalo.

The size of the radio source that we use here is estimated as $R_{\rm
  radio} = \sqrt{R_{\rm max} R_{\rm min}}$, where $R_{\rm max}$ and
$R_{\rm min}$ are the maximum and minimum radii of the $3\sigma$
surface brightness isocontour (Cassano et al.\ 2007). It obviously
depends on the $uv$\/ coverage of the data and the noise level of the
image. As detailed in \S\ref{sec:sample}, a large fraction of the
clusters in our statistical sample have already been reported to have
either a large radio halo (well above $0.2R_{500}$) or a non-detection
based on high-sensitivity low- and high-resolution images, which is
sufficient for our classification purpose.  For the remaining clusters
(except Perseus and Phoenix), which include all minihalos and ambiguous
classifications from the literature, as well as clusters with no published
radio images at the needed resolution, we have uniformly analyzed new and
archival radio data, with results presented in this paper
(\S\ref{sec:radio}), Giacintucci et al.\ (2014a) and G17. The halo
sizes come from those analyses. As shown in Appendices
\ref{sec:imagesr} and \ref{sec:size}, the data used for this paper,
while heterogeneous, have sufficient sensitivity and a range of angular
resolutions, as well as good sampling of short baselines in the
$uv$\/ plane, and thus would allow us to detect diffuse emission at the
typical brightness on the core scale and, in most cases, at the
larger scale of the giant halos. Furthermore, our measured minihalo sizes do
not correlate with the signal-to-noise ratio of the radio
images (Appendix \ref{sec:size}). This gives us assurance that the
radio extent is not determined by the image sensitivity --- the minihalos
are intrinsically smaller than halos (see also Murgia et al.\ 2009).
Of course, we cannot rule out the possibility that deeper data would
uncover large-scale emission much fainter than the current radio halos in some 
of the sources that we classify here as minihalos. However, in a few very
well-observed minihalos, we do see evidence for an abrupt drop of the radio
brightness at a certain radius (Giacintucci et al. 2011, 2014b), which
suggests that the minihalo extent has a physical significance.

\section{Cluster selection}
\label{sec:sample}

To quantify the earlier observation that mihihalos are preferentially found
in massive cool core systems \citep{2014ApJ...781....9G}, in this paper we
use a combination of two samples. One is a statistically complete,
mass-limited sample of massive clusters, while the other is a
similarly-sized arbitrary sample of clusters that do not satisfy some of the
criteria for the complete sample, but do have high-quality radio and X-ray
data and thus can increase the confidence of any correlations that we may
find. The latter sample extends to lower masses, which is obviously helpful
for investigating any correlation of the minihalo occurrence with the
cluster mass. As we will see, the {\em completeness}\/ of the sample at
lower masses is not critical for the conclusions of this work, while the
completeness of the high-mass sample, used for the statistical analysis, is.


\startlongtable
\begin{deluxetable*}{p{1.8cm}ccccrrcccc}
\tabletypesize{\scriptsize}  
\tablecaption{Statistical Sample}
 \tablehead{
    \colhead{Planck name} &
    \colhead{Alternative} &
    \colhead{RA$_{\rm J2000}$} &
    \colhead{DEC$_{\rm J2000}$} &
    \colhead{$z$} &
    \colhead{$M_{500}$} &
    \colhead{$R_{500}$} &
    \colhead{Central diffuse} &
    \colhead{$R_{\rm radio}/R_{500}$} &
    \colhead{Radio} 
    \\[0.5mm]
    \colhead{PSZ1}  &
    \colhead{name} & 
    \colhead{(h,m,s)} &
    \colhead{(deg, $^{\prime}, ^{\prime \prime}$)} &
    \colhead{} &
    \colhead{($10^{14}$ $M_{\odot}$)} &
    \colhead{(Mpc)} &
    \colhead{radio emission} &
    \colhead{} &
    \colhead{reference} 
}
\startdata
\multicolumn{10}{c}{Clusters with {\em Chandra} observations} \\
\hline\noalign{\smallskip}
G009.02$-$81.22 &  A\,2744             & 00 14 13.3 & $-$30 22 31 & 0.307 & $9.6^{+0.5}_{-0.5}$  & 1.35 &  halo         & 0.6      & 3, 4 \\
G116.90$-$53.55 &  A\,68               & 00 36 57.7 &   +09 08 37 & 0.255 & $6.2^{+0.6}_{-0.7}$  & 1.19 &  no detection & \nodata  & 5 \\
G106.84$-$83.24 &  A\,2813             & 00 43 27.4 & $-$20 37 27 & 0.292 & $9.2^{+0.5}_{-0.5}$  & 1.34 &  no detection & \nodata  & 6  \\
G212.97$-$84.04 &  A\,2895             & 01 18 10.8 & $-$26 58 28 & 0.228 & $6.1^{+0.5}_{-0.5}$  & 1.19 &  no detection & \nodata  & 6 \\
G159.81$-$73.47 &  A\,209              & 01 31 53.4 & $-$13 34 27 & 0.206 & $8.2^{+0.4}_{-0.4}$  & 1.33 &  halo         & 0.4      & 7, 8 \\ 
G138.35$-$39.80 &  RXC\,J0142.0+2131   & 01 42 11.6 & +21 32 32   & 0.280 & $6.1^{+0.8}_{-0.8}$  & 1.17 &  no detection & \nodata  & 9 \\
G210.08$-$60.96 & MACS\,J0257.6$-$2209 & 02 57 40.3 & $-$22 09 46 & 0.322 & $7.2^{+0.7}_{-0.7}$  & 1.22 & \phantom{000}halo (c) & \nodata & 10, 11 \\
G164.20$-$38.90 &  A\,401              & 02 58 54.8 & +13 32 24   & 0.074 & $6.8^{+0.3}_{-0.3}$  & 1.31 &  halo         & 0.3      & 2, 12 \\
G223.91$-$60.09 &  A\,3088             & 03 07 03.2 & $-$28 40 24 & 0.254 & $6.7^{+0.6}_{-0.6}$  & 1.22 &  no detection & \nodata  & 8 \\
G171.96$-$40.64 &   \nodata            & 03 13 00.3 & +08 22 53   & 0.270 & $11.1^{+0.6}_{-0.6}$ & 1.44 &  halo         & 0.4      & 13 \\
G182.42$-$28.28 &             A\,478   & 04 13 25.2 & +10 28 19   & 0.088 & $7.1^{+0.3}_{-0.4}$  & 1.32 &  minihalo     & 0.1      & 14 \\
G208.80$-$30.67 &             A\,521   & 04 54 05.0 & $-$10 13 35 & 0.248 & $6.9^{+0.6}_{-0.6}$  & 1.24 &  halo         & 0.6      & 15, 16 \\
G195.78$-$24.29 &             A\,520   & 04 54 15.9 & +02 57 10   & 0.203 & $7.1^{+0.6}_{-0.6}$  & 1.27 &  halo         & 0.4      & 17 \\ 
G208.59$-$26.00 &  RXC\,J0510.7$-$0801 & 05 10 44.3 & $-$08 01 12 & 0.220 & $7.4^{+0.6}_{-0.6}$  & 1.28 &  bad data     & \nodata  & 5 \\
G215.29$-$26.09 &  RXC\,J0520.7$-$1328 & 05 20 47.2 & $-$13 30 08 & 0.336 & $6.1^{+0.8}_{-0.8}$  & 1.15 &  bad data     & \nodata  & 18 \\ 
G139.61+24.20 &       \nodata          & 06 22 13.9 & +74 41 39   & 0.267 & $7.1^{+0.6}_{-0.6}$  & 1.24 &  minihalo (c) & 0.04     & 2 \\ 
G149.75+34.68 &             A\,665     & 08 30 50.9 & +65 52 01   & 0.182 & $8.2^{+0.4}_{-0.4}$  & 1.34 &  halo         & 0.5      & 19 \\
G186.37+37.26 &             A\,697     & 08 42 59.6 & +36 21 10   & 0.282 & $11.5^{+0.5}_{-0.5}$ & 1.45 &  halo         & 0.4      & 20, 6, 7 \\
G239.29+24.75 &             A\,754     & 09 08 56.2 & $-$09 40 21 & 0.054 & $6.7^{+0.2}_{-0.2}$  & 1.31 &  halo         & 0.5      & 21, 12, 22 \\
G166.11+43.40 &             A\,773     & 09 18 04.5 & +51 42 15   & 0.217 & $7.1^{+0.4}_{-0.5}$  & 1.26 &  halo         & 0.4      & 3 \\
G195.60+44.03 & A\,781                 & 09 20 16.0 & +30 29 56   & 0.295 & $6.4^{+0.6}_{-0.6}$  & 1.18 & \phantom{000}halo (c)    & 0.4 & 23, 24  \\
G135.03+36.03 &           RBS\,797     & 09 47 00.2 & +76 23 44   & 0.345 & $6.3^{+0.6}_{-0.7}$  & 1.16 &  minihalo     & 0.1      & 2, 25, 26 \\
G266.85+25.06 & A\,3444                & 10 23 54.8 & $-$27 17 09 & 0.254 & $7.6^{+0.5}_{-0.6}$  & 1.27 &  minihalo     & 0.1      & 2, 1, 5, 6 \\
G149.21+54.17 & A\,1132                & 10 58 25.9 & +56 48 09   & 0.137 & $6.2^{+0.3}_{-0.3}$  & 1.24 &  no detection & \nodata  & 19 \\
G257.13+55.63 &  RXC\,J1115.8+0129     & 11 15 54.9 & +01 29 56   & 0.350 & $6.4^{+0.7}_{-0.7}$  & 1.16 &  minihalo (c) & \nodata  & 27 \\
G278.58+39.15 & A\,1300                & 11 31 55.8 & $-$19 55 42 & 0.308 & $8.8^{+0.6}_{-0.6}$  & 1.31 &  halo         & 0.4      & 4, 28, 29 \\
G139.17+56.37 & A\,1351                & 11 42 24.5 & +58 31 41   & 0.322 & $7.1^{+0.5}_{-0.5}$  & 1.21 &  halo         & 0.4      & 30, 7 \\
G180.56+76.66 &            A\,1423     & 11 57 19.9 & +33 36 39   & 0.214 & $6.1^{+0.5}_{-0.5}$  & 1.20 &  no detection & \nodata  & 6 \\
G229.70+77.97 &  A\,1443               & 12 01 21.1 & +23 06 31   & 0.269 & $7.7^{+0.5}_{-0.6}$  & 1.27 &  halo         & 0.5      & 31 \\
G289.19+72.19 & RXC\,J1234.2+0947      & 12 34 31.8 & +09 46 23   & 0.229 & $6.0^{+0.6}_{-0.6}$  & 1.19 & \phantom{000}halo (c)    &  0.3 & 5 \\
G114.99+70.36 & A\,1682                & 13 06 54.9 & +46 31 33   & 0.226 & $6.2^{+0.4}_{-0.5}$  & 1.20 &  halo (c)     & 0.4      & 32, 4, 33 \\
G313.33+61.13 &            A\,1689     & 13 11 26.5 & $-$01 20 11 & 0.183 & $8.9^{+0.4}_{-0.4}$  & 1.38 &  halo         & 0.5      & 2, 34 \\
G323.30+63.65 & A\,1733                & 13 27 00.7 & +02 12 14   & 0.259 & $7.1^{+0.6}_{-0.7}$  & 1.24 &  no data      & \nodata  & \nodata \\ 
G107.14+65.29 &            A\,1758a    & 13 32 39.5 & +50 32 47   & 0.280 & $8.0^{+0.4}_{-0.5}$  & 1.28 &  halo         &  0.6     & 4,7\\
G092.67+73.44 &            A\,1763     & 13 35 18.1 & +41 00 10   & 0.228 & $8.3^{+0.4}_{-0.4}$  & 1.32 &  no detection & \nodata  & 6 \\
G340.37+60.57 &            A\,1835     & 14 01 02.7 & +02 51 56   & 0.253 & $8.5^{+0.5}_{-0.6}$  & 1.32 &  minihalo     & 0.18     & 2, 35 \\
G067.19+67.44 &            A\,1914     & 14 26 03.9 & +37 49 35   & 0.171 & $7.0^{+0.4}_{-0.4}$  & 1.28 &  halo         & 0.5      & 12 \\ 
G340.94+35.10 &            AS\,780     & 14 59 30.4 & $-$18 08 58 & 0.236 & $7.7^{+0.6}_{-0.6}$  & 1.29 &  minihalo     & 0.04     & 5, 2, 8 \\
G355.07+46.20 &  RXC\,J1504.1$-$0248   & 15 04 05.4 & $-$02 47 54 & 0.215 & $7.0^{+0.6}_{-0.6}$  & 1.26 &  minihalo     & 0.11     & 36  \\ 
G006.45+50.56 &            A\,2029     & 15 10 50.8 & +05 44 43   & 0.077 & $6.8^{+0.2}_{-0.2}$  & 1.30 &  minihalo     & 0.19     & 2, 35 \\
G346.61+35.06 & RXC\,J1514.9$-$1523    & 15 15 00.4 & $-$15 21 29 & 0.223 & $8.3^{+0.5}_{-0.6}$  & 1.33 &  halo         & 0.5      & 37 \\
G044.24+48.66 &            A\,2142     & 15 58 25.6 & +27 14 25   & 0.089 & $8.8^{+0.3}_{-0.3}$  & 1.42 &  halo         & 0.3      & 38, 39 \\
G006.76+30.45 &            A\,2163     & 16 15 49.2 & $-$06 09 09 & 0.203 & $16.4^{+0.4}_{-0.4}$ & 1.68 &  halo         & 0.7      & 40 \\
G021.10+33.24 &            A\,2204     & 16 32 47.8 & +05 35 32   & 0.151 & $8.0^{+0.4}_{-0.4}$  & 1.34 &  minihalo     & 0.04     & 14 \\
G097.72+38.13 &            A\,2218     & 16 35 52.0 & +66 11 44   & 0.171 & $6.4^{+0.3}_{-0.3}$  & 1.24 &  halo         & 0.3      & 2, 19 \\
G072.61+41.47 &            A\,2219     & 16 40 18.6 & +46 41 55   & 0.228 & $11.0^{+0.4}_{-0.4}$ & 1.45 &  halo         & 0.6      & 12 \\
G110.99+31.74 &            A\,2256     & 17 04 08.1 & +78 38 07   & 0.058 & $6.3^{+0.2}_{-0.2}$  & 1.28 &  halo         & 0.4      & 41, 42  \\
G049.22+30.84 &  RXC\,J1720.1+2637     & 17 20 12.6 & +26 37 23   & 0.164 & $6.3^{+0.4}_{-0.4}$  & 1.24 &  minihalo     & 0.1      & 43 \\
G055.58+31.87 &            A\,2261     & 17 22 21.9 & +32 07 58   & 0.224 & $7.4^{+0.4}_{-0.5}$  & 1.28 &  halo         & 0.4      & 44 \\
G094.00+27.41 & CL\,1821+643           & 18 22 00.4 & +64 20 34   & 0.332 & $6.3^{+0.4}_{-0.4}$  & 1.16 &  halo         & 0.4      & 45, 46 \\
G018.54$-$25.70 &  RXC\,J2003.5$-$2323 & 20 03 32.3 & $-$23 23 30 & 0.317 & $7.5^{+0.6}_{-0.7}$  & 1.24 &  halo         & 0.7      & 47 \\
G053.42$-$36.25 & MACS\,J2135.2$-$0102 & 21 35 10.1 & $-$01 03 15 & 0.330 & $7.6^{+0.6}_{-0.6}$  & 1.24 &  no data    & \nodata & \nodata  \\ 
G055.95$-$34.87 & A\,2355              & 21 35 13.6 & +01 25 40   & 0.231 & $6.9^{+0.5}_{-0.5}$  & 1.24 &  no data      & \nodata & \nodata \\
G073.98$-$27.83 &            A\,2390   & 21 53 44.0 & +17 41 35   & 0.233 & $9.5^{+0.4}_{-0.4}$  & 1.38 &  halo         & 0.3     & 44  \\
G073.85$-$54.94 &            A\,2537   & 23 08 28.1 & $-$02 12 00 & 0.297 & $6.2^{+0.6}_{-0.7}$  & 1.17 &  no detection & \nodata &  6  \\
G081.01$-$50.92 & A\,2552              & 23 11 36.3 & +03 38 38   & 0.300 & $7.5^{+0.6}_{-0.6}$  & 1.25 &  \phantom{000}halo (c)  & 0.3 & 5 \\
G087.03$-$57.37 &            A\,2631   & 23 37 43.7 & +00 16 06   & 0.278 & $7.0^{+0.6}_{-0.6}$  & 1.23 &   no detection& \nodata & 6 \\
G034.03$-$76.59 &            A\,2667   & 23 51 38.3 & $-$26 04 45 & 0.226 & $6.8^{+0.5}_{-0.5}$  & 1.24 &   minihalo    & 0.05    & 2 \\
\hline\noalign{\smallskip}
\multicolumn{10}{c}{Clusters without {\em Chandra} observations} \\ 
\hline\noalign{\smallskip}
G092.10$-$66.02 &  A\,2697             & 00 03 05.5 & $-$06 05 26 & 0.232 & $6.0^{+0.6}_{-0.6}$  & 1.19 & no detection&\nodata & 8       \\ 
G110.08$-$70.23 &  A\,56               & 00 34 01.6 & $-$07 47 45 & 0.300 & $6.2^{+0.7}_{-0.7}$  & 1.17 & no data     &\nodata & \nodata    \\
G114.34$-$60.16 &  RXC\,J0034.4+0225   & 00 34 23.6 &   +02 25 14 & 0.350 & $6.5^{+0.7}_{-0.8}$  & 1.17 & no data     &\nodata & \nodata        \\
G142.18$-$53.27 &  A\,220              & 01 37 22.0 & +07 52 31   & 0.330 & $6.7^{+0.8}_{-0.9}$  & 1.19 & no data     &\nodata & \nodata  \\ 
G222.97$-$65.69 &  A\,3041             & 02 41 27.6 & $-$28 38 51 & 0.232 & $6.1^{+0.5}_{-0.6}$  & 1.19 & no data     &\nodata & \nodata   \\
G205.07$-$62.94 &   \nodata            & 02 46 27.6 & $-$20 32 05 & 0.310 & $7.4^{+0.6}_{-0.7}$  & 1.24 & no detection &\nodata & 48  \\
G176.25$-$52.57 &  A\,384              & 02 48 13.1 & $-$02 14 21 & 0.236 & $6.4^{+0.6}_{-0.6}$  & 1.21 & no data     &\nodata &\nodata  \\ 
G172.93+21.31   &  \nodata             & 07 07 37.2 & +44 19 23   & 0.331 & $6.1^{+0.8}_{-0.8}$  & 1.15 & no data     &\nodata & \nodata\\
G169.64+33.84   &   \nodata            & 08 16 42.5 & +49 31 48   & 0.347 & $6.2^{+0.8}_{-0.9}$  & 1.15 & no data     &\nodata & \nodata\\
G227.55+54.88   &  ZwCl\,1028.8+1419   & 10 31 21.1 & +14 06 19   & 0.305 & $6.1^{+0.7}_{-0.7}$  & 1.16 & no data     &\nodata & \nodata\\
G288.26+39.94   &  RXC\,J1203.2$-$2131 & 12 03 14.4 & $-$21 33 02 & 0.192 & $7.3^{+0.5}_{-0.5}$  & 1.28 & no data     &\nodata & \nodata \\ 
G304.76+69.84   &    \nodata           & 12 53 58.8 & +06 58 40   & 0.346 & $6.2^{+0.7}_{-0.8}$  & 1.15 & no data     &\nodata & \nodata\\
G309.46+37.32   &  RXC\,J1314.4$-$2515 & 13 14 23.1 & $-$25 15 09 & 0.244 & $6.2^{+0.7}_{-0.7}$  & 1.19 & halo        &  0.3  & 8,49 \\ 
G068.32+81.81   &  RXC\,J1322.8+3138   & 13 22 48.0 & +31 39 06   & 0.308 & $6.6^{+0.6}_{-0.6}$  & 1.19 & no data     &\nodata & \nodata \\ 
G019.12+31.23   &    \nodata           & 16 36 29.4 & +03 08 51   & 0.280 & $7.1^{+0.6}_{-0.7}$  & 1.23 & no data     &\nodata & \nodata \\  
G049.83$-$25.22 &  RXC\,J2051.1+0216   & 20 51 20.3 & +02 16 40   & 0.321 & $6.1^{+0.7}_{-0.7}$  & 1.15 & no data     &\nodata & \nodata \\ 
G084.20$-$35.49 &            A\,2472   & 22 42 20.6 & +17 29 17   & 0.314 & $6.2^{+0.7}_{-0.8}$  & 1.16 & no data     &\nodata & \nodata \\ 
\enddata
\tablecomments{
Column 1: {\em Plank} cluster name. Column 2: alternative name. Columns 3--6: cluster coordinates, redshift and mass from Planck collaboration et al. (2014). 
Column 7: $R_{500}$, derived from $M_{500}$. Column 8: type of diffuse radio emission at the cluster center (c indicates a candidate detection). 
As defined in \S\ref{sec:def}, sources with $R_{\rm radio} \le 0.2 R_{\rm 500}$ are classified as minihalos. Column 9: radius of the central diffuse 
radio source, as defined in \S\ref{sec:def}, in units of $R_{500}$. Column 10: radio references. If more than one reference is given, 
the first one is the reference for the image used to measure the radio size. Reference code: 
(1) this work, 
(2) G17, 
(3) Govoni et al. (2001), 
(4) Venturi et al. (2013), 
(5) Kale et al. (2015),
(6) Venturi et al. (2008),
(7) Giovannini et al. (2009),
(8) Venturi et al. (2007),
(9) Kale et al. (2013),
(10) Venturi et al., in preparation,
(11) Bonafede et al., (private communication),
(12) Bacchi et al. (2013),
(13) Giacintucci et al. (2013),
(14) Giacintucci et al. (2014a),
(15) Brunetti et al. (2008),
(16) Dallacasa et al.(2209),
(17) Vacca et al. (2014),
(18) Macario et al. (2014),
(19) Giovannini \& Feretti (2000),
(20) Macario et al. (2011),
(21) Macario et al. (2011),
(22) Kassim et al. (2001),
(23) Govoni et al. (2011), 
(24) Venturi et al. (2011a),
(25) Gitti et al. (2006),
(26) Doria et al. (2012),
(27) Pandey-Pommier et al. (2016),
(28) Reid et al. (1999),
(29) Parekh et al. (2017),
(30) Giacintucci et al. (2009a),
(31) Bonafede et al. (2015),
(32) Venturi et al. (2011b),
(33) Macario et al. (2013),
(34) Vacca et al. (2011),
(35) Govoni et al. (2009),
(36) Giacintucci et al. (2011a),
(37) Giacintucci et al. (2011b),
(38) Venturi et al. (2017),
(39) Farnsworth et al. (2013),
(40) Feretti et al. (2001),
(41) Brentjens (2008),
(42) Clarke \& Ensslin (2006),
(43) Giacintucci et al. (2014b),
(44) Sommer et al. (2017),
(45) Bonafede et al. (2014),
(46) Kale \& Parekh (2016),
(47) Giacintucci et al. (2009b),
(48) Ferrari et al., (private communication),
(49) Feretti et al. (2005).}
\label{tab:statsample}
\end{deluxetable*}

\subsection{Statistical sample}

A complete essentially mass-limited cluster sample can be extracted from the
{\em Planck} SZ cluster catalog, since the cluster total SZ signal is a good
proxy for the total mass \citep[e.g.,][]{2006ApJ...650..538N}. We selected
all clusters with redshift $z\le 0.35$, Galactic latitude $|b|\ge
20^{\circ}$ and the \planck-estimated total mass within $R_{500}$ of
$M_{500} > 6\times10^{14}$ $M_{\odot}$. This mass cut is well above the
\planck\ completeness limit and, along with the $z$ cut, is a compromise
between the need to cover a range of cluster masses and the availability of
the radio and X-ray data. We also imposed a cut in declination of DEC$_{\rm
  J2000} \ge -30^{\circ}$ to ensure good visibility from the {\em VLA} and
{\em GMRT}, whose observations we use to investigate the presence of diffuse
radio emission at the cluster center.  Finally, we excluded the double
cluster A\,115; its total mass inferred by {\em Planck} is
$7.2\times10^{14}$ $M_{\odot}$ (so above our mass limit), however, based on
optical and X-ray estimates of the cluster mass ratio
\citep{2007A&A...469..861B,1997MNRAS.292..419W}, the mass of each individual
cluster falls below our mass threshold.

Our final \planck\ sample contains 75 clusters, listed in Table
\ref{tab:statsample}. This is essentially the same sample, apart from
slightly different selection criteria, as that used by Cuciti et al. (2015)
to study the occurrence of giant radio halos in clusters. {\em Chandra}\/
X-ray observations are currently available for 58 (77\%) of these clusters;
hereafter we will refer to these clusters as the {\em Chandra}\/ statistical
sample. 53 of these clusters ($91\%$) have published accurate radio
measurements from deep {\em Westerbork Synthesis Radio Telescope (WSRT)},
{\em VLA}, and/or {\em GMRT} observations (Table
\ref{tab:statsample}), most of the latter taken as part of the {\em GMRT}
Radio Halo Survey and its extension (Venturi et al.\ 2007, 2008, Kale et
al.\ 2013, 2015). Two of the remaining clusters --- RXC\,J0510.7--0801 and
RXC\,J0520.7--1328 --- have pointed {\em GMRT} observations; however, the
resulting images are not sensitive enough to investigate the presence of
diffuse radio emission in these systems
\citep{2015A&A...579A..92K,2014A&A...565A..13M}. We will see in
\S\ref{sec:disc} that these two clusters, along with those with no available
radio data (A\,1733, MACS\,J2135.2-0102 and A\,2355), do not possess a cool
core and the absence of radio information will not affect our main findings
that are based on the cool-core part of the sample.

Our statistical sample (Table \ref{tab:statsample}) contains 12 minihalos,
of which 9 are previously known and three --- A\,2667, PSZ1\,G139.61+24.2
(candidate) and RXC\,J1115.8+0129 (candidate) --- are new detections.
The former two are reported in the forthcoming paper G17 and the latter
in \cite{2016arXiv161200225P}. The sample includes 26 radio-halo
clusters and 5 candidates. Many of the clusters also contain 
peripheral radio relics, which is a distinct phenomenon
\citep[e.g.,][]{2010Sci...330..347V} and we do not discuss it in this paper.

Among the 17 clusters in our \planck\ sample that do not have \chandra\ data
(Table \ref{tab:statsample}), 11 have {\em XMM-Newton} observations.
However, most of them do not have high-sensitivity radio data at present.
We inspected the {\em XMM-Newton} images and found that only 3 out of 11 can
possibly have cool cores, thus omitting them does not significantly affect
our statistical conclusions.

\subsection{Supplementary sample}\label{sec:supp}

We supplement our statistical sample with additional 48 clusters drawn from
ACCEPT, which includes clusters observed with \chandra\ as of 2008 and
presents uniform X-ray analysis for them (C09), suitable for our work.  The
clusters were required to have deep, pointed {\em VLA}, {\em WSRT} and
{\em GMRT} observations in
the literature and/or in the data archives (see \S6), be at $z\le 0.5$,
DEC$_{\rm J2000}\ge -30^{\circ}$ and an average temperature of $kT>3.5$ keV.
This temperature cut corresponds to a lower total mass ($\sim
2\times10^{14}$ $M_{\odot}$, based on an $M_{\rm 500}-T_{\rm X}$\/ relation,
Vikhlinin et al.\ 2009) than the lower limit of our statistical sample. To
those clusters we add the Phoenix cluster ($z=0.6$), which hosts the most
distant minihalo found to date \citep{2014ApJ...786L..17V}. This
supplementary sample is given in Table\ 2. Twelve of these clusters possess
a minihalo, including a recent {\em GMRT} detection in A907 (G17), and 4
host a candidate minihalo. The presence of a central diffuse radio source
has been reported in 3 more clusters in this sample --- ZwC\,1742.1+3306,
MACS\,J1931.8--2634 (Giacintucci et al. 2014a) and A\,2626 (Gitti 2013) ---
but their classification as a minihalo is uncertain: the size ($<50$ kpc 
in A\,2626 and ZwC\,1742.1+3306), radio morphology and ultra-steep
spectrum ($\alpha_{\rm radio}\sim2$) of these sources, as well as the
possible association with X-ray cavities in MACS\,J1931.8--2634, suggest
that they could instead be dying/restarted radio galaxies (Giacintucci et
al. 2014a), whose aged emission, no longer fed by the central nucleus, is
rapidly fading. For 9 clusters with no published radio results on the
minihalo angular scales, we analyzed {\em VLA}\/ archival observations
(\S\ref{sec:radio}) and present the results here (no minihalos were detected
among them).

Our final combined sample (statistical + supplementary) consists of 106
clusters and includes all 28 known minihalos, including 6 candidates.

\startlongtable
\begin{deluxetable*}{p{2cm}ccccrrcccc}
\tabletypesize{\scriptsize}  
\tablecaption{Supplementary Sample}
 \tablehead{
    \colhead{Cluster} &
    \colhead{Planck name} &
    \colhead{RA$_{\rm J2000}$} &
    \colhead{DEC$_{\rm J2000}$} &
    \colhead{$z$} &
    \colhead{$M_{500}$} &
    \colhead{$R_{500}$} &
    \colhead{Central diffuse} &
    \colhead{$R_{\rm radio}/R_{500}$} &
    \colhead{Radio} 
    \\[0.5mm]
    \colhead{name}  &
    \colhead{PSZ1} & 
    \colhead{(h,m,s)} &
    \colhead{(deg, $^{\prime}, ^{\prime \prime}$)} &
    \colhead{} &
    \colhead{($10^{14}$ $M_{\odot}$)} &
    \colhead{(Mpc)} &
    \colhead{radio emission} &
    \colhead{} &
    \colhead{reference} 
}
\startdata
Z348               &\nodata         &  01 06 50.3  & +01 03 17    & 0.255  & $2.5\pm0.3^{\star}$              & 0.88 &  no detection  & \nodata & 3  \\
A\,119             &G125.68$-$64.12 &  00 56 14.5  & $-$01 16 55  & 0.044  & $3.34^{+0.22}_{-0.23}$\phantom{0} & 1.04 &  no detection  & \nodata & 4 \\
A\,141             &G175.59$-$85.95 &  01 05 34.6  & $-$24 38 00  & 0.230  & $4.48^{+0.6}_{-0.73}$\phantom{0}  & 1.08 &  no detection  & \nodata & 3 \\ 
A\,193             &G136.90$-$53.31 &  01 24 59.4  & +08 38 43    & 0.049  & $1.8^{+0.27}_{-0.29}$\phantom{0}  & 0.85 &  no detection  & \nodata & 1 \\
A\,267             &G153.07$-$58.27 &  01 52 41.9  & +00 58 01    & 0.227  & $4.95^{+0.67}_{-0.72}$\phantom{0} & 1.11 &  no detection  & \nodata & 3 \\
MACS\,J0159.8-0849 &G167.63$-$65.57 &  01 59 54.5  & $-$08 50 14  & 0.405  & $6.88^{+0.90}_{-0.98}$\phantom{0} & 1.16 &  minihalo      & 0.1     & 5, 2 \\
A\,383             &G177.64$-$53.52 &  02 47 46.5  & $-$03 29 56  & 0.188  & $4.47^{+0.60}_{-0.64}$\phantom{0} & 1.09 &  no detection  & \nodata & 1 \\
A\,399             &G164.31$-$39.43 &  02 57 52.7  & +13 04 11    & 0.072  & $5.29^{+0.34}_{-0.35}$\phantom{0} & 1.20 &  halo          & 0.25    & 6 \\
Perseus            &\nodata         &  03 19 47.2  & +41 30 47    & 0.018  & $6.1\pm0.6^{\star}$              & 1.28 &  minihalo      & 0.1     & 7, 8, 9 \\
MACS\,J0329.6-0211 &\nodata         &  03 29 41.5  & $-$02 11 46  & 0.450  & $4.9\pm0.7^{\star}$              & 1.02 &  minihalo      & 0.1     & 2, 5 \\
2A\,0335+096       &G176.30$-$35.06 &  03 38 44.4  & +09 56 34    & 0.035  & $2.27^{+0.24}_{-0.25}$\phantom{0} & 0.92 &  minihalo      & 0.1     & 2, 10  \\
MACS\,J0417.5-1154 &G205.94$-$39.46 &  04 17 36.2  & $-$11 54 12  & 0.443  & $11.70^{+0.64}_{-0.66}$\phantom{0}& 1.37 &  halo          & 0.3    & 11 \\
MACS\,J0429.6-0253 &\nodata             &  04 29 36.0  & $-$02 53 08  & 0.399  & $4.1\pm0.8^{\star}$          & 0.98 &  no detection  & \nodata & 1 \\
RX\,J0439.0+0715   & G189.52$-$25.10 & 04 39 01.2  & +07 15 36    & 0.244  & $5.75^{+0.70}_{-0.74}$            & 1.16 &  no detection  & \nodata & 3 \\
MS\,0440.5+0204    &\nodata             &  04 43 09.7  & +02 10 19    & 0.190  & $5.0\pm1.2^{\star}$          & 1.13 &  no detection  & \nodata & 1 \\
A\,611             &G184.70+28.92   &  08 01 01.7  & +36 05 06    & 0.288  & $5.85^{+0.60}_{-0.64}$\phantom{0} & 1.15 &  no detection  & \nodata & 4  \\
MS\,0839.8+2938    &\nodata             &  08 42 55.9  & +29 27 26    & 0.194  & $3.4\pm0.5^{\star}$          & 1.00 &  no detection  & \nodata & 1 \\
Z\,2089            &\nodata             &  09 00 37.9  & +20 54 58    & 0.235  & $3.2\pm0.4^{\star}$          & 0.96 &  no detection  & \nodata & 3, 4 \\
ZwCl\,2701         &\nodata             &  09 52 49.2  & +51 53 05    & 0.214  & $4.0\pm0.5^{\star}$          & 1.04 &  no detection  & \nodata & 3, 4 \\
A\,907             &G249.38+33.27   &  09 58 22.2  & $-$11 03 35  & 0.167  & $5.18^{+0.47}_{-0.50}$            & 1.16 &  minihalo      & 0.05    & 2 \\
ZWCL\,3146         &\nodata             &  10 23 39.6  & +04 11 10    & 0.291  & $6.7\pm0.8^{\star}$          & 1.20 &  minihalo      & 0.07    & 12, 5 \\
A\,1068            &G179.13+60.14   &  10 40 48.7  & +39 56 05    & 0.137  & $3.55^{+0.38}_{-0.41}$\phantom{0} & 1.03 & \phantom{000}minihalo (c) & 0.10 & 13  \\
A\,1204            &\nodata             &  11 13 32.2  & +17 35 40    & 0.171  & $2.4\pm0.3^{\star}$          & 0.89 &  no detection  & \nodata & 1 \\
A\,1240            &\nodata             &  11 23 32.1  & +43 06 32    & 0.159  & $2.6\pm0.4^{\star}$          & 0.92 &  no detection  & \nodata & 14 \\ 
A\,1413            &G226.19+76.78   &  11 55 19.4  & +23 24 26    & 0.143  & $5.98^{+0.38}_{-0.40}$\phantom{0} & 1.22 &  \phantom{000}minihalo (c) & 0.09 & 13 \\
A\,1576            &G125.72+53.87   &  12 36 48.9  & +63 10 40    & 0.302  & $5.98^{+0.48}_{-0.50}$\phantom{0} & 1.16 &  no detection  & \nodata & 3 \\
A\,1650            &G306.71+61.04   &  12 58 45.6  & $-$01 46 11  & 0.085  & $4.00^{+0.34}_{-0.36}$\phantom{0} & 1.09 &  no detection  & \nodata & 13 \\
RX\,J1347.5--1145  &G324.05+48.79   &  13 47 33.5  & $-$11 45 42  & 0.452  & $10.61^{+0.74}_{-0.77}$\phantom{0}& 1.32 &  minihalo      & 0.2    & 2, 15, 16 \\
A\,1795            &G033.84+77.17   &  13 48 55.0  & +26 36 01    & 0.062  & $4.54^{+0.21}_{-0.21}$\phantom{0} & 1.15 &  \phantom{000}minihalo (c) & 0.09 & 5  \\
A\,1995            &G096.87+52.48   &  14 52 56.4  & +58 03 35    & 0.318  & $5.15^{+0.49}_{-0.52}$\phantom{0} & 1.09 &  halo          & 0.3    & 17 \\
MS\,1455.0+2232    &\nodata             &  14 57 15.1  & +22 20 34    & 0.258  & $3.46\pm0.35^{\star}$        & 0.98 &  minihalo      & 0.1    & 2, 4, 18 \\
A\,2034            &G053.52+59.52   &  15 10 12.6  & +33 29 21    & 0.113  & $4.92^{+0.35}_{-0.36}$\phantom{0} & 1.16 &  halo          & 0.3    & 17 \\
RX\,J1532.9+3021   &\nodata         &  15 32 53.8  & +30 20 58    &\phantom{0}0.363\tablenotemark{a}& $4.7\pm0.6^{\star}$ & 1.04 & minihalo & 0.10 & 5, 3, 19 \\
A\,2111            &G054.99+53.42   &  15 39 34.9  & +34 25 46    & 0.229  & $5.46^{+0.60}_{-0.64}$\phantom{0} & 1.15 &  no detection  & \nodata & 4 \\
A\,2125            &\nodata             &  15 40 58.3  & +66 18 28    & 0.247  & $1.6\pm0.3^{\star}$          & 0.76 &  no detection  & \nodata & 1 \\
Ophiuchus          &\nodata             &  17 12 25.9  & $-$23 22 33  & 0.028  & $12.4\pm1.2^{\star}$         & 1.62 &  minihalo      &  0.15   & 2, 13, 20 \\
A\,2255            &G093.93+34.92   &  17 12 48.4  & +64 04 03    & 0.081  & $5.19^{+0.19}_{-0.19}$\phantom{0} & 1.19 &  halo          &  0.7    & 21, 22, 23  \\ 
RX\,J1720.2+3536   &G059.51+33.06   &  17 20 20.6  & +35 37 42    & 0.387  & $6.04^{+0.69}_{-0.74}$\phantom{0} & 1.12 & \phantom{000}minihalo (c)& 0.18 & 24 \\
ZwCl\,1742.1+3306  &G057.91+27.62   &  17 44 19.6  & +32 59 19    & 0.076  & $2.63^{+0.27}_{-0.29}$\phantom{0} & 0.95 &  uncertain    & 0.04     & 5 \\
A\,2319            &G075.71+13.51   &  19 21 09.6  & +43 58 30    & 0.056  & $8.59^{+0.22}_{-0.22}$\phantom{0} & 1.42 &  halo         & 0.4      & 25, 26, 27  \\
MACS\,J1931.8-2634 &G012.58$-$20.07 &  19 31 46.0  & -26 33 51    & 0.352  & $6.19^{+0.77}_{-0.83}$\phantom{0} & 1.15 &  uncertain    & 0.1      & 5, 28 \\
RX\,J2129.6+0005   &G053.65$-$34.49 &  21 29 42.5  & +00 04 51    & 0.235  & $4.24^{+0.55}_{-0.59}$\phantom{0} & 1.06 &  minihalo     & 0.08      & 12, 2 \\
A\,2420            &G046.48$-$49.42 &  22 10 12.9  & $-$12 09 51  & 0.085  & $4.48^{+0.26}_{-0.27}$\phantom{0} & 1.13 &  no detection & \nodata  & 1 \\
MACS\,J2228.5+2036 &G083.30$-$31.01 &  22 28 29.1  & +20 38 22    & 0.412  & $7.82^{+0.71}_{-0.75}$\phantom{0} & 1.21 &  no detection & \nodata  & 4, 11 \\
MACS\,J2245.0+2637 &\nodata         &  22 45 04.7  & +26 38 04    & 0.304  & $4.8\pm0.8^{\star}$              & 1.07 &  no detection & \nodata  & 29  \\
A\,2556            &\nodata         &  23 13 00.9  & $-$21 37 55  & 0.087  & $2.4\pm0.2^{\star}$              & 0.92 &  no detection & \nodata  & 1 \\
A\,2626            &\nodata         &  23 36 30.3  & +21 08 33    & 0.055  & $2.4\pm0.5^{\star}$              & 0.93 &  uncertain    & 0.03     & 30 \\
Phoenix\tablenotemark{b}& \nodata   &  23 44 42.2  & $-$42 43 08  & 0.597  & $12.6^{+2.0}_{-1.5}$\phantom{0}   & 1.32 &  minihalo     & 0.17     & 31 \\
\enddata
\tablenotetext{a}{Crawford et al. (1999).}
\tablenotetext{b}{SPT-CL J2344-4243: coordinates, redshift and mass are from McDonald et al. (2015) and references therein.}

\tablecomments{Column 1: cluster name. 
Column 2: {\em Planck} name. 
Columns 3--5: cluster coordinates and redshift from Planck collaboration et al. (2014) for the {\em Planck} clusters and 
NASA/IPAC Extragalactic Database for the others.  
Column 6: cluster mass from Planck collaboration et al. (2014). Values marked with 
$^\star$ were estimated from the $M_{\rm 500}-T_X$ relation of Vikhlinin et
al. (2009) using the core-excised temperatures in Table \ref{tab:k0}; errors
were calculated from the temperature uncertainties and include statistical
and systematic uncertainties for the $M_{\rm 500}-T_{\rm X}$\/ relation
itself (\S\ref{sec:xrayprop}).
Column 8: $R_{\rm 500}$, derived from $M_{\rm 500}$. Column 7: type
of diffuse radio emission at the cluster center (c indicates a candidate
detection). As defined in \S\ref{sec:def}, sources with $R_{\rm radio}
  \le 0.2 R_{\rm 500}$ are classified as minihalos. Clusters marked as {\em
  uncertain} host central extended radio sources whose classification as a
minihalo is uncertain; the radio size, morphology and ultra-steep spectrum
of these sources suggest that they could be instead dying/restarted radio
galaxies (\S\ref{sec:supp}; Giacintucci et al. 2014a).
Column 9: radius of the central diffuse radio source, as defined in \S\ref{sec:def}, in units of $R_{\rm 500}$.
Column 10: Radio references. If more than one reference is given, the
  first one is the reference for the image used to measure the radio size.
Reference code:
(1) this work, 
(2) G17,
(3) Kale et al. (2013),
(4) Venturi et al. (2008),
(5) Giacintucci et al. (2014a),
(6) Murgia et al. (2010a),
(7) Sijbring (1993),
(8) Burns et al. 1992,
(9) Gendron-Marsolais et al. (2017),
(10) Sarazin et al. (1995),
(11) Parekh et al. (2017),
(12) Kale et al. (2015),
(13) Govoni et al. (2009),
(14) Bonafede et al. (2009),
(15) Gitti et al. (2007),
(16) Ferrari et al. (2011),
(17) Giovannini et al. (2009),
(18) Mazzotta \& Giacintucci (2008),
(19) Hlavacek-Larrondo et al. (2013),
(20) Murgia et al. (2010b), 
(21) Pizzo \& de Bruyn (2009),
(22) Govoni et al. (2005),
(23) Feretti et al. (1997a),
(24) Giacintucci et al. (2014b), 
(25) Storm et al. (2015), 
(26) Farnsworth et al. (2013),
(27) Feretti et al. (1997b),
(28) Ehlert et al. (2011),
(29) Venturi et al., in preparation, 
(30) Gitti (2013),
(31) van Weeren et al. (2014).}
\label{tab:supplsample}
\end{deluxetable*}

\section{Cluster X-ray properties}
\label{sec:xrayprop}

The purpose of this study is to quantify the occurrence of radio minihalos
in clusters of different total masses with and without cool cores.  To
identify cool-core clusters in the sample, we follow C09 and use the
specific entropy ``floor'' in the cluster centers. They find that the radial
dependence of the specific entropy, defined in C09 and here as
\begin{equation}
K\equiv kTn_e^{-2/3},
\label{eq:entrodef}
\end{equation}
where T is the gas temperature and $n_e$ is the electron number density, can
be described, in the cluster central regions, as a function of radius, $r$,
by
\begin{equation}
K(r)=K_0+K_{100}\left(\frac{r}{100\,{\rm kpc}}\right)^\alpha
\label{eq:entroprof}
\end{equation}
where $K_0$ is the so-called core entropy, $K_{100}$ is a normalization for
entropy at 100 kpc, and $\alpha$ is the power-law index.

Clusters with $K_0<30-50$ keV\,cm$^2$ (which represents a deep minimum of
the entropy at the cluster center) invariably exhibit all the attributes of
a cool core.  For the cluster entropy profiles in the ACCEPT database, C09
combined projected gas temperatures with 3-dimensional gas densities, which
does not yield physically meaningful entropy values.  However, their values
serve our current purpose of identifying the cool cores well, and have been
used for similar purposes in the literature.  Therefore, for the present
analysis, we chose to use the $K_0$ values from C09 for those clusters that
already had high-quality \chandra\ data (81 out of 106 clusters in our
combined sample), and emulate the C09 derivation (with some technical
differences) for the 27 clusters with new or significantly-improved
\chandra\ data that appeared since C09. This will be described in
\S\ref{sec:tprof}.

For cluster total masses, we use the \planck\ SZ-based estimates where
available, otherwise we obtain a mass estimate from the $M_{\rm 500}-T_{\rm
  X}$\/ relation in \cite{2009ApJ...692.1033V}, using an X-ray measured,
core-excised temperature.  $M_{\rm 500}-T_{\rm X}$\/ estimates are required
for 17 out of 48 clusters (35\%) in the supplementary sample.  Core-excised
temperatures are taken from Cavagnolo et al.\ (2008, hereafter C08) if
available, otherwise we derive them here (\S\ref{sec:t}) in the same manner
for uniformity. Errors for those masses are calculated from the temperature
statistical uncertainties and include statistical and systematic
uncertainties for the $M_{\rm 500}-T_{\rm X}$\/ relation itself, as
estimated by \cite{2009ApJ...692.1033V}.

\section{{\em CHANDRA}\/ data analysis}

The ACCEPT sample of C09 contains clusters in the \chandra\ archive as of
2008.  Since then, \chandra\ observed new clusters, e.g., those in our
statistical sample that were discovered by \planck\ (2 clusters), and
reobserved others, including several borderline clusters (``warm cores'')
for which it is important for us to have an accurate core temperature
profile.  Here, we have analyzed \chandra\ data for such new clusters and
important improvements.  Seven of the new observations are part of the
\chandra\ Visionary Program to study the X-ray properties of {\em
  Planck}-selected clusters, including the fraction of cool cores 
(Andrade-Santos et al. 2017).

We will use C08 and C09 entropy and temperature values for those clusters
that have been reobserved but for which the old data were accurate enough.
Updates of \chandra\ calibration since C08 and C09 have negligible effects
for our current qualitative purposes.

\subsection{Data reduction and image preparation}
\label{sec:xray}

The {\em Chandra} observations analyzed in this paper are listed in Table
\ref{tab:xray}, which includes the observation identifiers, clean exposure
times and the adopted Galactic absorption column density $N_H$
(\S\ref{sec:spec}). The Level-1 ACIS event files from the archive were
reprocessed following the procedure described in \cite{2005ApJ...628..655V}
using the \chandra\ Calibration Database (CALDB) 4.6.3. Exclusion of time
intervals with elevated background and background modeling were done as
described in \cite{2003ApJ...586L..19M}. To model the detector and sky
background, we used the blank-sky datasets from the CALDB appropriate for
the date of each observation, normalized using the ratio of the observed to
blank-sky count rates in the 9.5--12 keV band. Following
\cite{2000ApJ...541..542M}, we also subtracted the ACIS readout artifact,
which is an important effect on the radial temperature profiles in the
presence of a sharp brightness peak.

We used images in the 0.5--4 keV and 2--7 kev energy bands to detect the
unrelated X-ray point sources and small-scale extended sources in each
observation.  These sources were masked from the image and spectral
analysis. For each cluster, we need to extract and fit a radial surface
brightness profile and a radial temperature profile.

\startlongtable
\begin{deluxetable*}{lcccr}
\tablewidth{0pt}
\tablecaption{{\em Chandra} observations analyzed in this work
\label{tab:xray}}
\tablehead{
\colhead{Cluster name} & \colhead{Observation} & \colhead{Detector} & \colhead{Exposure} & \colhead{$N_{H}$}  \\
\colhead{} & \colhead{ID} & \colhead{(ACIS)}  &  \colhead{(ksec)} &  \colhead{($10^{20}$ cm$^{-2}$)}   \\
}
\startdata
A\,2813                  &  9409                    & I     & 20.0              & $1.84^\star$           \\
A\,2895                  &  9429                    & I     & 19.7              & $1.37^\star$           \\
RXC\,J0142.0+2131        &  10440                   & I     & 20.0              & $6.20^\star$           \\
A\,401                   &  14024                   & I     & 111.5             & $13.04^{+0.30}_{-0.30}$\phantom{0} \\
A\,3088                  &  9414                    & I     & 19.2              & $1.27^\star$           \\
PSZ1\,G171.96-40.64      &  15302                   & I     & 25.9              & $31.69^{+1.59}_{-1.57}$\phantom{0} \\ 
RXC\,J0510.7-0801        &  14011                   & I     & 20.7              & $6.44^\star$  \\
PSZ1\,G139.61+24.20      &  15139,15297             & I, I   & 17.9,9.3          & $13.11^{+1.7}_{-1.6}$\phantom{0} \\
A\,781                   &   15128                  & I     & 33.7              & $1.65^\star$ \\ 
A\,3444                  &   9400                   & S     & 36.3              & $5.55^\star$ \\ 
A\,1132                  &   13376                  & I     &  9.1              & $8.52^{+1.9}_{-1.6}$\phantom{0} \\
A\,1351                  &   15136                  & I     & 32.9              & $0.98^\star$ \\
A\,1423                  &   11724                  & I     & 25.7              & $1.81^\star$ \\
A\,1443                  &   16279                  & I     & 18.9              & $5.18^{+1.9}_{-1.4}$\phantom{0} \\
RXC\,J1234.2+0947        &   11727                  & I     & 20.7              & $1.61^\star$ \\
A\,1682                  &   11725                  & I     & 19.7              & $1.04^\star$ \\
A\,1689                  &   6930, 7289             & I, I  & 76,75             & $3.48^{+0.29}_{-0.29}$\phantom{0} \\
A\,1733                  &   11771                  & I     &  6.7              & $1.81^\star$ \\
AS\,780                  &   9428                   & S     & 39.4              & $7.39^\star$ \\
RXC\,J1504.1--0248       &   17197,17669,17670      & I, I, I & 29.8,28.5,44.6    & $14.35^{+4.6}_{-4.6}$\phantom{0} \\
RXC\,J1514.9-1523        &   15175                  & I     & 59.1              & $8.28^\star$  \\
A\,2142                  &   5005,15186             & I, S   & 39.4,87.1         & $6.16^{+0.12}_{-0.12}$\phantom{0} \\  
CL\,1821+643             &   9398                   & S     & 34.2              & $3.44^{\star}$ \\ 
RXC\,J2003.5-2323        &   7916                   & I     & 50.0              & $7.57^\star$ \\
MACS\,J2135.2-0102       &   11710                  & I     & 26.9              & $3.88^\star$  \\
A\,2355                  &   15097                  & I     & 14.5              & $8.98^{+2.41}_{-2.18}$\phantom{0}  \\
A\,2552                  &   11730                  & I     & 22.8              & $4.60^\star$ \\
A\,2667                  &    2214                  & S     &  9.3              & $1.73^\star$ \\ 
Z348                     &   10465                  & S     & 48.7              & $2.50^\star$ \\
A\,141                   &    9410                  & I     & 19.7              & $1.79^\star$ \\
Z2089                    &   10463                  & S     & 40.4              & $2.88^\star$\\ 
A\,1240                  &    4961                  & I     & 51.8              & $2.85^\star$ \\
A\,1413                  &    5002                  & I     & 37.1              & $1.83^\star$ \\
A\,1576                  &   15127                  & I     & 28.8              & $1.08^\star$ \\
A\,1650                  &    7242                  & I     & 37.6              & $1.35^\star$ \\
ZwCl\,1742.1+3306        &    8267                  & I     &  7.8              & $3.83^\star$\\ 
RX\,J2129.6+0005         &    9370                  & I     & 29.8              & $3.64^\star$\\
A\,2420                  &    8271                  & I     &  7.8              & $3.70^\star$ \\
A\,2626                  &    3192                  & S     & 24.9              & $3.83^\star$ \\
Phoenix                   &   16545                  & I     & 59.9              & $1.52^\star$ \\
\enddata 
\tablecomments{   
Column 1: cluster name. Column 2: observation identification number. Column 3: {\em Chandra} ACIS detector.
Column 4: Total clean exposure. Column 5: Galactic absorption column density adopted in this paper;
values marked with $^\star$ are from LAB (\S\ref{sec:spec}).}
\end{deluxetable*}

\begin{figure*}
\centering
\epsscale{1.1}
\includegraphics[height=5.7cm,bb=18 155 548 657,clip]{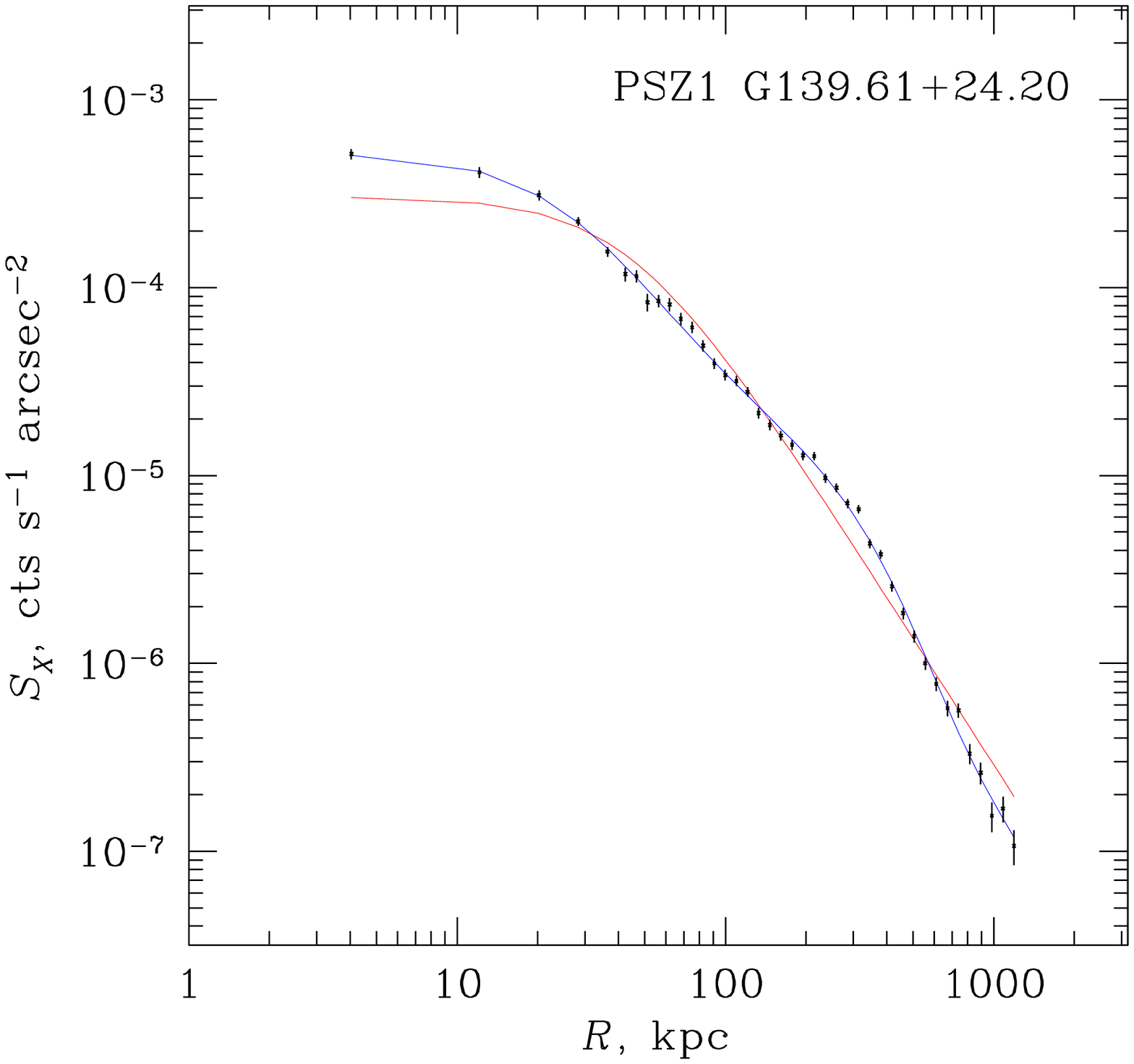}
\hspace{0.1cm}
\includegraphics[height=5.7cm,bb=50 155 548 657,clip]{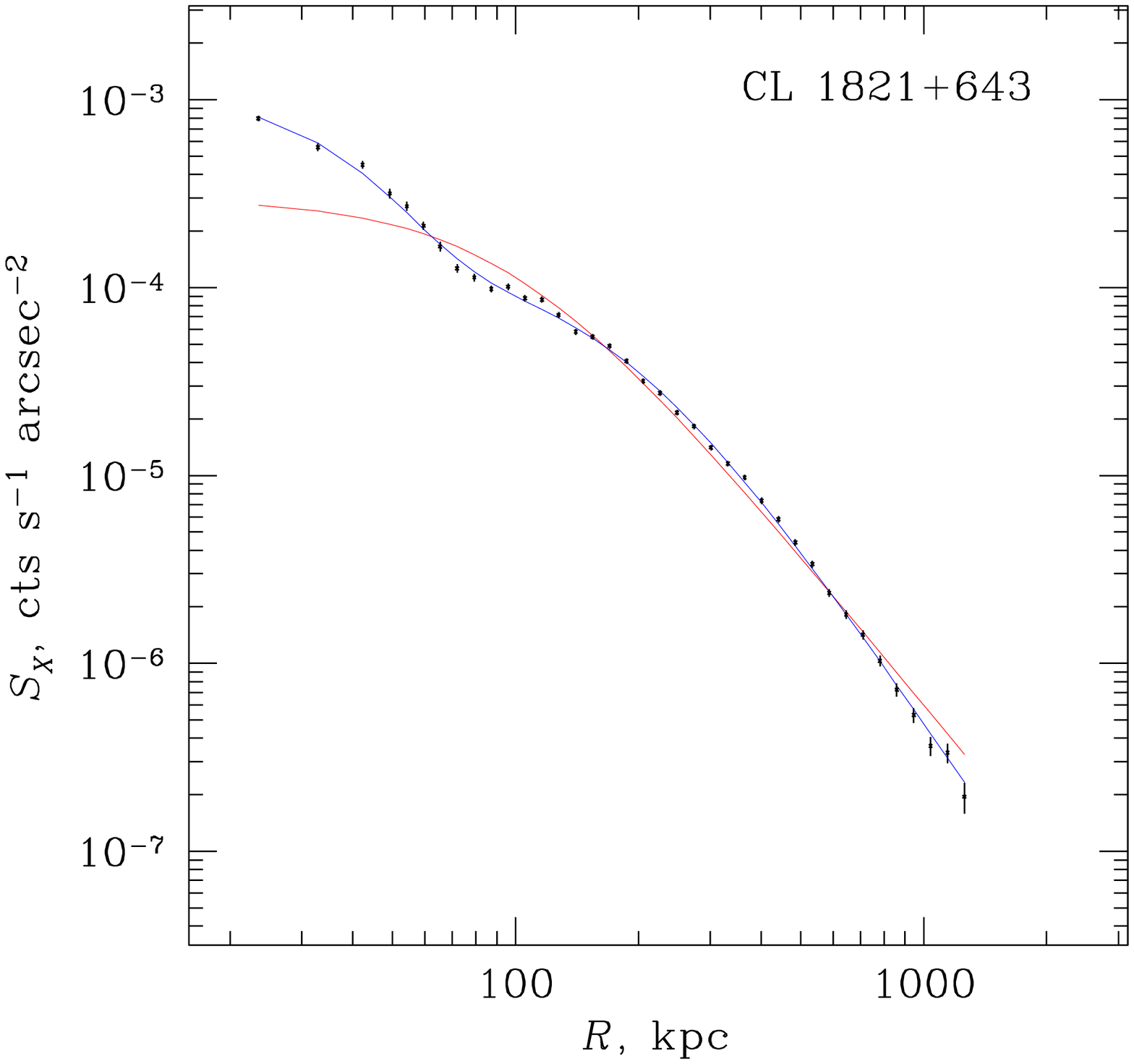}
\hspace{0.1cm}
\includegraphics[height=5.7cm,bb=50 155 548 657,clip]{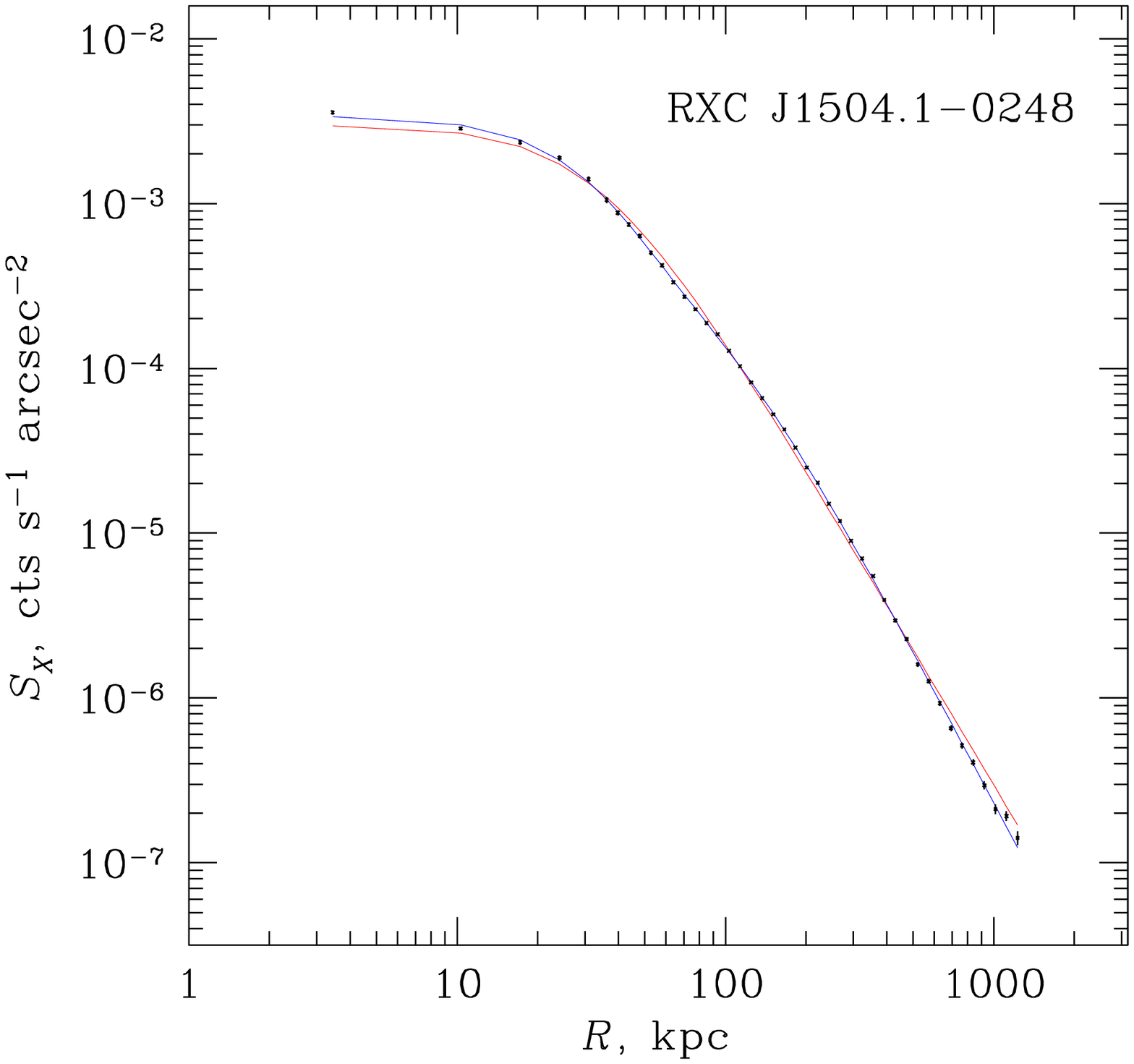}
\smallskip
\caption{{\em Chandra} X-ray surface brightness profiles of
  PSZ1\,G139.61+24.20, CL\,1821+643 and RXC\,J1504.1-0248. Solid lines show the
  best-fit single (red) and double (blue) $\beta$-models.}
\label{fig:sbprof}
\end{figure*}

\subsection{Gas density profiles}
\label{sec:bri}

For each cluster, we obtained a background-subtracted, exposure-corrected
image in the 0.7--2 keV band. For clusters with multiple (typically offset)
ACIS observations, we first coadded the individual background-subtracted
images and then divided the counts images by the sum of the corresponding
exposure maps. We then extracted radial surface brightness profiles, 
centered on the cluster X-ray centroid.

While a simple spherically-symmetric $\beta$-model for the gas density
\citep{1978A&A....70..677C} is a reasonable description for non-cool-core
clusters, it does not describe cool cores well
\citep[e.g.,][]{1984ApJ...276...38J}.  We therefore fit the brightness
profiles by projecting a spherically-symmetric double $\beta$-model,
\begin{equation}
n_e(r)=\frac{n_0}{1+f} 
   \left[\left(1+\frac{r^2}{r_{c1}^2}\right)^{\!-\frac{3}{2}\beta_1} \!\!\!+  
         f\left(1+\frac{r^2}{r_{c2}^2}\right)^{\!-\frac{3}{2}\beta_2}
   \right],
\label{eq:beta}
\end{equation}
where $n_{0}$\/ (central density for the sum of the two components),
$r_{c1}$, $r_{c2}$, $\beta_1$, $\beta_2$, and $f$\/ were free parameters. We
ignore the very mild temperature dependence of the X-ray emissivity in this
\chandra\ band and the relevant range of temperatures.  Fits to the
brightness profile were done out to $R_{500}$ or as far in radius as the
\chandra\ coverage allowed, which in all cases was far beyond the cores.
While such a model is not physically motivated (and no physical significance
should be assigned to the particular values of $r_c$ and $\beta$), it fits
all our clusters well, which is what we need for determining the central
entropy. For those clusters where a single $\beta$-model provided a good
fit, we set $f=0$.  Example cool-core fits with single and double
$\beta$-models are shown in Fig.\ \ref{fig:sbprof}, and fit results for all
clusters are given in Table 4. We do not derive errors on best-fit
quantities, because statistical errors are negligible and uncertainties of
the geometry dominate. The entropy uncertainties will be dominated by the
temperature accuracy.

\subsection{Spectral analysis}\label{sec:spec}

We derived the intracluster medium temperatures using the \chandra\ data as
follows. For each observation, we extracted a spectrum of the cluster for
each region of interest, and generated the instrument responses (ARF and
RMF) using the current calibration files (version N0008 for the telescope
effective area, N0006 for the CCD quantum efficiency, and N0009 for the ACIS
time-dependent low-energy contamination model).

The background spectra were extracted for the same regions from the
corresponding blank-sky datasets, normalized as in \S\ref{sec:xray}. To
ensure the sky-variable soft cosmic X-ray background is subtracted
correctly, we checked for the presence of significant excess/deficit of soft
X-ray thermal emission in the spectrum using local background regions. In
those cases where significant systematic residuals were found, they were
modeled with a low-temperature APEC or absorbed power-law and then included
as a fixed additional background component in the spectral fits, normalized
by the region area, as in Vikhlinin et al.\ (2005). This has little effect
in the cores, but affects the fits in the outskirts.

The cluster X-ray emission was modeled with an absorbed, single-temperature
APEC model in the the 0.7--7 keV energy band, with the metal abundance free
to vary. For the full-cluster spectra, the absorption column density $N_H$
was allowed to vary. If the best-fit $N_H$\/ value was consistent with that
from the Leiden/Argentine/Bonn (LAB) radio survey of Galactic HI
\citep{2005A&A...440..775K}, we fixed it at the database value for
subsequent analysis. The adopted $N_H$ values are listed in Table
\ref{tab:xray}

For several clusters, we combined multiple {\em Chandra}\/ observations
(Table \ref{tab:xray}) by fitting their spectra simultaneously with the
temperature and metal abundance of the hot APEC components tied together.
For RXC\,J1504.1-0248, for which the observations had the same pointing 
position and roll angle 
and thus had the same responses, we instead coadded the spectra.


\startlongtable
\begin{deluxetable*}{lcccccccc}
\tablecaption{Gas density profile $\beta$-model fits}
\label{tab:sb}
\tablehead{
\colhead{Cluster name} & \colhead{$n_0$} & \colhead{$r_{c1}$}  & \colhead{$\beta_1$}  & \colhead{$r_{c2}$}   & \colhead{$\beta_2$} & \colhead{$f$}   \\
    \colhead{}         & \colhead{($10^{-3}$ cm$^{-3}$)}   & \colhead{(kpc)}    & \colhead{}           & \colhead{(kpc)}    &   \colhead{}        & \colhead{}  \\
}
\startdata
 A\,2895            &  5.6  & 212 & 0.7 & ...  & ...  & 0  \\
 RXC\,J0142.0+2131  & 16.1  &  32 & 0.6 & 212  & 0.7  & 0.50 \\
 A\,401             &  8.6  &  51 & 0.3 & 277  & 1.1  & 0.65 \\
 PSZ1G171.96-40.64  & 11.4  & 183 & 3.0 & 376  & 0.8  & 0.73 \\
 RXC\,J0510.7-0801  &  8.9  & 143 & 1.2 & 647  & 1.1  & 0.40 \\
 PSZ1\,G139.61+24.20& 77.3  & 22  & 0.5 & 804  & 3.0  & 0.03 \\
 A\,781             &  4.2  & 237 & 2.3 & 644  & 0.9  & 0.66 \\
 A\,3444            & 77.9  & 37  & 0.7 & 190  & 0.6  & 0.07 \\
 A\,1132            &  4.3  & 256 & 0.7 & ...  & ...  & 0 \\
 A\,1351            &  3.0  & 683 & 1.4 & ...  & ...  & 0 \\ 
 A\,1423            & 18.5  & 56  & 0.6 & 422  & 0.6  & 0.05 \\
 A\,1443            &  4.3  & 276 & 0.6 & ...  & ...  & 0 \\
 RXC\,J1234.2+0947  &  1.6  & 549 & 0.8 & ...  & ...  & 0 \\ 
 A\,1682            &  6.4  & 156 & 3.0 & 329  & 0.8  & 0.81 \\
 A\,1689            &  34.3 & 123 & 3.0 & 184  & 0.7  & 0.44 \\ 
 A\,1733            &  5.1  & 176 & 0.6 & ...  & ...  & 0  \\
 AS\,780            & 86.5  &  21 & 0.5 & ...  & ...  & 0 \\
 RXC\,J1504.1$-$0248& 170.2 &  68 & 3.0 &  75  & 0.7  & 0.41 \\
 RXC\,J1514.9$-$1523&   2.0 & 592 & 0.8 & ...  & ...  & 0 \\
 A\,2142            &  27.6 &  90 & 3.0 & 139  & 0.6  & 0.81 \\
 CL\,1821+643       &  80.5 &  86 & 3.0 & 186  & 0.7  & 0.18 \\
 RXC\,J2003.5$-$2323&   2.2 & 698 & 1.0 & ...  & ...  & 0 \\ 
 MACS\,J2135.2$-$0102&  5.6 & 191 & 0.6 & ...  & ...  & 0 \\
 A\,2355            &   2.1 & 615 & 1.0 & ...  & ...  & 0 \\
 A\,2552            &  18.1 &  54 & 0.7 & 238  & 0.7  & 0.40 \\
 Z348               &  75.2 &  33 & 0.7 & ...  & ...  & 0 \\
 Phoenix            & 209.4 &  34 & 0.6 & ...  & ...  & 0 \\
\enddata
\tablecomments{Column 1: cluster name. Columns 2--7: parameters of a
$\beta$-model fit, see eq.\ (\ref{eq:beta}). Fits were done out to $R_{500}$ or
as far in radius as the {\em Chandra} coverage allowed, which in all cases was far beyond the core region.}
\end{deluxetable*}

\subsection{Average cluster temperatures}
\label{sec:t}

One of the quantities that we correlate with the presence of a radio halo is
the global cluster temperature. In addition, for those clusters in the
supplementary sample that do not have \planck\ masses, we estimate total
masses from the $M_{\rm 500}-T_{\rm X}$\/ relation (\S\ref{sec:xrayprop}).
For 29 clusters in the statistical sample and for most in the supplementary
one, we used core-excised global temperatures from C08. For the remaining
clusters in the supplementary sample and for 12 in the statistical one with
significantly better recent {\em Chandra}\/ observations, we derived global
temperatures (\S\ref{sec:spec}) using a spectrum extracted from an annulus with
70 kpc$<r<R_{\rm 2500}$%
\footnote{$R_{2500}$ is the radius that encloses a mean overdensity of 2500
  with respect to the critical density at the cluster redshift.}
as in C08. 
The core-excised temperatures, $T_{\rm X,\,ce}$,
of all clusters are summarized in Tables \ref{tab:k0} and 7. 
The exceptions are 2A0335+096, Perseus and Ophiuchus, for which we
used average temperatures from {\em ASCA}\/%
\footnote{{\em Advanced Satellite for Cosmology and Astrophysics.}}
because of \chandra's limited radial coverage. Another nearby cluster for
which we used the total temperature from the literature is A193, a clear
non-cool-core cluster, for which core excision would not change the
temperature significantly. For cool cores, this should result in slightly
underestimated masses from the $M_{\rm 500}-T_{\rm X}$\/ relation. This will
not affect our qualitative conclusions.

\startlongtable
\begin{deluxetable*}{lccrrccrcrc}
\tablecaption{Best-fit parameters for temperature and entropy profiles}
\label{tab:tmodel}
\tablehead{
\colhead{Cluster}            & \colhead{$T_0$}    & \colhead{$r_t$}   & \colhead{$a$\phantom{00}} & \colhead{$T_{\rm min}/T_0$} & \colhead{$r_{\rm cool}$} & \colhead{$a_{\rm cool}$} & \colhead{$K_0$\phantom{00}} & \colhead{$K_{100}$} & \colhead{$\alpha$\phantom{0}} & \colhead{$r_{\rm max}$}\\
  \colhead{name}   & \colhead{(keV)}    & \colhead{(Mpc)}   &  \colhead{}   & \colhead{}    &  \colhead{(kpc)}     & \colhead{}   & \colhead{(keV cm$^2$)} &\colhead{(keV cm$^2$)}&\colhead{} & \colhead{(Mpc)} \\
}
\startdata
 A\,2895              & 11.1    & 0.3  & $-0.3$\phantom{0}  & 1.0$^{\star}$    & ... & ... & $173\pm65$\phantom{0}  & 106 & 1.0\phantom{0} & 1.4 \\ 
 RXC\,J0142.0+2131    &  7.5    & 1.9  & 0.0$^{\star}$       & 1.0$^{\star}$    & ... & ... & $131\pm51$\phantom{0}  & 90  & 1.2\phantom{0} & 1.8 \\
 A\,401               &  7.9    & 3.1  & 0.0$^{\star}$       & 1.0$^{\star}$    & ... & ... & $180\pm6$\phantom{0}   & 82  & 1.3\phantom{0} & 1.0 \\
 PSZ1G171.96$-$40.64  & 11.6    & 0.9  & 0.0$^{\star}$       & 1.0$^{\star}$    & ... & ... & $329\pm74$\phantom{0}  & 42  & 1.4\phantom{0} & 2.0 \\
 RXC\,J0510.7$-$0801  &  9.2    & 1.0  & 0.0$^{\star}$       & 1.0$^{\star}$    & ... & ... & $158\pm99$\phantom{0}  & 129 & 0.9\phantom{0} & 1.0 \\
 PSZ1\,G139.61+24.20  & 13.8    & 0.4  & 0.0$^{\star}$       & 0.04\phantom{0} & 62  & 0.8 & $10\pm10$\phantom{0}   & 186 & 1.1\phantom{0} & 0.1 \\ 
 A\,781               & 8.2     & 1.0  & 0.0$^{\star}$       & 1.0$^{\star}$    & ... & ... & $170\pm36^{(a)}$           & 196 & 0.7\phantom{0} & 1.8 \\
 A\,3444              & 12.6    & 0.8  & 0.0$^{\star}$       & 0.31\phantom{0} & 299 & 1.1 & $18\pm2$\phantom{0}    & 100 & 1.3\phantom{0} & 0.5  \\ 
 A\,1132              & 9.5     & 0.4  & $-0.33$\phantom{0} & 1.0$^{\star}$    & ... & ... & $154\pm31^{(a)}$           & 111 & 0.9\phantom{0} & 1.1 \\
 A\,1351              & 13.4    & 0.6  & 0.0$^{\star}$       & 1.0$^{\star}$    & ... & ... & $620\pm93$\phantom{0}  &   3 & 2.6\phantom{0} & 1.3 \\
 A\,1423              & 9.7     & 0.6  & $-0.28$\phantom{0} & 1.0$^{\star}$    & ... & ... & $27\pm18$\phantom{0}   & 170 & 1.0\phantom{0} & 1.2 \\ 
 A\,1443              & 8.7     & 5.0  & 0.0$^{\star}$       & 1.0$^{\star}$    & ... & ... & $283\pm57^{(a)}$           & 61  & 1.4\phantom{0} & 1.6 \\ 
 RXC\,J1234.2+0947    &  6.5    & 0.9  & 0.0$^{\star}$       & 1.0$^{\star}$    & ... & ... & $404\pm93$\phantom{0}  & 23  & 1.3\phantom{0} & 1.2  \\ 
 A\,1682              & 10.4    & 0.4  & $-0.37$\phantom{0} & 1.0$^{\star}$    & ... & ... & $143\pm26^{(a)}$           & 139 & 1.0$^{\star}$   & 1.1 \\ 
 A\,1689              & 13.0    & 0.6  & $-0.16$\phantom{0} & 1.0$^{\star}$    & ... & ... & $59\pm4$\phantom{0}    & 109 & 1.7\phantom{0}   & 0.1 \\ 
 AS\,780              &  9.5    & 5.8  & 0.0$^{\star}$       & 0.37\phantom{0} & 146 & 3.2 & $19\pm2$\phantom{0}    & 110 & 1.7\phantom{0} & 0.2 \\
 RXC\,J1504.1$-$0248  & 19.2    & 0.4  & 0.0$^{\star}$       & 0.15\phantom{0} & 348 & 1.0 & $11.1\pm0.3$\phantom{0}& 86  & 1.6\phantom{0} & 0.2 \\
 RXC\,J1514.9$-$1523  & 11.4    & 0.8  & $-0.10$\phantom{0} & 1.0$^{\star}$    & ... & ... & $490\pm108^{(a)}$          & 52  & 1.2\phantom{0} & 0.8 \\
 A\,2142              & 12.8    & 1.3  & $-0.21$\phantom{0} & 1.0$^{\star}$    & ... & ... & $58\pm2$\phantom{0}    & 127 & 1.2\phantom{0} & 1.0 \\   
 CL\,1821+643         & 9.5     & 5.0  & 0.0$^{\star}$       & 0.38\phantom{0} & 88  & 4.2 & $8\pm5$\phantom{0}     & 125 & 1.3\phantom{0} & 0.2 \\
 RXC\,J2003.5$-$2323  & 17.6    & 0.4  & 0.0$^{\star}$       & 1.0$^{\star}$    & ... & ... & $708\pm85$\phantom{0}  & 11  & 1.3$^{\star}$   & 1.1 \\
 MACS\,J2135.2-0102   & 12.4    & 0.7  & $-0.32$\phantom{0} &  1.0$^{\star}$   & ... & ... & $142\pm18$\phantom{0}  & 145 & 1.0\phantom{0} & 1.9 \\
 A\,2355              & 10.5    & 1.4  & 0.0$^{\star}$       & 1.0$^{\star}$    & ... & ... & $519\pm117$\phantom{0} & 57  & 1.3$^{\star}$   & 0.8 \\
 A\,2552              & 12.8    & 0.7  & $-0.21$\phantom{0} &  1.0$^{\star}$   & ... & ... & $78\pm33$\phantom{0}   & 164 & 1.0\phantom{0} & 1.6 \\
 Z348                 &  4.7    & 0.6  & 0.0$^{\star}$       & 0.53\phantom{0} &  95 & 2.6 & $13\pm1$\phantom{0}      & 73  & 1.7\phantom{0} & 0.2 \\
 Phoenix              & 32.6    & 0.4  & 0.0$^{\star}$       & 0.21\phantom{0} & 255 & 1.5 & $19\pm3$\phantom{0}    & 123 & 1.7\phantom{0} & 0.3 \\
\enddata
\tablenotetext{a}{uncertainty from error in the first bin of the temperature profile.}
$^{\star}$ fixed in the fit.
\tablecomments{Column 1: cluster name. Columns 2--7: best-fit parameters of the temperature fit, see eq.\ (4).
Columns 8--10: best-fit parameters of the entropy fit (see \S\ref{sec:tprof} for details). Column 11:
maximum radius used for the entropy fit.}
\end{deluxetable*}

\subsection{Temperature and specific entropy profiles}
\label{sec:tprof}

We obtained radial temperature profiles by fitting spectra in annuli
centered on the X-ray surface brightness peak or the centroid for disturbed
clusters without a well-defined peak. The radial bins were selected as a
compromise between the need to sample well the temperature decline in the
cores and to keep the uncertainties small. The maximum radius was as far as
the detector coverage and the background uncertainties allowed; this is
unimportant as we are interested in the cores. The spectra were fit as
described in \S\ref{sec:t}, with metallicities allowed to vary at small
radii (since we expect strong gradients in cool cores) and fixed at 0.2
solar at large radii. The $N_H$ was fixed at the full-cluster values given
in Table \ref{tab:xray}. The resulting projected temperature profiles for
the clusters analyzed in this work are shown in Figs.\ 
\ref{fig:tprof1}--\ref{fig:tprof5}, except for A\,1733, which only has a 7
ks {\em Chandra}\/ exposure (see below).

Our goal here is to derive the specific entropy floor in cluster centers,
defined in the same way as in C09 (\S\ref{sec:xrayprop}). For this purpose,
we take our best-fit 3D density profile in the form of eq.\ (\ref{eq:beta}),
assume the entropy radial dependence given by eq.\ (\ref{eq:entroprof}),
calculate a temperature at a given radius from eq.\ (\ref{eq:entrodef}).
These temperature values are then averaged within each radial bin of the
measured temperature profiles using the weighting of
\cite{2004MNRAS.354...10M}.  The average values are compared with the
observed temperature profiles, and the parameters $K_0$, $K_{100}$ and
$\alpha$ are fit in such a way.  Because of limited modeling freedom of eq.\ 
(\ref{eq:entroprof}), such a temperature profile usually described the cool
cores well but often not the regions outside the core. Therefore, we only
used the radii where it fit well, similarly to C09. The modeled projected
temperature profiles are shown as blue lines overlaid on the measured
profiles in Figs.\ \ref{fig:tprof1}--\ref{fig:tprof5}, and the best-fit
entropy profile parameters, as well as the maximum radius used for the fit,
are given in Table 5.
For A\,1733, the central entropy
is estimated using the best-fit density profile and the temperature 
within the central $r=120$ kpc region, which is enough to determine that
it is not a cool core (Table 6).

For reference in our future work, and to verify the above model, we also fit
the observed temperature profiles with a more complex phenomenological 3D
temperature model used by \cite{2006ApJ...640..691V}:
%
\begin{equation}
T(r) = T_0\; t_{\rm cool}(r)\; t(r),
\end{equation}
where 
\begin{equation}
t(r)=\left(\frac{r}{r_t}\right)^{-\alpha} 
     \left[1+\left(\frac{r}{r_t}\right)^b\right]^{-c/b}
\end{equation} 
describes the temperature profile outside the central cooling region and
\begin{equation}
t_{\rm cool}(r)=\frac{x+T_{\rm min}/T_0}{x+1},\; 
    x\equiv \left(\frac{r}{r_{\rm cool}}\right)^{a_{\rm cool}}
\end{equation} 

describes the temperature decline in the central region.  $T_0$, $r_t$,
$\alpha$, $b$, $c$, $r_{\rm cool}$, $a_{\rm cool}$ and $T_{\rm min}$ are
free parameters.  The term $t_{\rm cool}(r)$ was required only in several
cases with a strong radial increase of the temperature profile in the core;
in most cases, setting $t_{\rm cool}(r)\equiv 1$ was sufficient for a good
fit, as the term $t(r)$ alone can describe a temperature dip in the center.
We also fixed $b=2$ and $c=1$, which provides a good fit in all cases. The
model was projected along the line of sight and within each radial bin and
fit to the observed projected temperature profiles. The best-fit profile
parameters are given in Table 5. The 3D and the corresponding projected
profiles are shown by red curves (dashed and solid, respectively) in Figs.\ 
\ref{fig:tprof1}--\ref{fig:tprof5}. Because this model is more flexible, it
can describe the whole radial range of the measured temperatures. Comparison
with the profiles for the entropy model (blue lines) shows that the entropy
model provides an adequate temperature fit in the cores of all clusters.

Because we will combine the central entropy values from C09 and from our new
analysis, we checked their consistency using RXJ1504.1--0248, one of the
clusters in the C09 sample for which we used a more recent \chandra\ 
dataset. Our best-fit central density (from a double $\beta$-model fit, see
Fig.\ \ref{fig:sbprof}) is only 7\% higher than that in C09 (who derived it
using the deprojection method), and our entropy floor value of
$K_0=11.1\pm0.3$ keV~cm$^2$ is consistent with, and more accurate than, the
C09 value of $K_0=13\pm1$ keV~cm$^2$. Other parameters of the core entropy
profile are in agreement as well. The accuracy of $K_0$ depends on the
number of temperature bins in the core, which is where the new data helped.

\begin{figure*}
\centering
\epsscale{1.1}
\includegraphics[width=7cm]{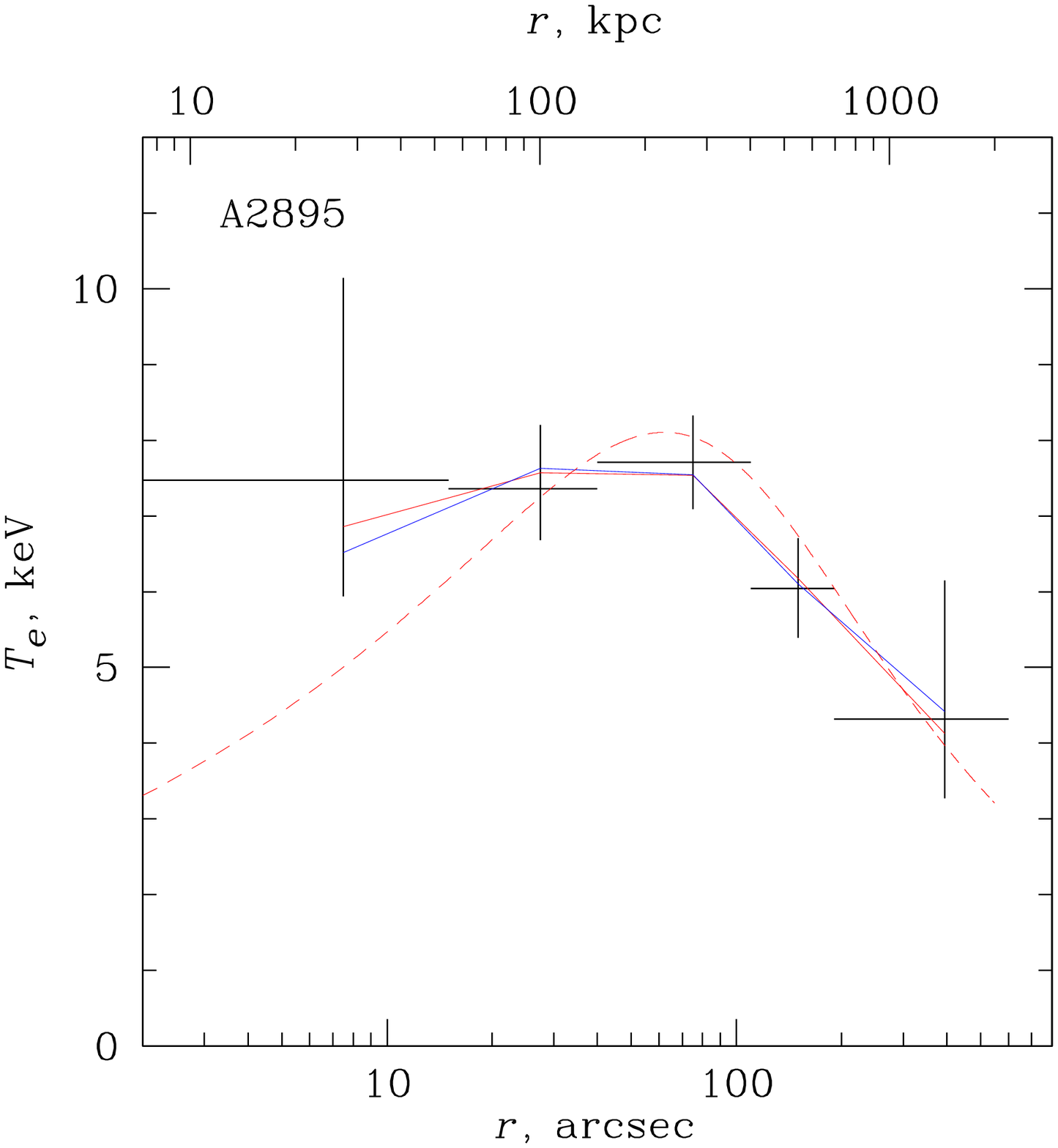}
\includegraphics[width=7cm]{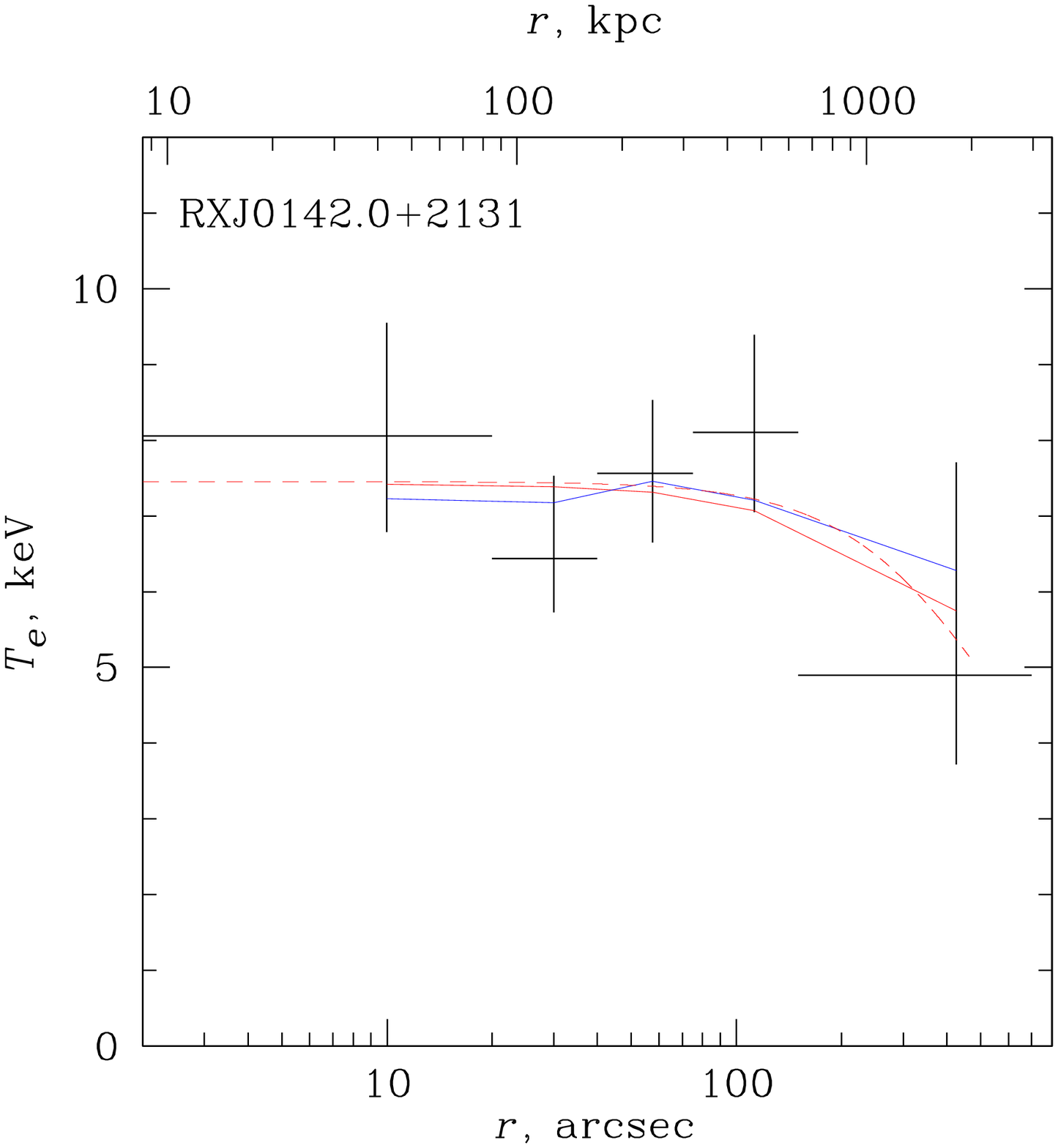}
\includegraphics[width=7cm]{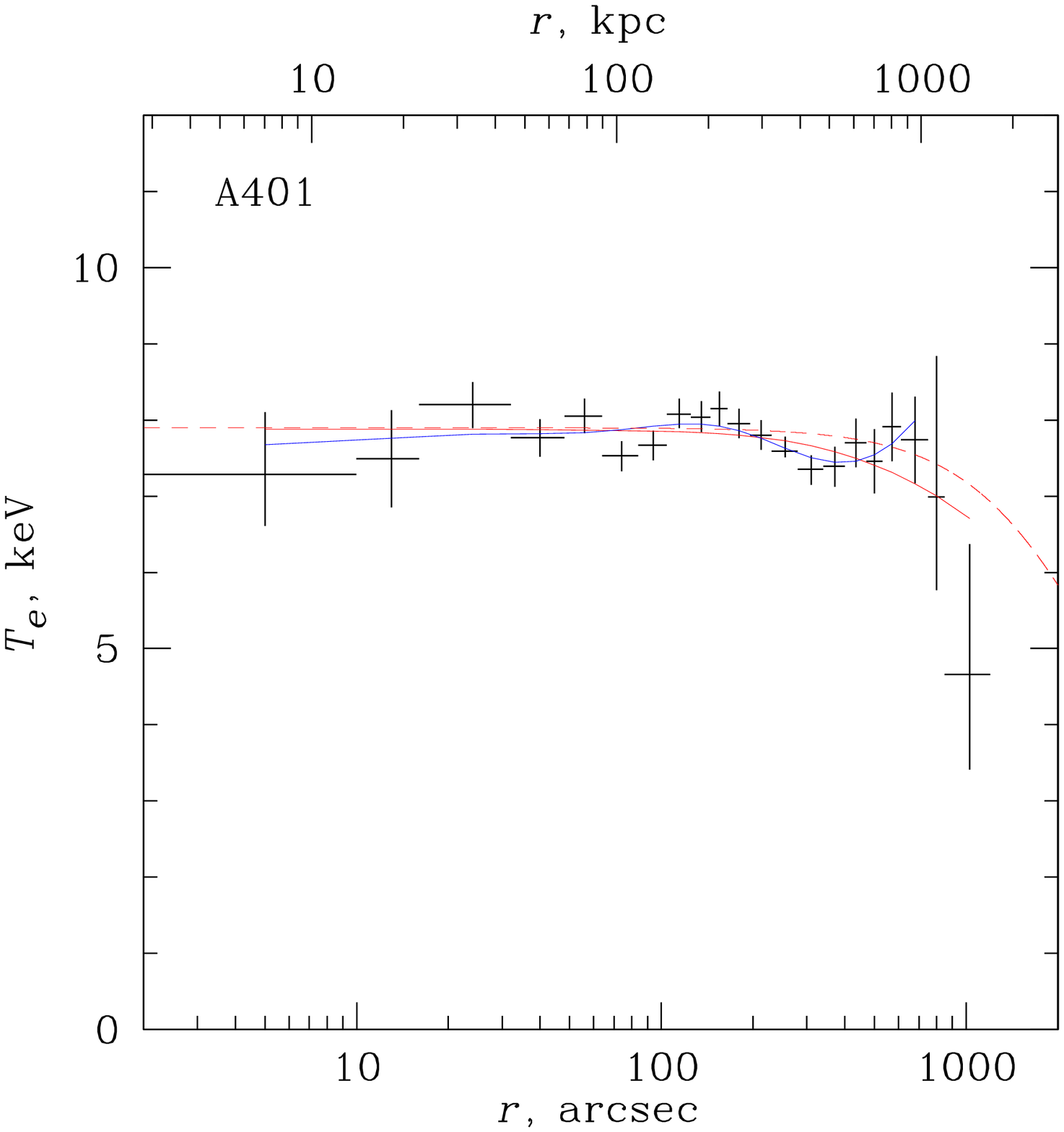}
\includegraphics[width=7cm]{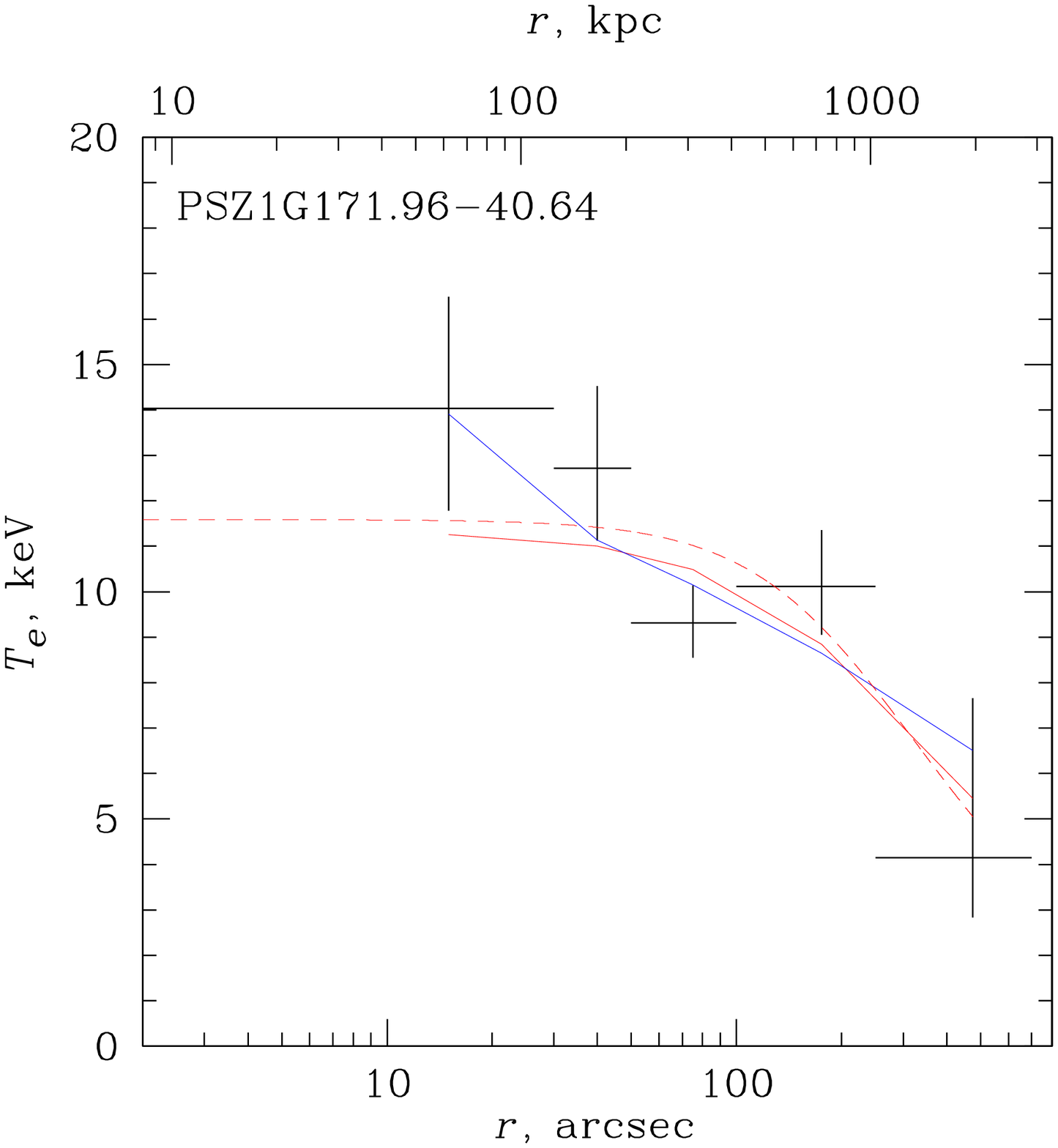}
\includegraphics[width=7cm]{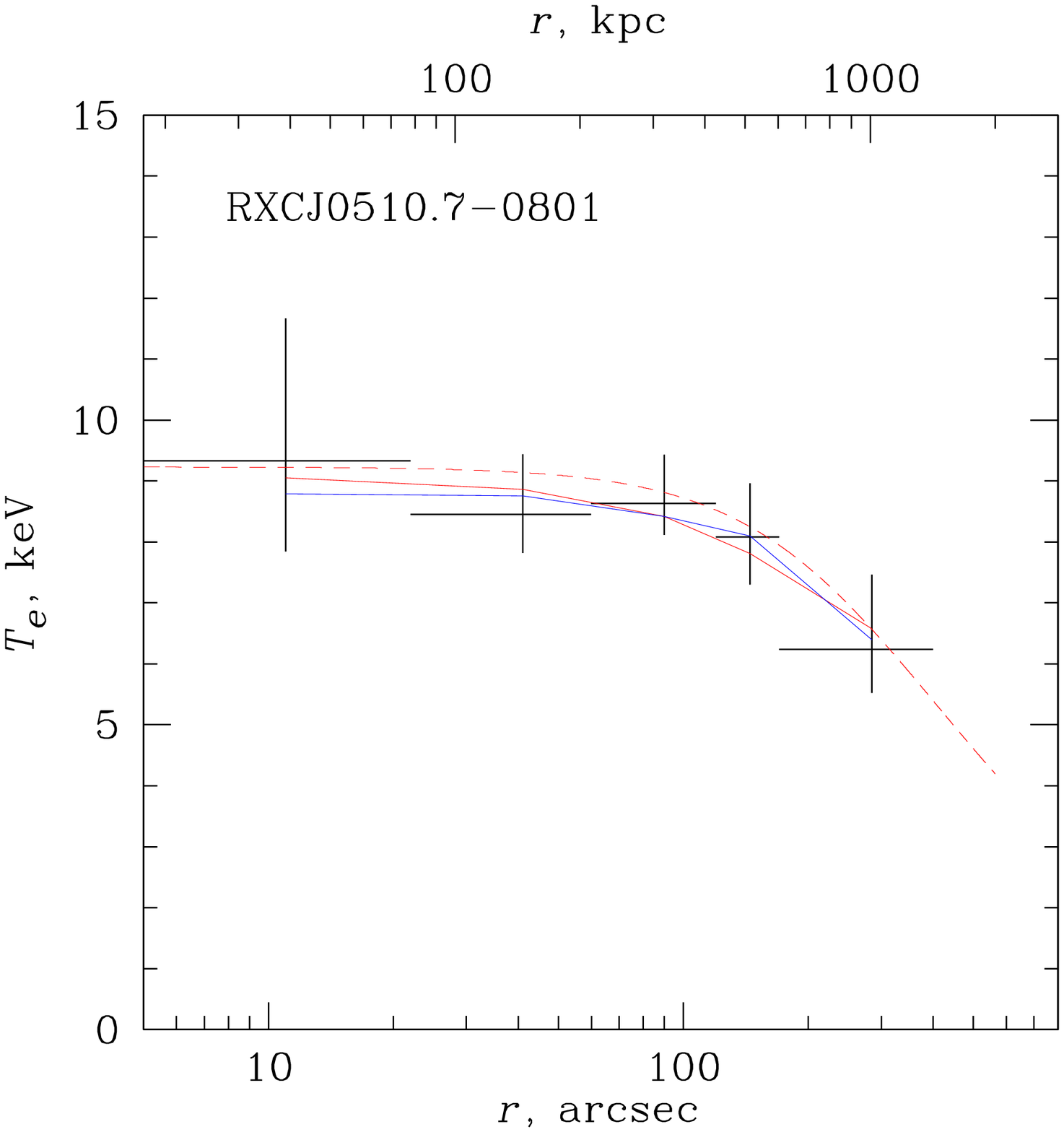}
\includegraphics[width=7cm]{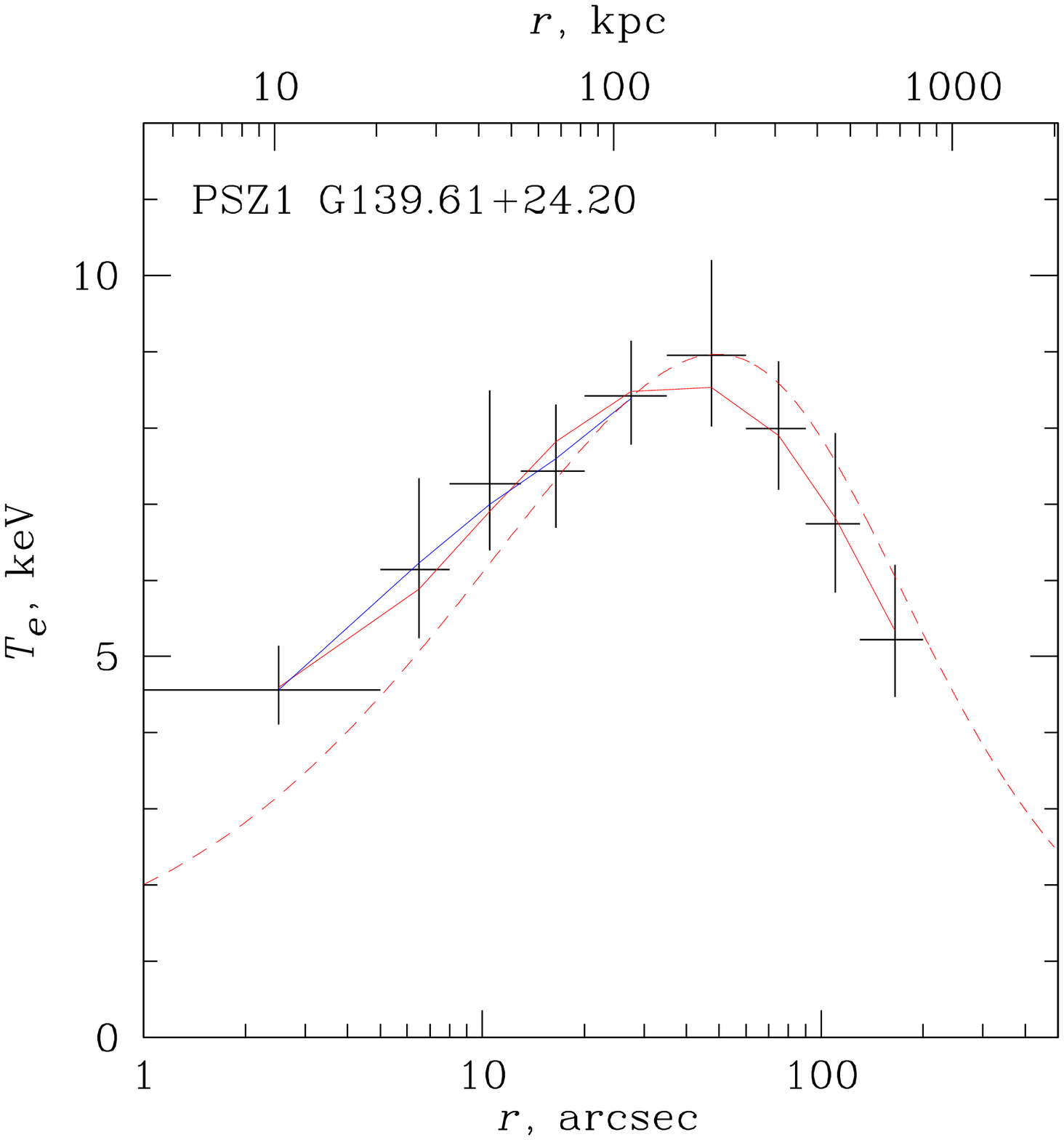}
\smallskip
\caption{Temperature profiles for A\,2895, RXC\,J0142.0+2131, A\,401, PSZ1G171.96-40.64, RXC\,J0510.7-0801
and PSZ1G139.61+24.20. Crosses are the observed projected
temperatures. Solid and dashed red lines show the best-fit
three-dimensional model and the corresponding projected profile,
respectively, and solid blue lines show the best-fit entropy model
(\S\ref{sec:tprof} and Table 5).
}
\label{fig:tprof1}
\end{figure*}

\begin{figure*}
\centering
\epsscale{1.1}
\includegraphics[width=7cm]{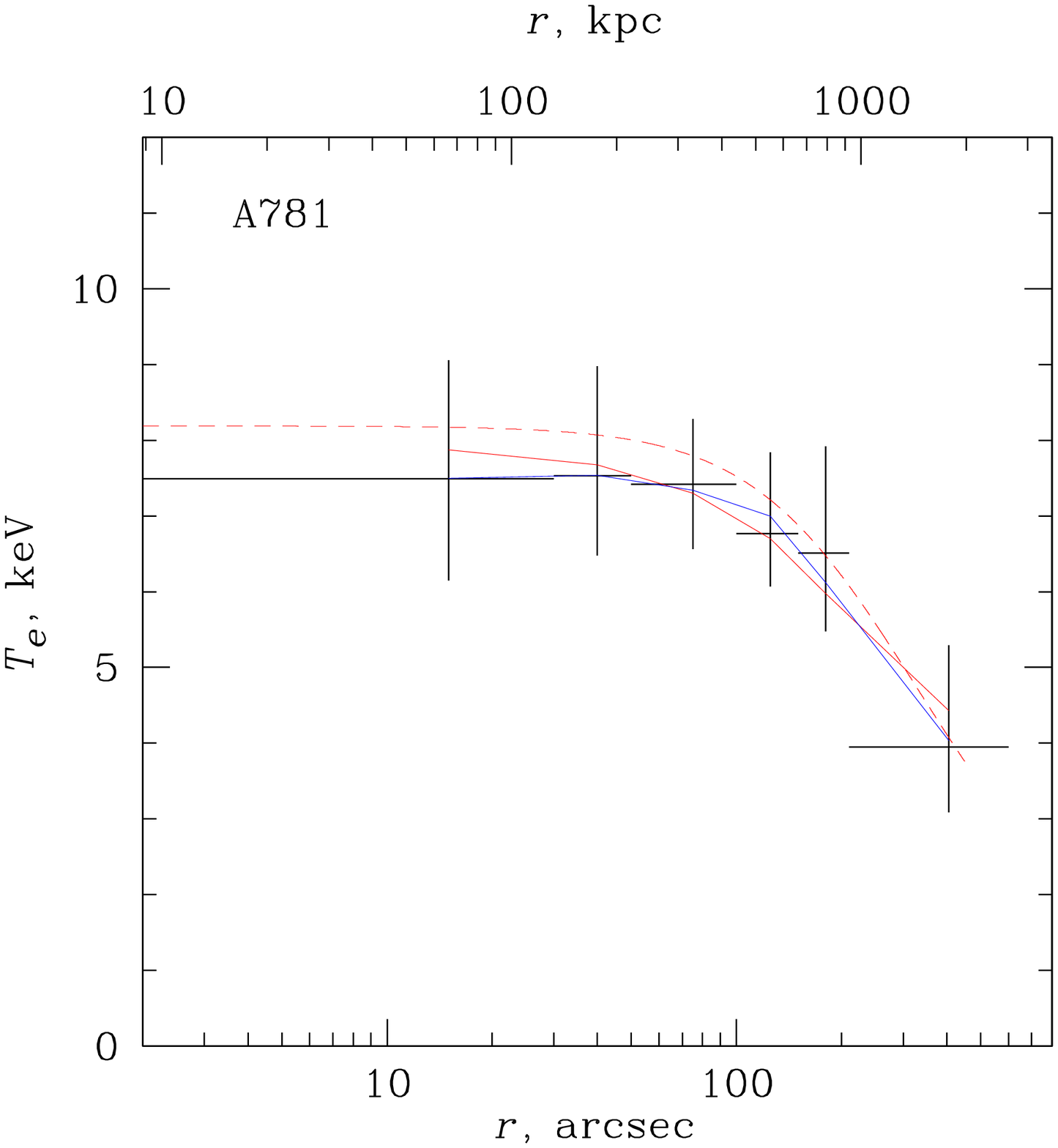}
\includegraphics[width=7cm]{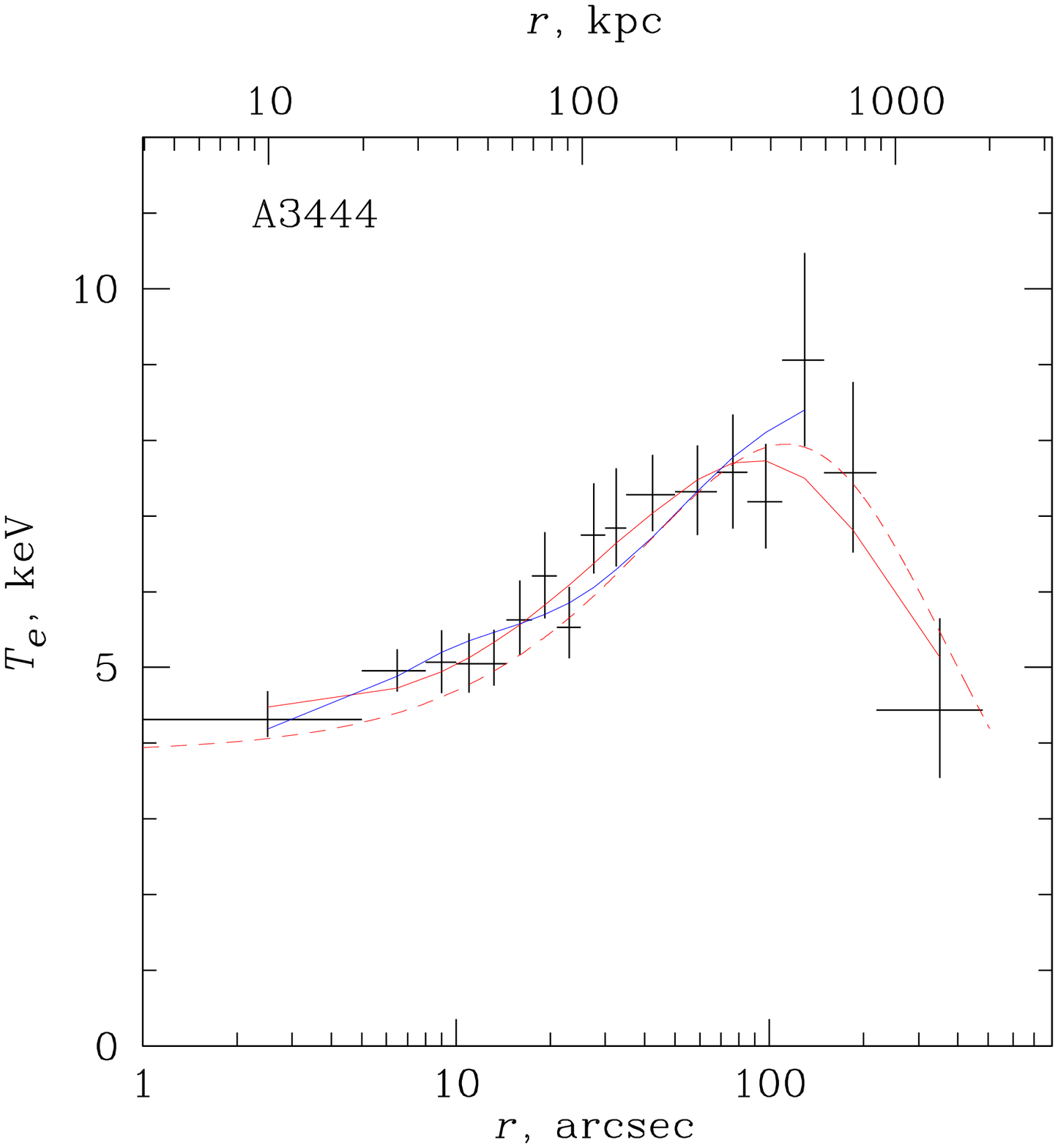}
\includegraphics[width=7cm]{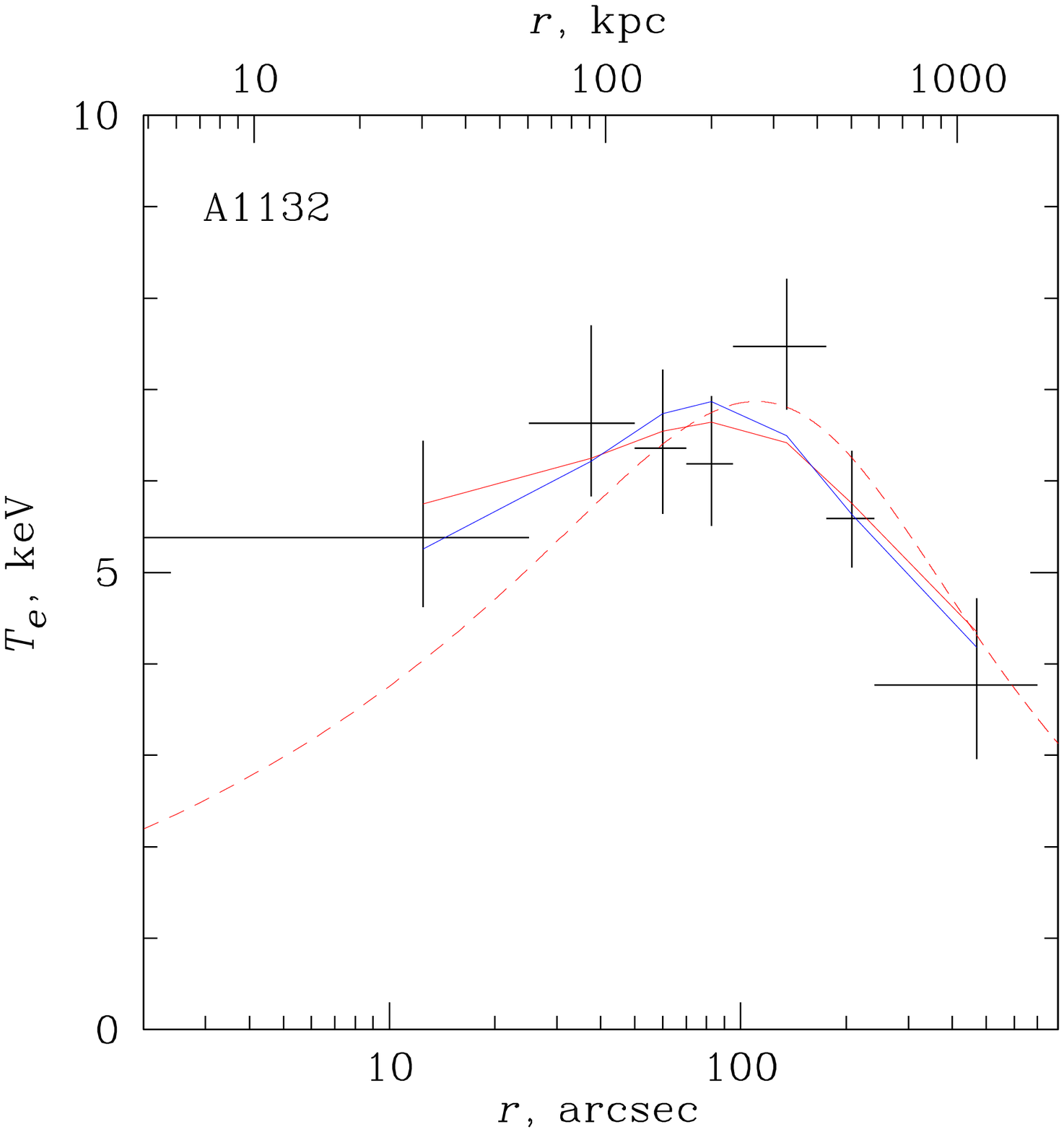}
\includegraphics[width=7cm]{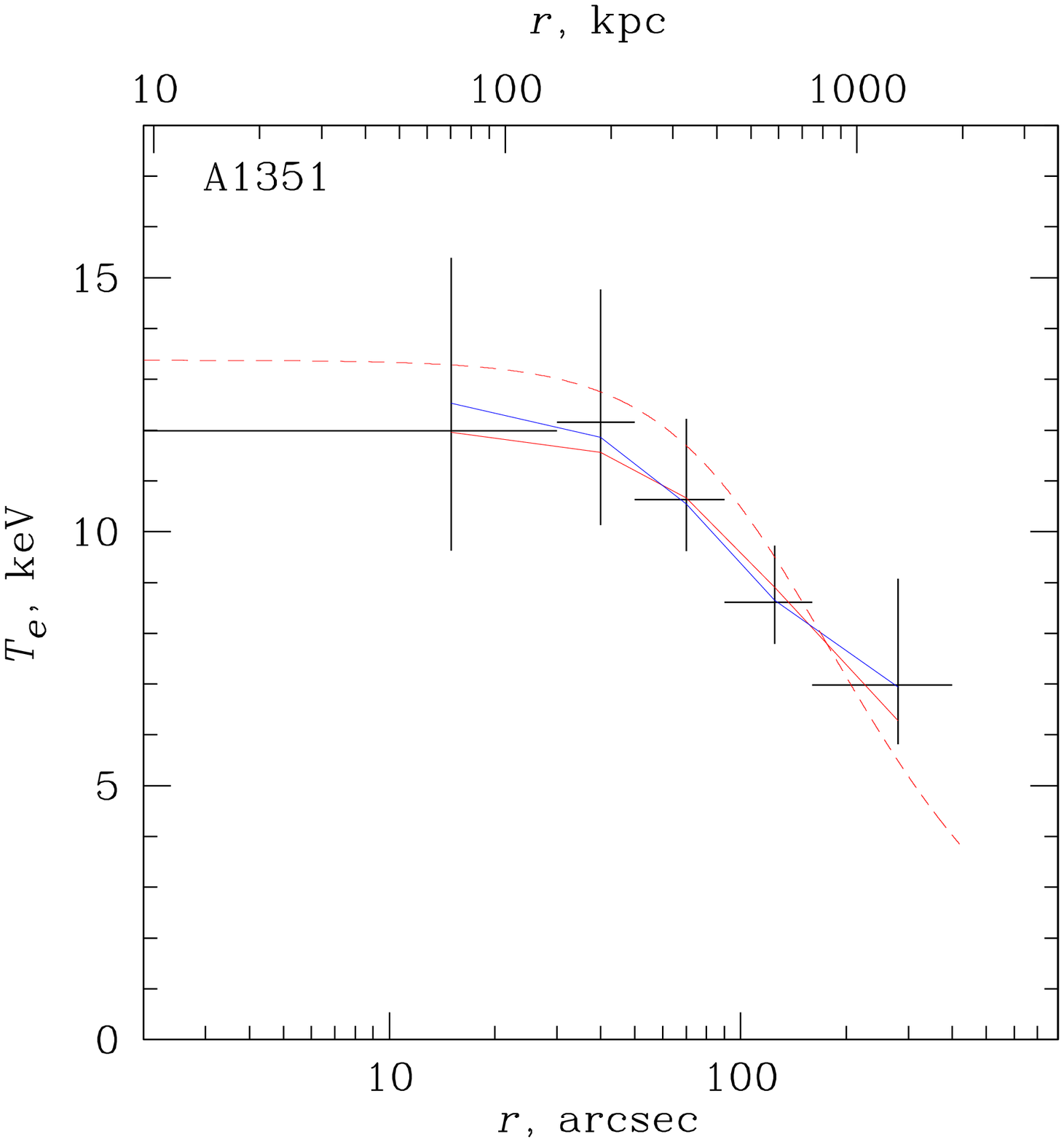}
\includegraphics[width=7cm]{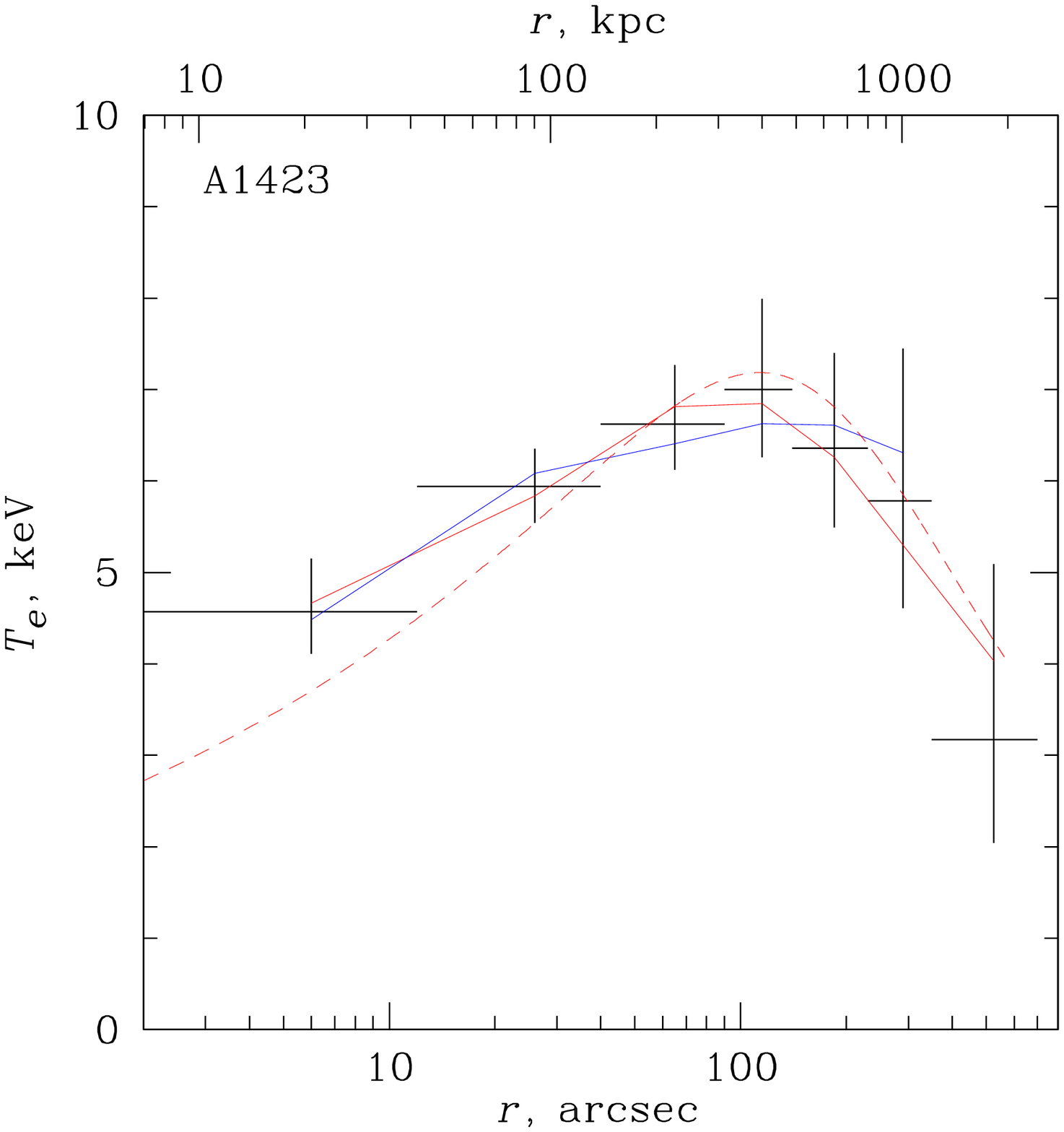}
\includegraphics[width=7cm]{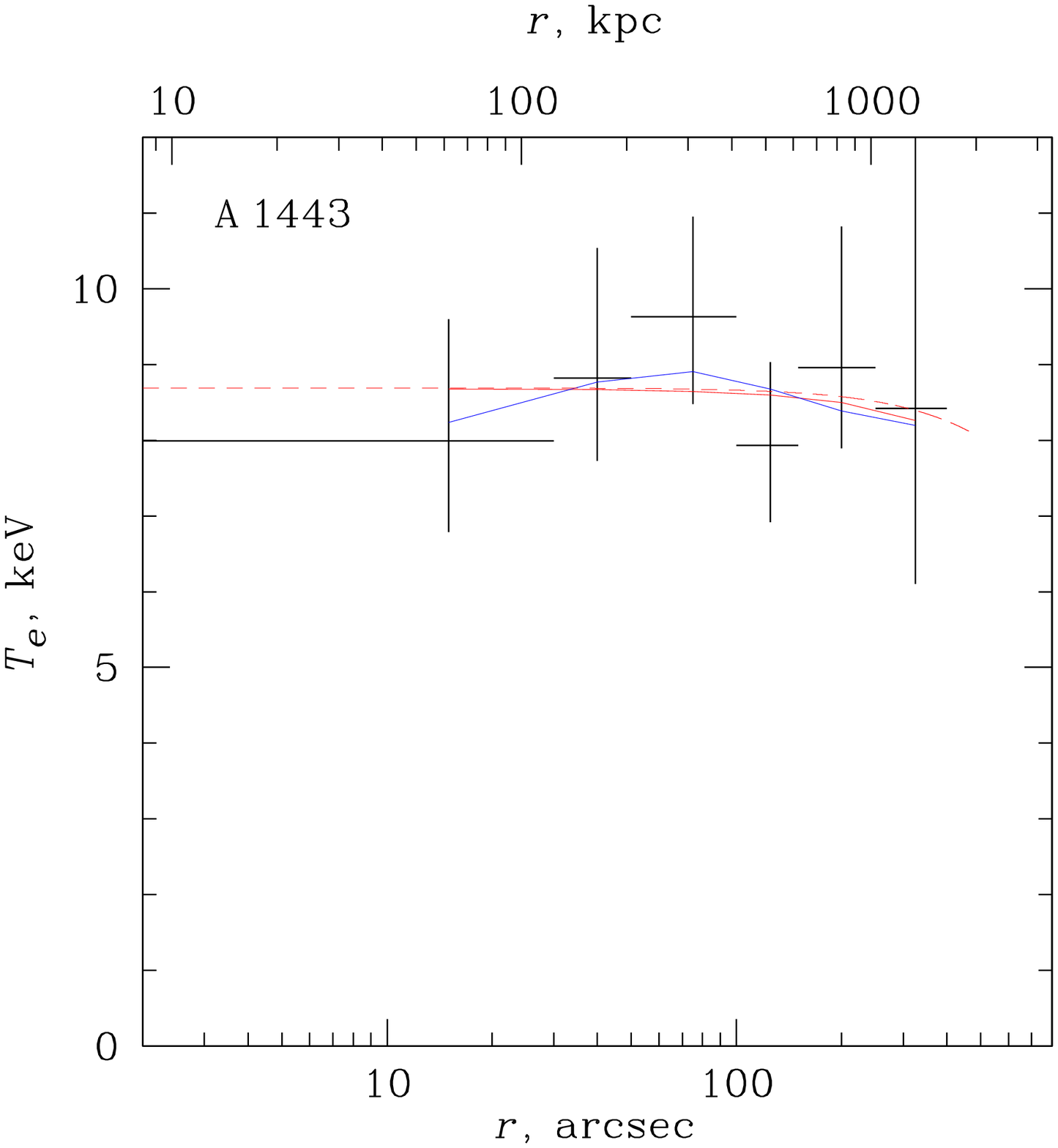}
\smallskip
\caption{Same as Fig.~\ref{fig:tprof1}, but for A\,781, A\,3444, A\,1132, A\,1351, A\,1423 and A\,1443.}
\label{fig:tprof2}
\end{figure*}

\begin{figure*}
\centering
\epsscale{1.1}
\includegraphics[width=7cm]{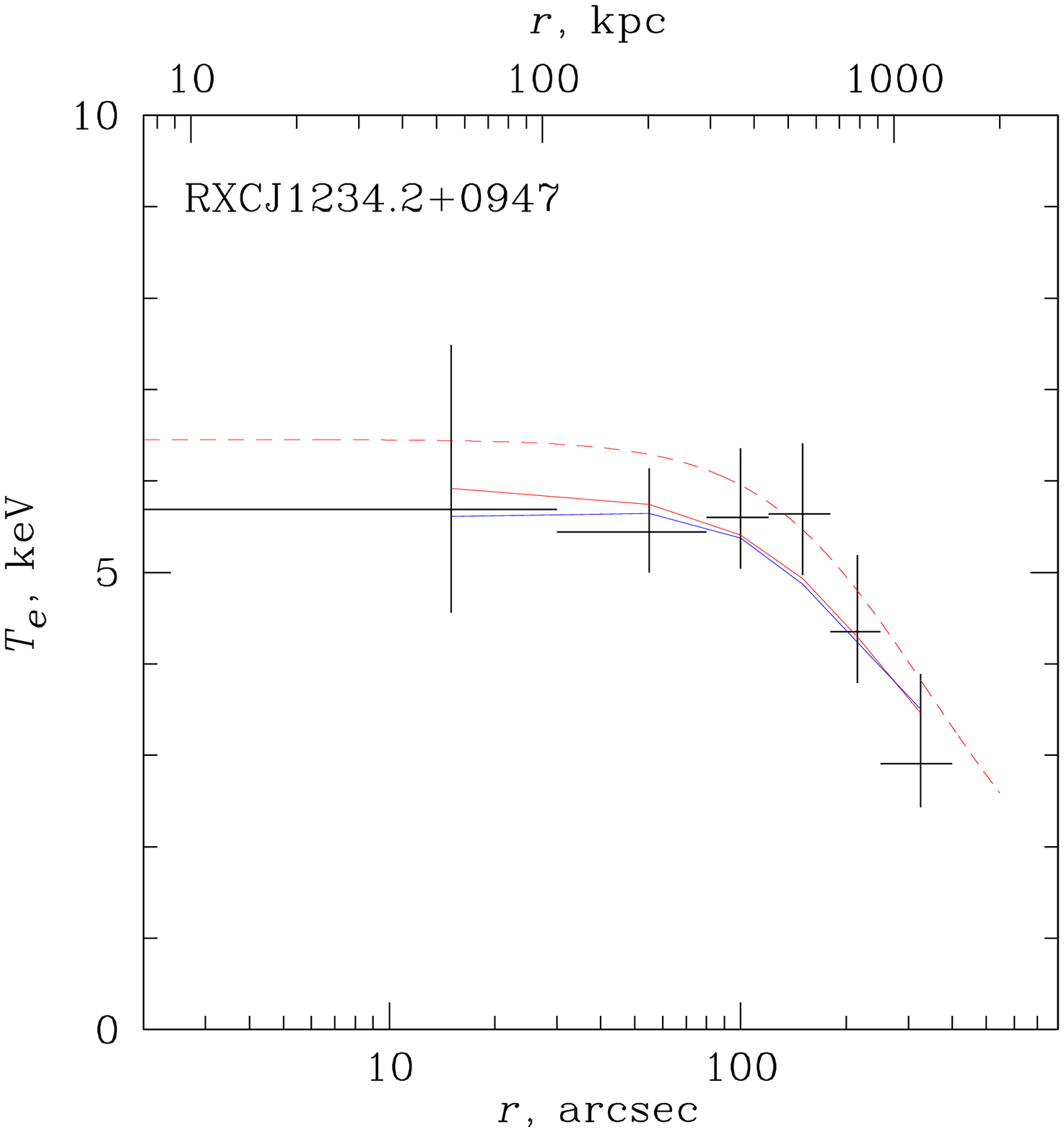}
\includegraphics[width=7cm]{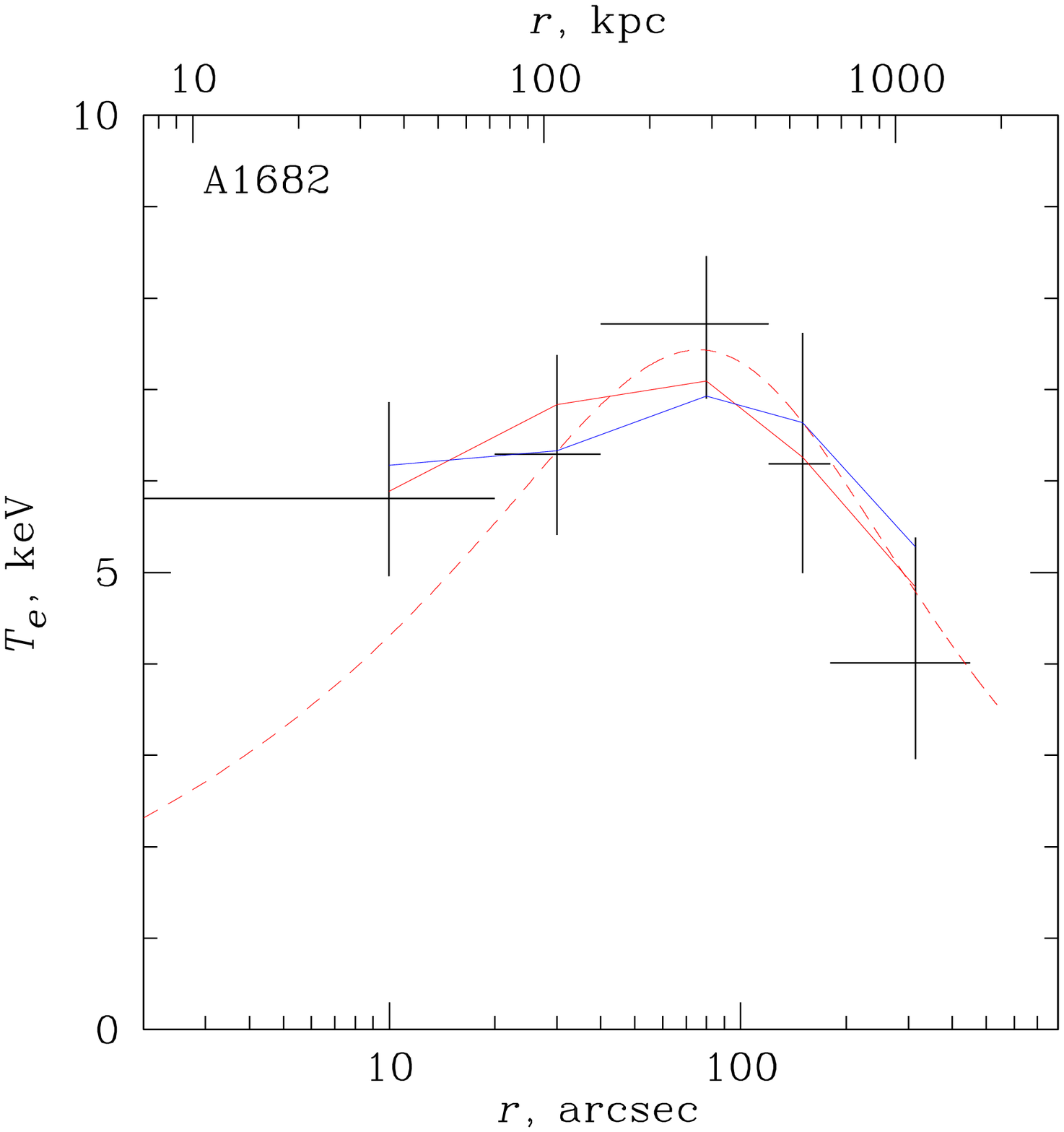}
\includegraphics[width=7cm]{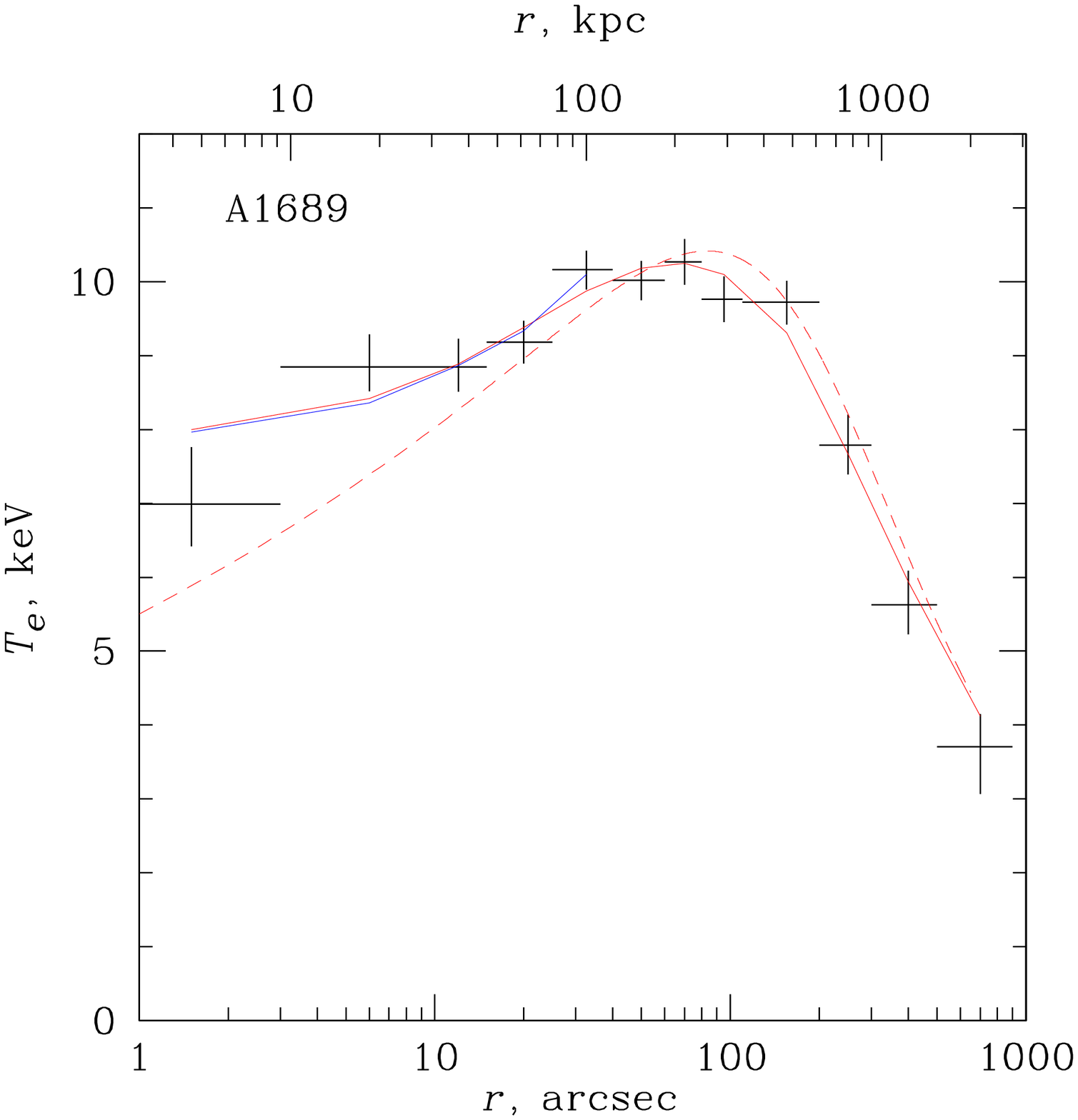}
\includegraphics[width=7cm]{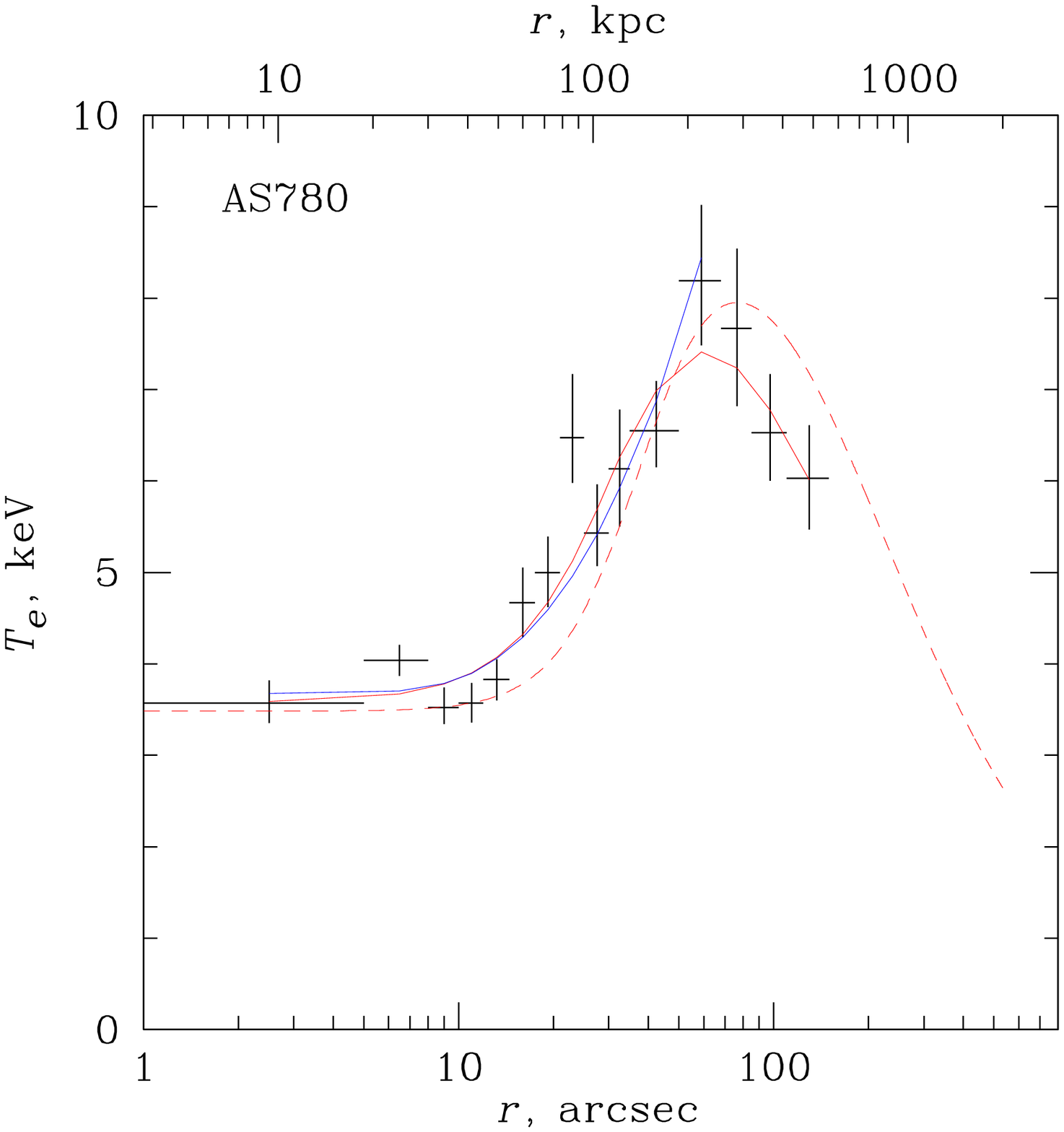}
\includegraphics[width=7cm]{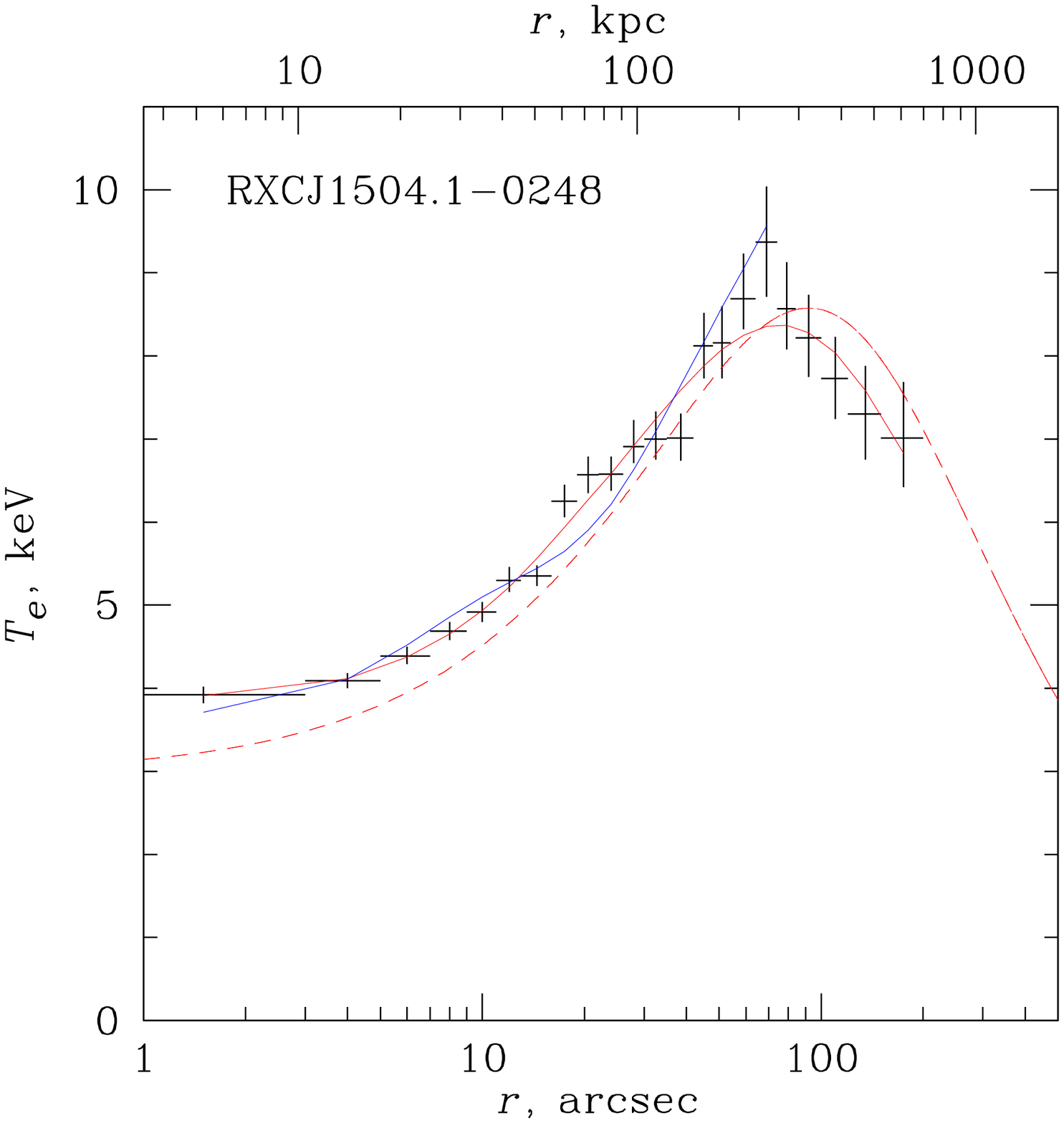}
\includegraphics[width=7cm]{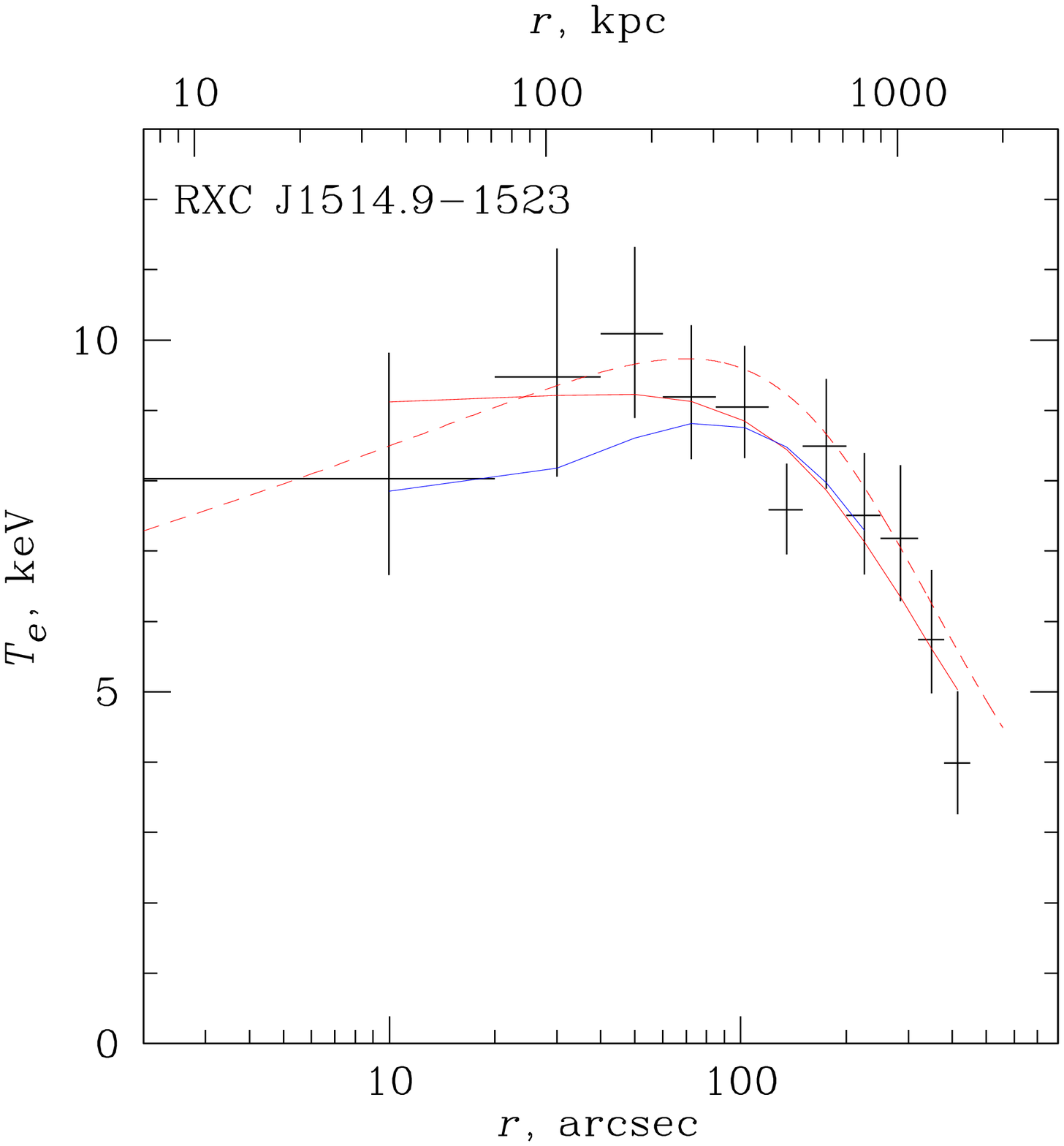}
\smallskip
\caption{Same as Fig.~\ref{fig:tprof1}, but for RXC\,J1234.2+0947, A\,1682, A\,1689, AS\,780, RXC\,J1504.1-0248 and RXC\,J1514.9-1523.}
\label{fig:tprof3}
\end{figure*}

\begin{figure*}
\centering
\epsscale{1.1}
\includegraphics[width=7cm]{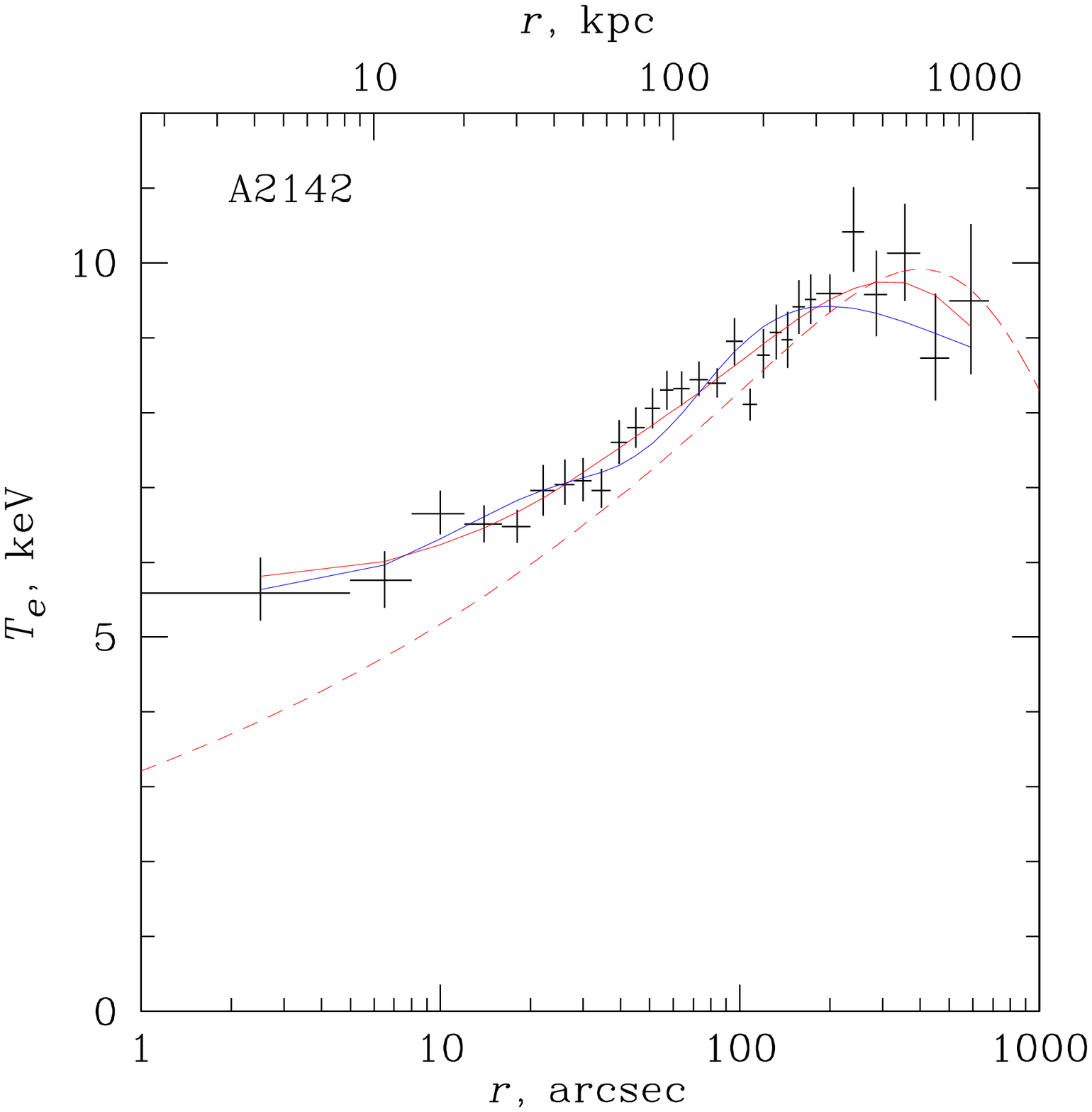}
\includegraphics[width=7cm]{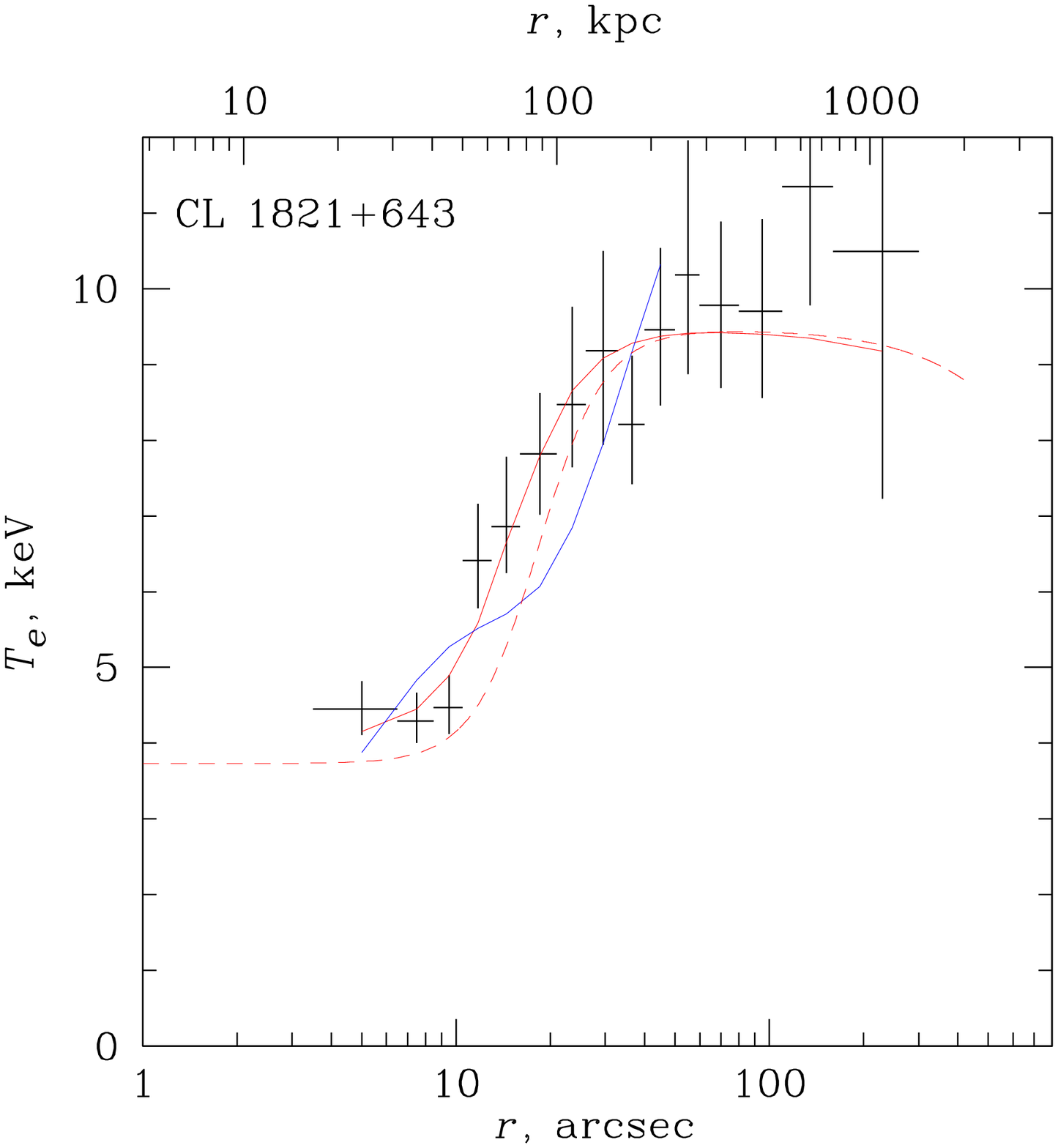}
\includegraphics[width=7cm]{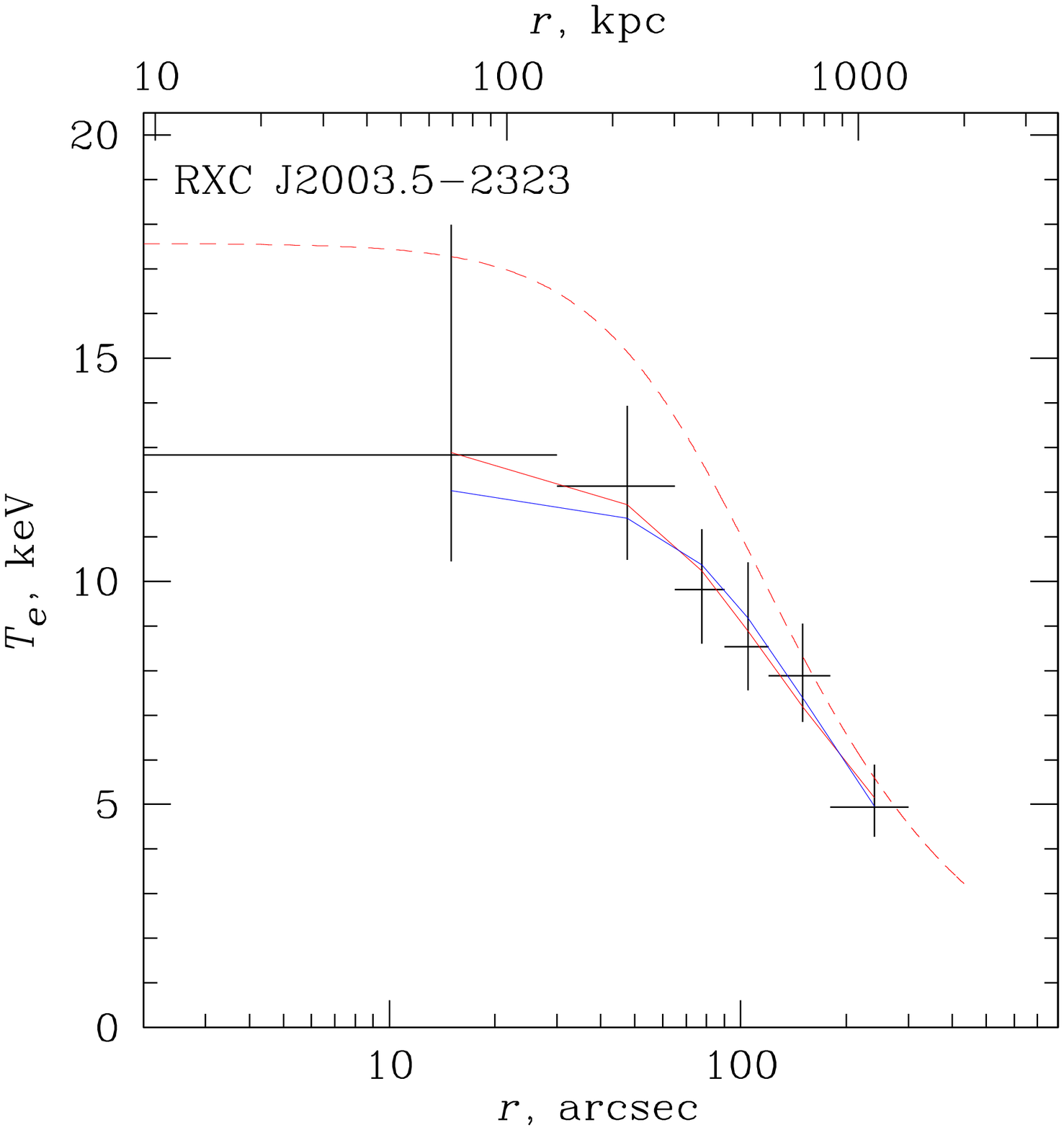}
\includegraphics[width=7cm]{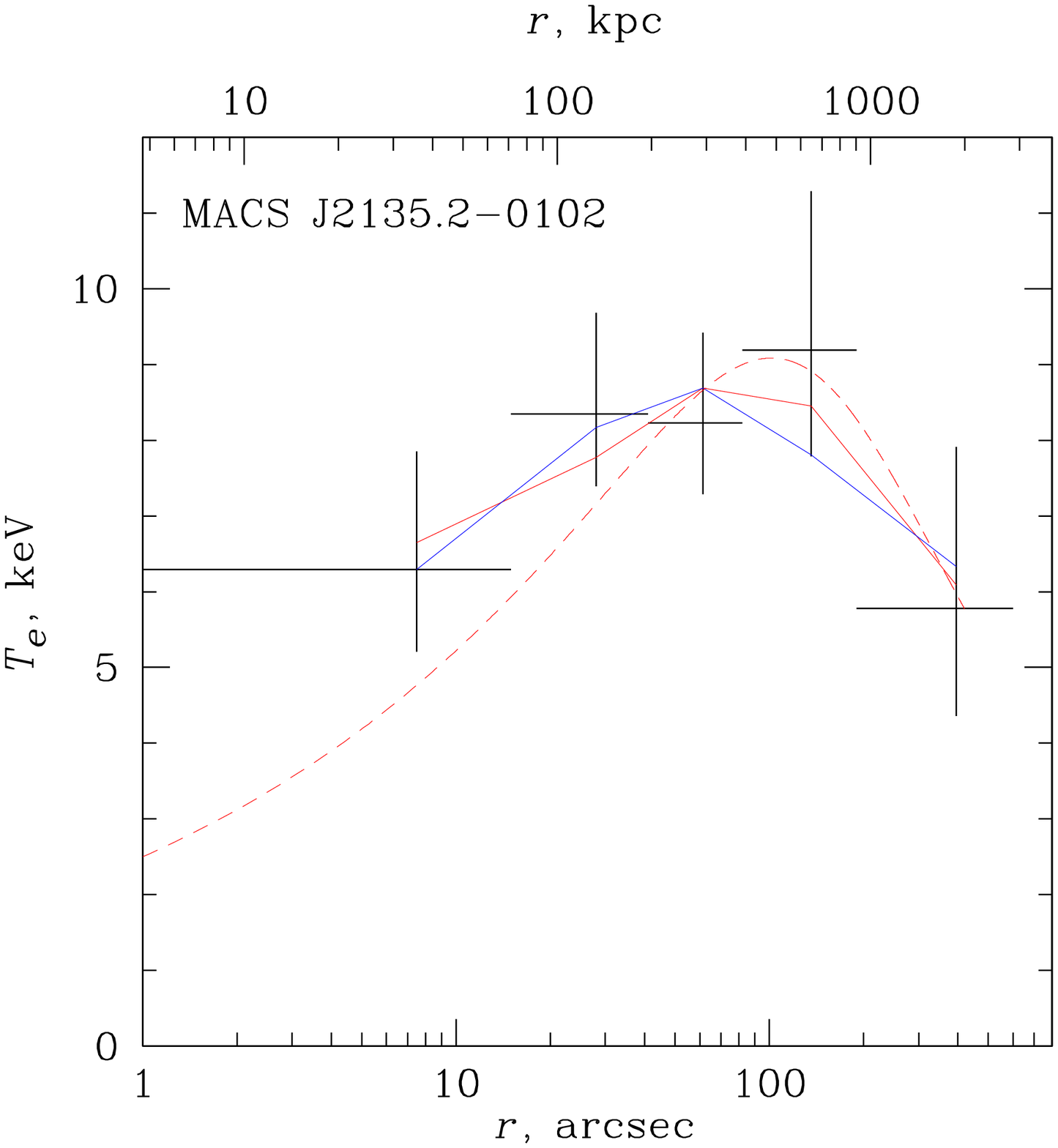}
\includegraphics[width=7cm]{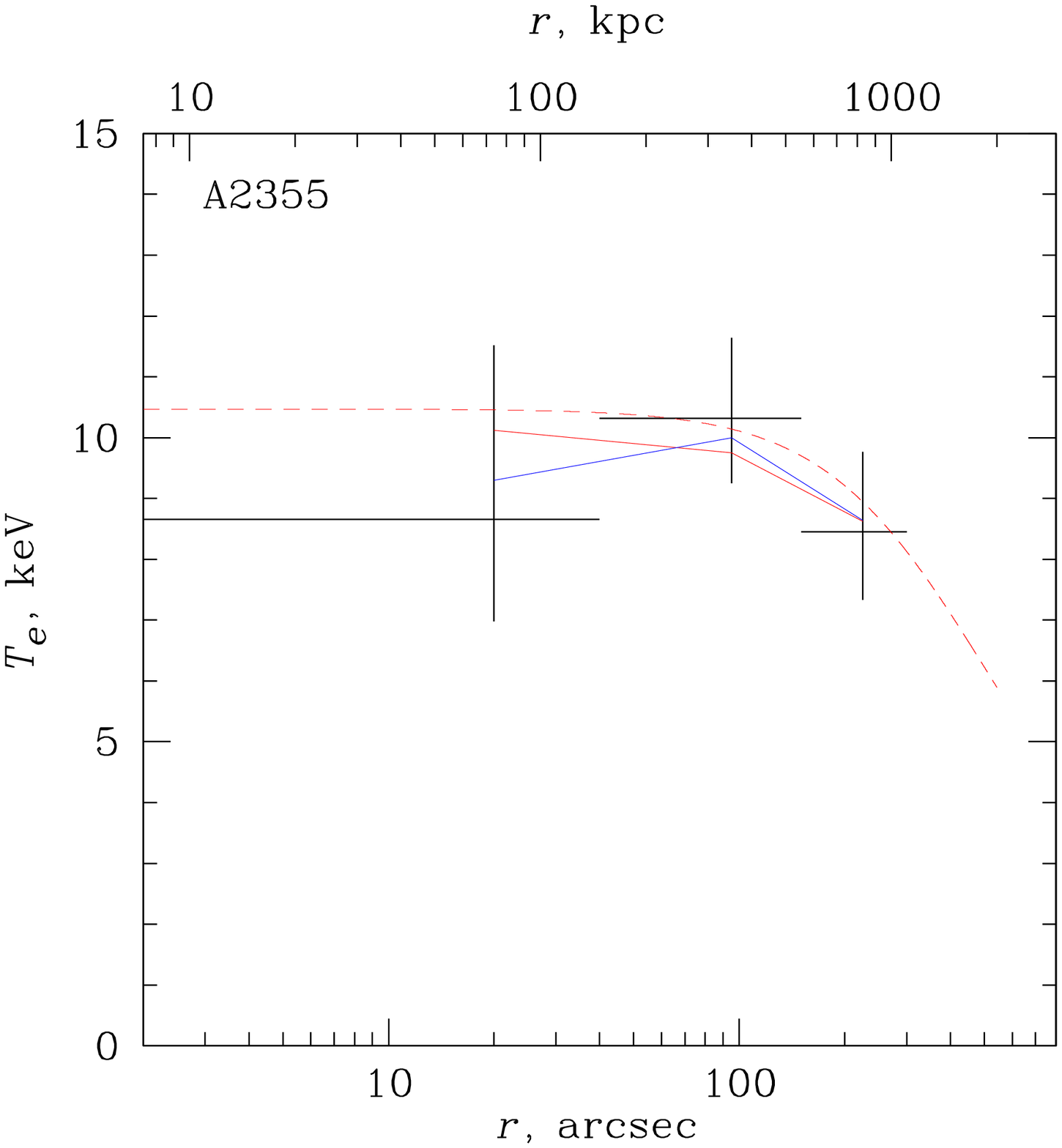}
\includegraphics[width=7cm]{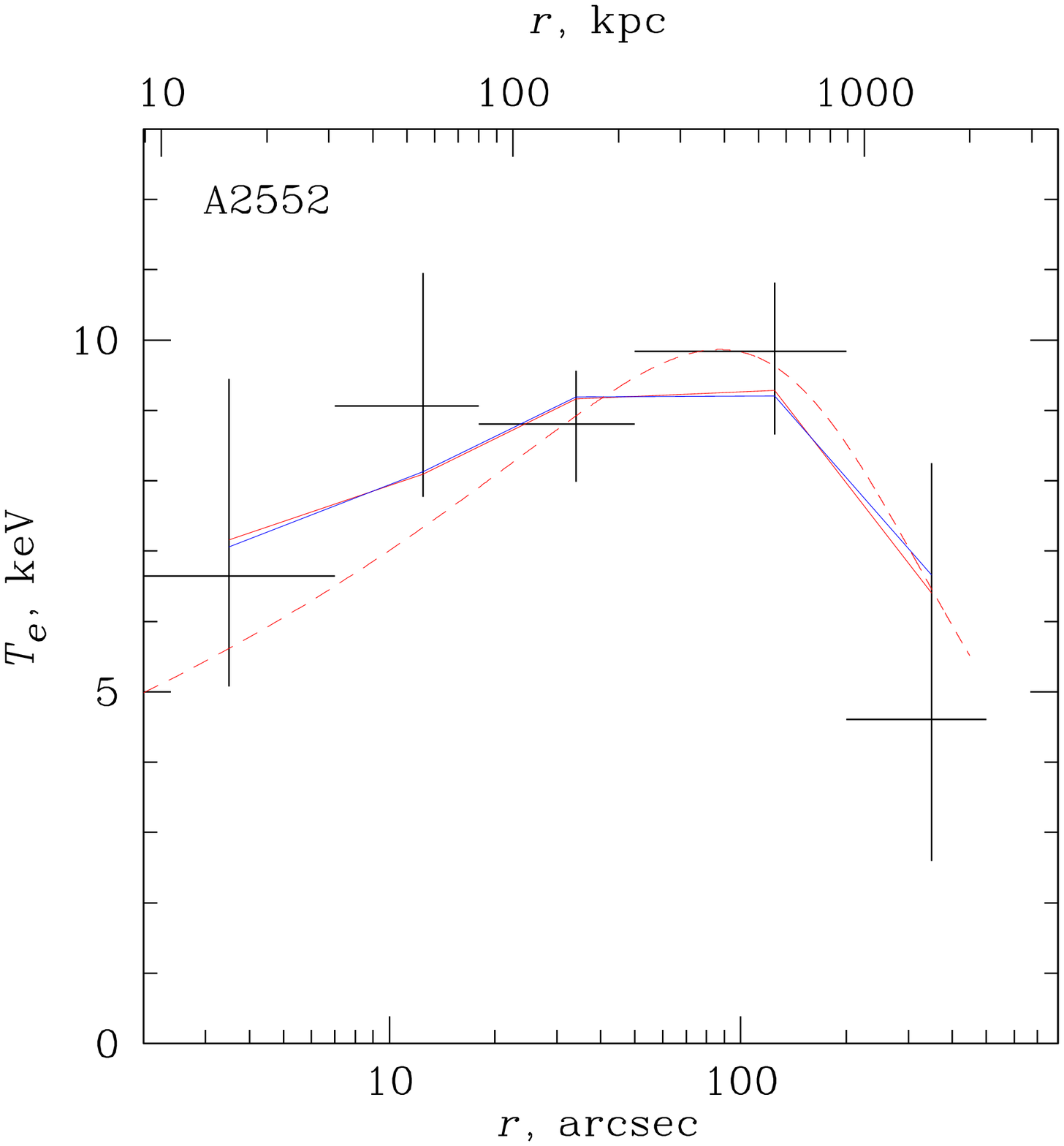}
\smallskip
\caption{Same as Fig.~\ref{fig:tprof1}, but for A\,2142, CL\,1821+643, RXC\,J2003.5-2323, MACS\,J2135.2-0102, A\,2355 and A\,2552.}
\label{fig:tprof4}
\end{figure*}

\begin{figure*}
\centering
\epsscale{1.1}
\includegraphics[width=7cm]{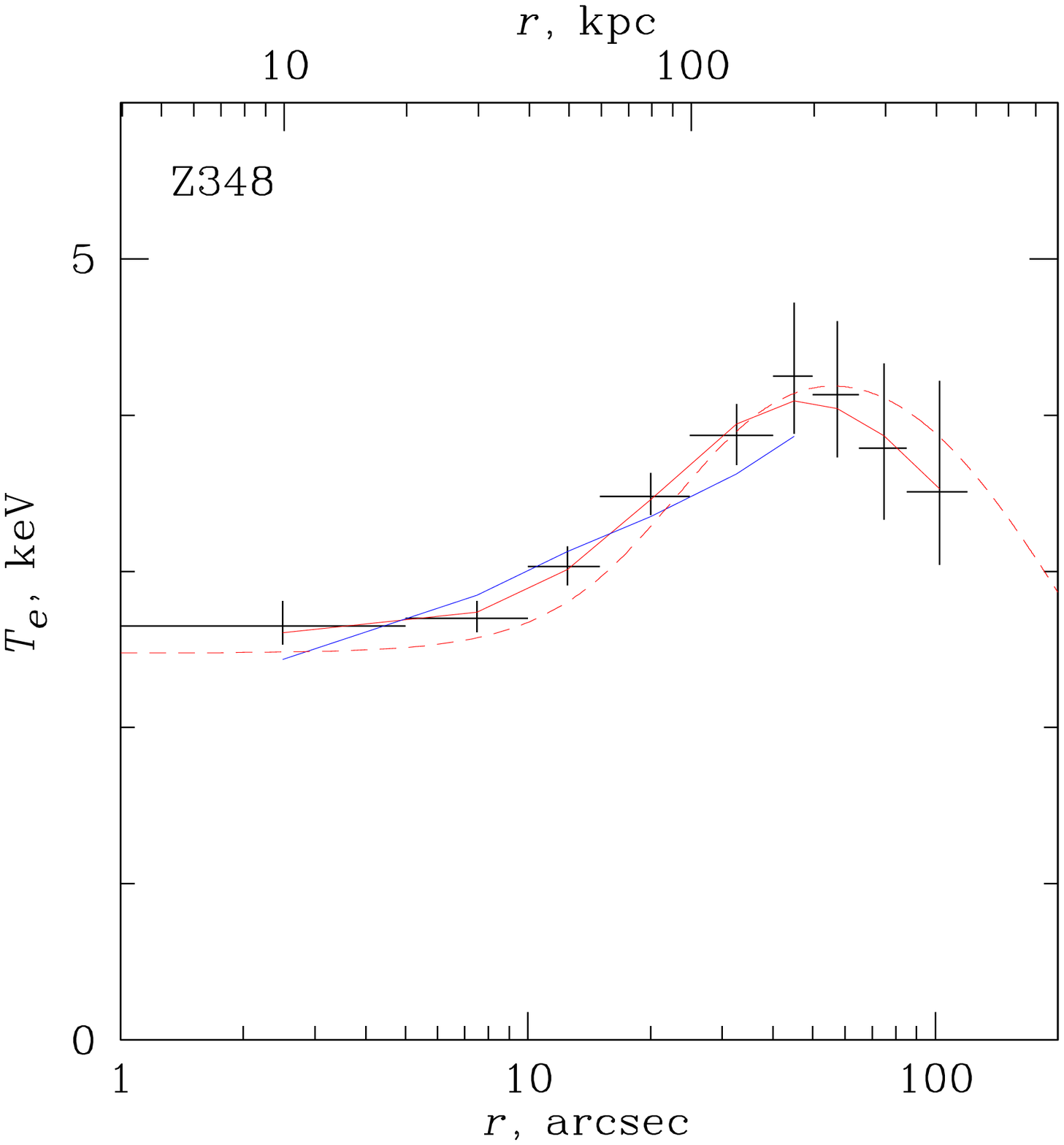}
\includegraphics[width=7cm]{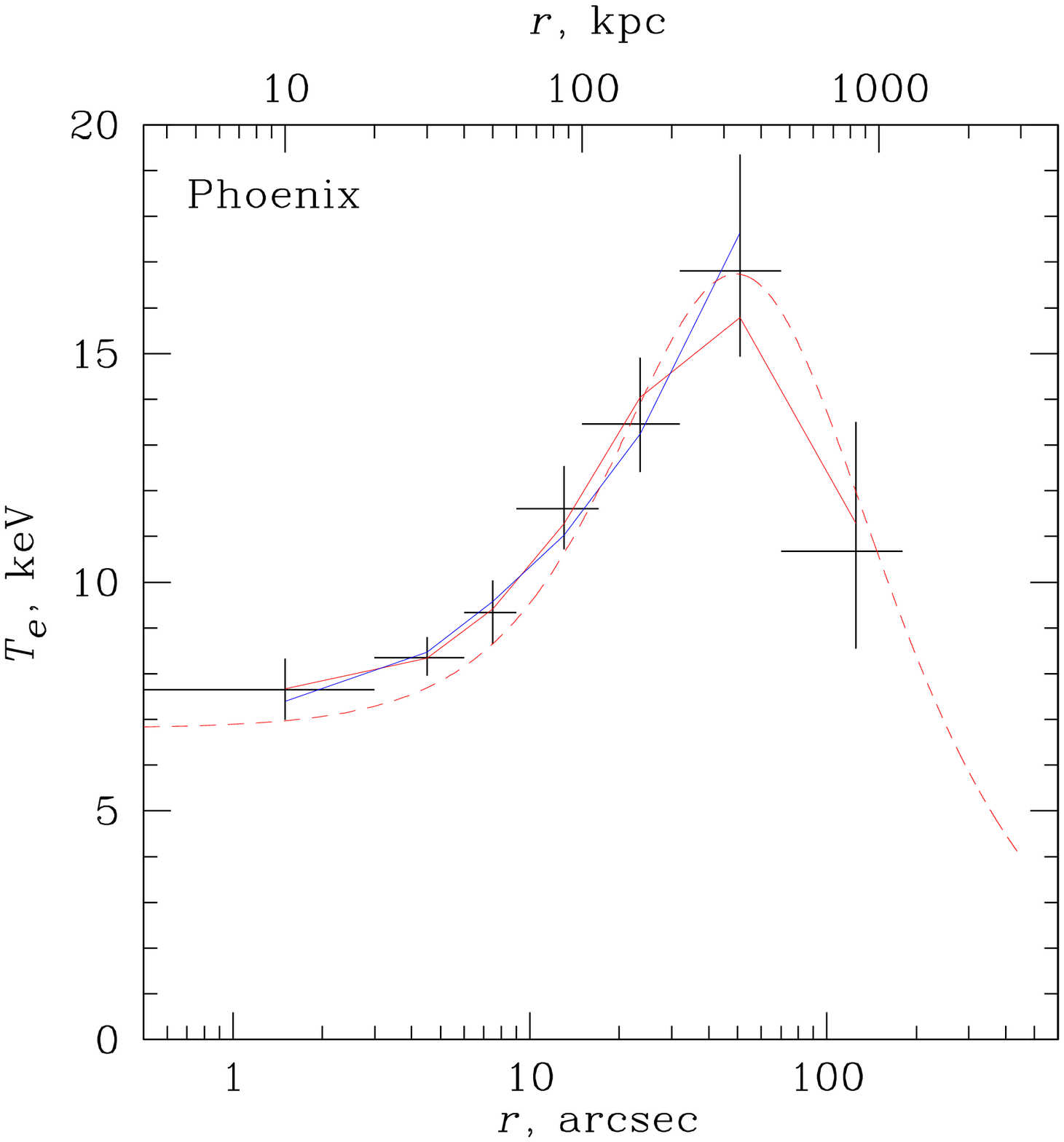}
\smallskip
\caption{Same as Fig.~\ref{fig:tprof1}, but for Z348 and the Phoenix cluster.}
\label{fig:tprof5}
\end{figure*}


\begin{deluxetable*}{lrrr|lrrr}\label{tab:k0}
\tablecaption{Temperatures and core entropies for statistical sample}
\tablehead{
\colhead{Cluster} & 
\colhead{$T_{\rm X, \,ce}$\phantom{0}} & 
\colhead{$K_{\rm 0}$\phantom{000}} & 
\colhead{Ref.} & 
\colhead{Cluster} & 
\colhead{$T_{\rm X, \,ce}$\phantom{00}} & 
\colhead{$K_{\rm 0}$\phantom{000}} & 
\colhead{Ref.}\\
\colhead{name} & 
\colhead{(keV)\phantom{0}}& 
\colhead{(keV cm$^2$)} &  
\colhead{}   &  
\colhead{name}   & 
\colhead{(keV)\phantom{00}}  & 
\colhead{(keV cm$^2$)}& \colhead{}  \\
}
\startdata
A\,2744              & $9.2^{+0.7}_{-0.6}$   & $438\pm59$   \phantom{0}& 2, 3 &\hspace{0.5cm}RXC\,J1234.2+0947   & $5.7^{+0.4}_{-0.4}$\phantom{0} &  $404\pm93$\phantom{0} & 1, 1   \\          
A\,68                & $9.0^{+1.5}_{-1.1}$   & $217\pm89$   \phantom{0}& 2, 3 &\hspace{0.5cm}A\,1682             & $7.2^{+0.6}_{-0.5}$\phantom{0} &  $143\pm26$\phantom{0} & 1, 1 \\   
A\,2813              & $8.2^{+0.5}_{-0.5}$   & $268\pm44$   \phantom{0}& 1, 3 &\hspace{0.5cm}A\,1689             &  $9.9^{+0.2}_{-0.2}$\phantom{0} &  $59\pm4$\phantom{0}   & 1,1 \\          
A\,2895              & $7.8^{+0.5}_{-0.5}$   & $173\pm65$   \phantom{0}& 1, 1 &\hspace{0.5cm}A\,1733             & $9.1^{+1.8}_{-1.3}$\phantom{0} &  $332\pm148$\tablenotemark{a} & 1, 1 \\    
A\,209               & $7.3^{+0.6}_{-0.5}$   & $106\pm27$   \phantom{0}& 2, 3 &\hspace{0.5cm}A\,1758a            & $12.1^{+1.2}_{-0.9}$\phantom{0}&  $231\pm37$\phantom{0} & 2, 3 \\         
RXC\,J0142.0+2131    & $7.1^{+0.6}_{-0.5}$   & $131\pm51$   \phantom{0}& 1, 1 &\hspace{0.5cm}A\,1763             & $7.8^{+0.7}_{-0.6}$\phantom{0} &  $215\pm33$\phantom{0} & 2, 3 \\         
MACS\,J0257.6$-$2209 & $8.0^{+1.1}_{-0.9}$   & $156\pm25$   \phantom{0}& 2, 3 &\hspace{0.5cm}A\,1835 (MH)         & $9.8^{+0.6}_{-0.5}$\phantom{0} &  $11\pm3$\phantom{0}   & 2, 3\\          
A\,401               & $7.8^{+0.6}_{-0.6}$   & $180\pm6$    \phantom{0}& 1, 1 &\hspace{0.5cm}A\,1914             & $9.6^{+0.6}_{-0.5}$\phantom{0} &  $107\pm18$\phantom{0} & 2, 3\\          
A\,3088              & $9.6^{+0.7}_{-0.7}$   & $83\pm8$     \phantom{0}& 1, 3 &\hspace{0.5cm}AS\,780 (MH)         & $7.1^{+0.3}_{-0.3}$\phantom{0} &  $19\pm2$\phantom{0}   & 1, 1\\             
PSZ1G171.96$-$40.64  & $11.0^{+0.9}_{-0.5}$  & $329\pm74$   \phantom{0}& 1, 1 &\hspace{0.5cm}RXC\,J1504.1$-$0248 (MH)& $7.6^{+0.1}_{-0.1}$\phantom{0} &  $11.1\pm0.3$\phantom{0}& 1, 1\\             
A\,478 (MH)           & $7.3^{+0.3}_{-0.2}$   &  $8\pm1$     \phantom{0}& 2, 3 &\hspace{0.5cm}A\,2029 (MH)         & $8.2^{+0.3}_{-0.3}\star$       &  $11\pm1$\phantom{0}   & 2, 3\\          
A\,521               & $7.0^{+0.6}_{-0.5}$   &  $260\pm36$  \phantom{0}& 2, 3 &\hspace{0.5cm}RXC\,J1514.9$-$1523 & $8.6^{+0.4}_{-0.3}$\phantom{0} &  $490\pm108$\phantom{0}& 1, 1 \\             
A\,520               & $9.3^{+0.7}_{-0.6}$   &  $326\pm29$  \phantom{0}& 2, 3 &\hspace{0.5cm}A\,2142             & $8.8^{+0.1}_{-0.1}$\phantom{0} &  $58\pm2$\phantom{0}   & 1, 1 \\            
RXC\,J0510.7$-$0801  & $8.4^{+0.5}_{-0.4}$   &  $158\pm99$  \phantom{0}& 1, 1 &\hspace{0.5cm}A\,2163             & $19.2^{+0.9}_{-0.8}$\phantom{0}&  $438\pm83$\phantom{0} & 2, 3\\          
RXCJ\,0520.7$-$1328  & $6.4^{+0.8}_{-0.7}$   &  $89\pm22$   \phantom{0}& 2, 3 &\hspace{0.5cm}A\,2204 (MH)         & $8.7^{+0.6}_{-0.5}$\phantom{0} &  $10\pm1$\phantom{0}   & 2, 3 \\         
PSZ1G139.61+24.20 (cMH)& $7.5^{+0.4}_{-0.4}$   &  $10\pm10$   \phantom{0}& 1, 1 &\hspace{0.5cm}A\,2218             & $7.3^{+0.4}_{-0.4}$\phantom{0} &  $289\pm20$\phantom{0} & 2, 3\\          
A\,665               & $7.5^{+0.4}_{-0.3}$   &  $135\pm24$  \phantom{0}& 2, 3 &\hspace{0.5cm}A\,2219             & $12.6^{+0.7}_{-0.6}\star$      &  $412\pm43$\phantom{0} & 2, 3\\          
A\,697               & $9.5^{+0.9}_{-0.8}$   &  $167\pm24$  \phantom{0}& 2, 3 &\hspace{0.5cm}A\,2256             & $5.7^{+0.2}_{-0.2}\star$       &  $350\pm12$\phantom{0} & 2, 3\\          
A\,754               & $10.0^{+0.3}_{-0.3}$  &  $270\pm24$  \phantom{0}& 4, 3 &\hspace{0.5cm}RXC\,J1720.1+2637 (MH)& $6.4^{+0.3}_{-0.3}$\phantom{0} &  $21\pm2$\phantom{0}   & 2, 3\\          
A\,773               & $7.8^{+0.7}_{-0.6}$   &  $244\pm32$  \phantom{0}& 2, 3 &\hspace{0.5cm}A\,2261             & $7.6^{+0.5}_{-0.4}$\phantom{0} &  $61\pm8$\phantom{0}   & 2, 3\\          
A\,781               & $8.2^{+0.7}_{-0.6}$   &  $170\pm36$  \phantom{0}& 1, 1 &\hspace{0.5cm}CL\,1821+643        & $9.5^{+0.4}_{-0.4}$\phantom{0} &  $8\pm5$\phantom{0}    & 1, 1\\             
RBS\,797 (MH)         & $7.7^{+0.9}_{-0.8}$   &  $21\pm2$    \phantom{0}& 2, 3 &\hspace{0.5cm}RXC\,J2003.5$-$2323 & $10.8^{+0.8}_{-0.6}$\phantom{0}&  $708\pm85$\phantom{0} & 1, 1 \\  
A\,3444 (MH)          & $7.1^{+0.2}_{-0.2}$   &  $18\pm2$    \phantom{0}& 1, 1 &\hspace{0.5cm}MACS\,J2135.2$-$0102& $8.6^{+0.8}_{-0.6}$\phantom{0} &  $142\pm18$\phantom{0}& 1, 1 \\  
A\,1132              & $6.8^{+0.6}_{-0.5}$   &  $154\pm31$  \phantom{0}& 1, 1 &\hspace{0.5cm}A\,2355             & $9.4^{+0.9}_{-0.9}$\phantom{0} &  $519\pm117$\phantom{0}& 1, 1 \\   
RXC\,J1115.8+0129 (cMH) & $6.8^{+1.2}_{-0.9}$   &  $23\pm5$    \phantom{0}& 2, 3 &\hspace{0.5cm}A\,2390             & $10.9^{+0.3}_{-0.3}$\phantom{0}&  $15\pm7$\phantom{0}   & 2, 3\\          
A\,1300              & $8.6^{+1.2}_{-1.0}$   &  $97\pm23$   \phantom{0}& 2, 3 &\hspace{0.5cm}A\,2537             & $8.4^{+0.8}_{-0.7}$\phantom{0} &  $110\pm19$\phantom{0} & 2, 3\\          
A\,1351              & $9.9^{+0.7}_{-0.7}$   &  $620\pm93$  \phantom{0}& 1, 1 &\hspace{0.5cm}A\,2552             & $9.2^{+0.7}_{-0.7}$\phantom{0} &  $78\pm33$\phantom{0}  & 1, 1\\             
A\,1423              & $6.4^{+0.3}_{-0.3}$   &  $27\pm18$   \phantom{0}& 1, 1 &\hspace{0.5cm}A\,2631             & $7.1^{+1.1}_{-0.8}$\phantom{0} &  $309\pm37$\phantom{0} & 2, 3\\          
A\,1443              & $8.6^{+0.6}_{-0.4}$   &  $283\pm57$  \phantom{0}& 1, 1 &\hspace{0.5cm}A\,2667 (MH)         & $7.3^{+0.4}_{-0.4}$\phantom{0} &  $19\pm3$\phantom{0}   & 1, 3\\   
\enddata
\tablenotetext{a}{from the temperature measured within the central $r=120$
  kpc.}
\tablecomments{Column 1: cluster name. The presence of a minihalo or a
  candidate is indicated as MH and cMH, respectively. Column 2: temperature
  within R$_{2500}$ (R$_{5000}$ for those clusters marked with $\star$)
  measured in the $0.7-7$ keV band; the central $r=70$ kpc region has been
  excised for all clusters except A\,754. Column 3: core entropy. Column 4:
  references for temperature and the core entropy floor, respectively: (1)
  this work, (2) C08, (3) C09, (4) Markevitch et al. (2003).}
\end{deluxetable*}


\begin{deluxetable*}{lrrr|lrrr}\label{tab:k02}
\tablecaption{Temperatures and core entropies for supplementary sample}
\tablehead{
\colhead{Cluster} & 
\colhead{$T_{\rm X, \,ce}$\phantom{0}} & 
\colhead{$K_{\rm 0}$\phantom{000}} & 
\colhead{Ref.} & 
\colhead{Cluster} & 
\colhead{$T_{\rm X, \,ce}$\phantom{00}} & 
\colhead{$K_{\rm 0}$\phantom{000}} & 
\colhead{Ref.}\\
\colhead{name} &
\colhead{(keV)\phantom{0}}& 
\colhead{(keV cm$^2$)} &  
\colhead{}   &  
\colhead{name}   & 
\colhead{(keV)\phantom{00}}  & 
\colhead{(keV cm$^2$)}& \colhead{}  \\
}
\startdata
Z348                 & $3.9^{+0.1}_{-0.1}$\phantom{0} &  $13\pm1$    \phantom{0}& 1,1   & A\,1413 (cMH)        & $8.3^{+0.2}_{-0.2}$\phantom{0} &  $64\pm8$    \phantom{0}& 1,3    \\
A\,119               & $5.9^{+0.3}_{-0.3}\star$       &  $234\pm88$  \phantom{0}& 2,3   & \hspace{0.5cm}A\,1576              & $9.5^{+0.7}_{-0.7}$\phantom{0}   &  $186\pm49$  \phantom{0}& 1,3 \\ 
A\,141               & $7.2^{+0.6}_{-0.5}$\phantom{0} &  $205\pm27$  \phantom{0}& 1,3   & \hspace{0.5cm}A\,1650              & $6.1^{+0.1}_{-0.1}$\phantom{0}   &  $38\pm10$   \phantom{0}& 1,3 \\   
A\,193               & $4.2^{+1.0}_{-0.6}$\phantom{0} &  $186\pm13$  \phantom{0}& 5,3   & \hspace{0.5cm}RX\,J1347.5$-$1145 (MH)  & $14.6^{+1.0}_{-0.8}$\phantom{0}  &  $13\pm21$   \phantom{0}& 2,3  \\ 
A\,267               & $6.7^{+0.6}_{-0.5}$\phantom{0} &  $169\pm18$  \phantom{0}& 2,3   & \hspace{0.5cm}A\,1795 (cMH)        & $6.1^{+0.2}_{-0.2}\star$         &  $19\pm1$    \phantom{0}& 2,3  \\      
MACS\,J0159.8$-$0849 (MH) & $9.2^{+0.7}_{-0.6}$\phantom{0} &  $19\pm4$    \phantom{0}& 2,3   &  \hspace{0.5cm}A\,1995              & $8.4^{+0.7}_{-0.6}$\phantom{0}   &  $374\pm60$  \phantom{0}& 2,3  \\  
A\,383               & $4.9^{+0.3}_{-0.3}$\phantom{0} &  $13\pm2$    \phantom{0}& 2,3   &  \hspace{0.5cm}MS\,1455.0+2232 (MH) & $4.8^{+0.1}_{-0.1}$\phantom{0}   &  $17\pm2$    \phantom{0}& 2,3  \\
A\,399               & $8.0^{+0.4}_{-0.3}$\phantom{0} &  $153\pm19$  \phantom{0}& 2,3   &  \hspace{0.5cm}A\,2034              & $7.2^{+0.2}_{-0.2}$\phantom{0}   &  $233\pm23$  \phantom{0}& 2,3  \\      
Perseus (MH)         & $6.4^{+0.1}_{-0.1}$\phantom{0} & $19.4\pm0.2$ \phantom{0}& 4,3   &  \hspace{0.5cm}RX\,J1532.9+3021 (MH) & $6.0^{+0.4}_{-0.4}$\phantom{0}   &  $17\pm2$    \phantom{0}& 2,3  \\     
MACS\,J0329.6$-$0211 (MH) & $6.3^{+0.5}_{-0.4}$\phantom{0}&  $11\pm3$\phantom{0}& 2,3   &  \hspace{0.5cm}A\,2111              & $7.1^{+1.3}_{-1.0}$\phantom{0}   &  $107\pm97$  \phantom{0}& 2,3  \\      
2A\,0335+096  (MH)       & $3.6^{+0.1}_{-0.1}$\phantom{0}&$7.1\pm0.1$\phantom{0}& 5,3   &  \hspace{0.5cm}A\,2125         & $2.9^{+0.3}_{-0.3}$\phantom{0}   &  $225\pm32$  \phantom{0}& 2,3  \\           
MACS\,J0417.5$-$1154 & $11.1^{+2.0}_{-1.5}$\phantom{0}&  $27\pm7.3$  \phantom{0}& 2,3   &  \hspace{0.5cm}Ophiuchus (MH)   & $10.3^{+0.2}_{-0.2}$\phantom{0}  &  $9\pm1$     \phantom{0}& 4,3  \\          
MACS\,J0429.6$-$0253 & $5.7^{+0.6}_{-0.5}$\phantom{0} &  $17\pm4.3$  \phantom{0}& 2,3   &  \hspace{0.5cm}A\,2255              & $6.1^{0.2}_{-0.2}$\phantom{0}    &  $529\pm28$  \phantom{0}& 2,3  \\      
RX\,J0439.0+0715     & $5.6^{+0.4}_{-0.3}$\phantom{0} &  $67\pm19$   \phantom{0}& 2,3   &  \hspace{0.5cm}RX\,J1720.2+3536 (cMH) & $7.2^{+0.5}_{-0.5}$\phantom{0}   &  $24\pm3$    \phantom{0}& 2,3 \\         
MS\,0440.5+0204      & $6.0^{+0.9}_{-0.7}\star$       &  $26\pm8$    \phantom{0}& 2,3   & \hspace{0.5cm}ZwCl\,1742.1+3306 (u)   & $4.4^{+0.2}_{-0.2}$\phantom{0}   &  $24\pm2$    \phantom{0}& 1,3  \\    
A\,611               & $7.1^{+0.6}_{-0.5}$\phantom{0} &  $125\pm19$  \phantom{0}& 2,3   & \hspace{0.5cm}A\,2319              & $8.8^{+0.3}_{-0.2}$\phantom{0}   &  $270\pm5$   \phantom{0}& 4,3  \\    
MS\,0839.8+2938      & $4.7^{+0.3}_{-0.3}$\phantom{0} &  $19\pm3$    \phantom{0}& 2,3   & \hspace{0.5cm}MACS\,J1931.8$-$2634 (u) & $7.0^{+0.7}_{-0.6}$\phantom{0}   &  $15\pm4$    \phantom{0}& 2,3  \\   
Z2089                & $4.4^{+0.2}_{-0.2}$\phantom{0} &  $24\pm5$    \phantom{0}& 1,3   & \hspace{0.5cm}RX\,J2129.6+0005 (MH) & $7.3^{+0.3}_{-0.3}$\phantom{0}   &  $21\pm4$    \phantom{0}& 1,3  \\   
ZwCl\,2701           & $5.2^{+0.3}_{-0.3}$\phantom{0} &  $40\pm4$    \phantom{0}& 2,3   & \hspace{0.5cm}A\,2420              & $6.6^{+0.2}_{-0.2}$\phantom{0}   &  $333\pm68$  \phantom{0}& 1,3  \\    
A\,907 (MH)          & $5.6^{+0.2}_{-0.2}$\phantom{0} &  $23\pm3$    \phantom{0}& 2,3   & \hspace{0.5cm}MACS\,J2228.5+2036    & $7.9^{+1.1}_{-0.9}$\phantom{0}   &  $119\pm39$  \phantom{0}& 2,3  \\   
ZWCL\,3146 (MH)      & $7.5^{+0.3}_{-0.3}$\phantom{0} &  $11\pm2$    \phantom{0}& 2,3   & \hspace{0.5cm}MACS\,J2245.0+2637    & $6.1^{+0.6}_{-0.5}$\phantom{0}   &  $42\pm7$    \phantom{0}& 2,3  \\   
A\,1068 (cMH)        & $4.7^{+0.2}_{-0.2}\star$       &  $9\pm1$     \phantom{0}& 2,3   & \hspace{0.5cm}A\,2556               & $3.6^{+0.2}_{-0.2}\star$         &  $12\pm1$    \phantom{0}& 2,3  \\   
A\,1204              & $3.6^{+0.2}_{-0.2}$\phantom{0} &  $15\pm1$    \phantom{0}& 2,3   & \hspace{0.5cm}A\,2626  (u)          & $3.6^{+0.1}_{-0.1}$\phantom{0}   &  $23\pm3$    \phantom{0}& 1,3  \\   
A\,1240              & $4.4^{+0.3}_{-0.3}$\phantom{0} &  $462\pm42$  \phantom{0}& 1,3   & \hspace{0.5cm}Phoenix (MH)          & $12.9^{+0.7}_{-0.7}$\phantom{0}  &  $19\pm3$    \phantom{0}& 1,1  \\   
\enddata
\tablecomments{Column 1: cluster name. The presence of a minihalo, a
  candidate minihalo or a central diffuse source with uncertain
  classification (\S\ref{sec:supp}) are indicated as MH, cMH and u
  respectively.  Column 2: cluster global temperature within R$_{2500}$
  (R$_{5000}$ for those clusters marked with $\star$) measured in the
  $0.7-7$ keV band; the central $r=70$ kpc region has been excised for all
  clusters except Perseus, 2A\,0335+096, Ophiuchus, A\,193 and A\,2319.
  Column 3: core entropy.
Column 4: references for temperature and the core entropy floor: (1) this
work, (2) C08, (3) C09, (4) Ikebe et al. (2002), (5) David et al. (1993).}
\end{deluxetable*}

\begin{figure*}
\centering
\includegraphics[width=6.2cm]{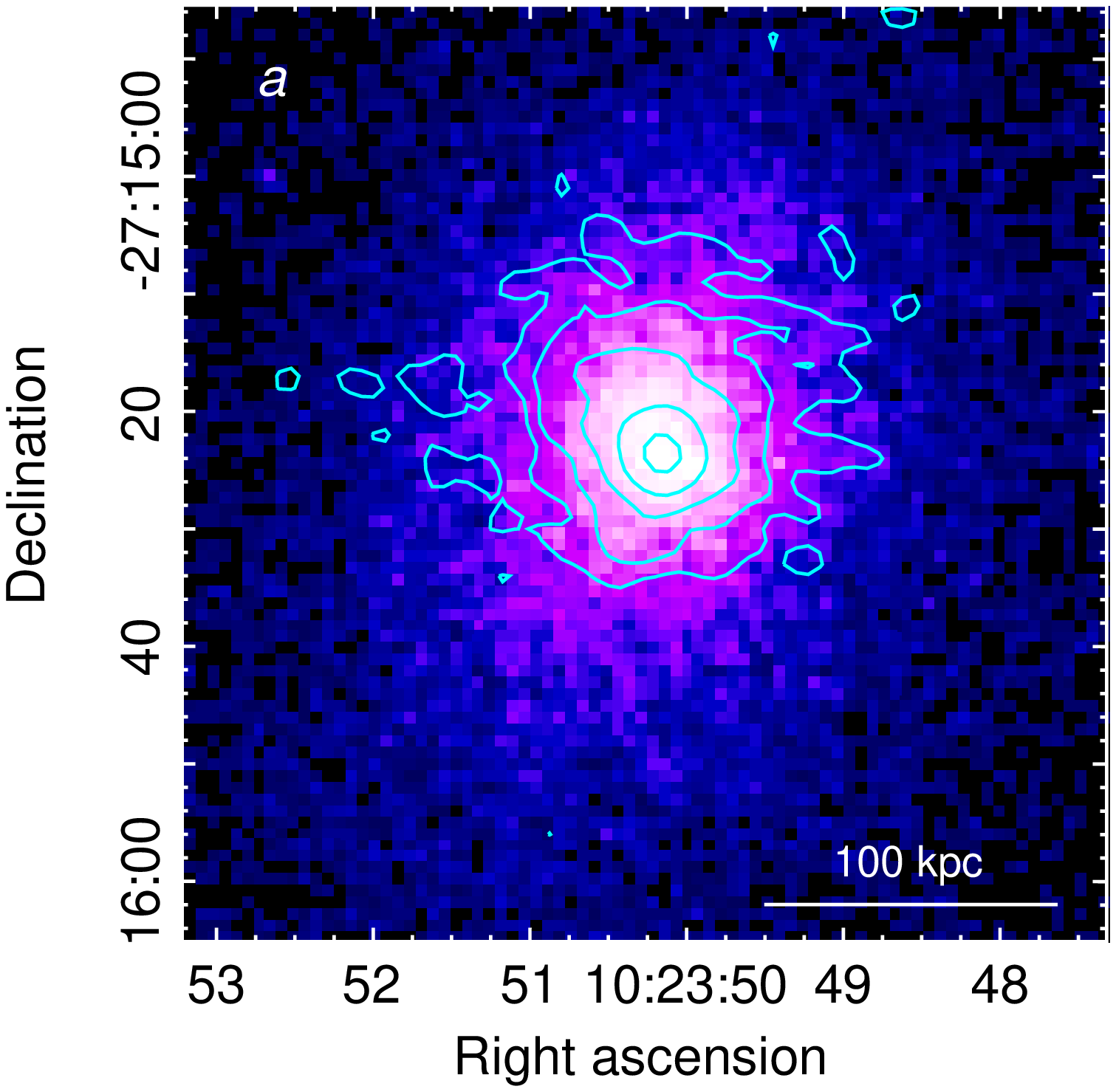}
\hspace{-0.5cm}
\includegraphics[width=6.2cm]{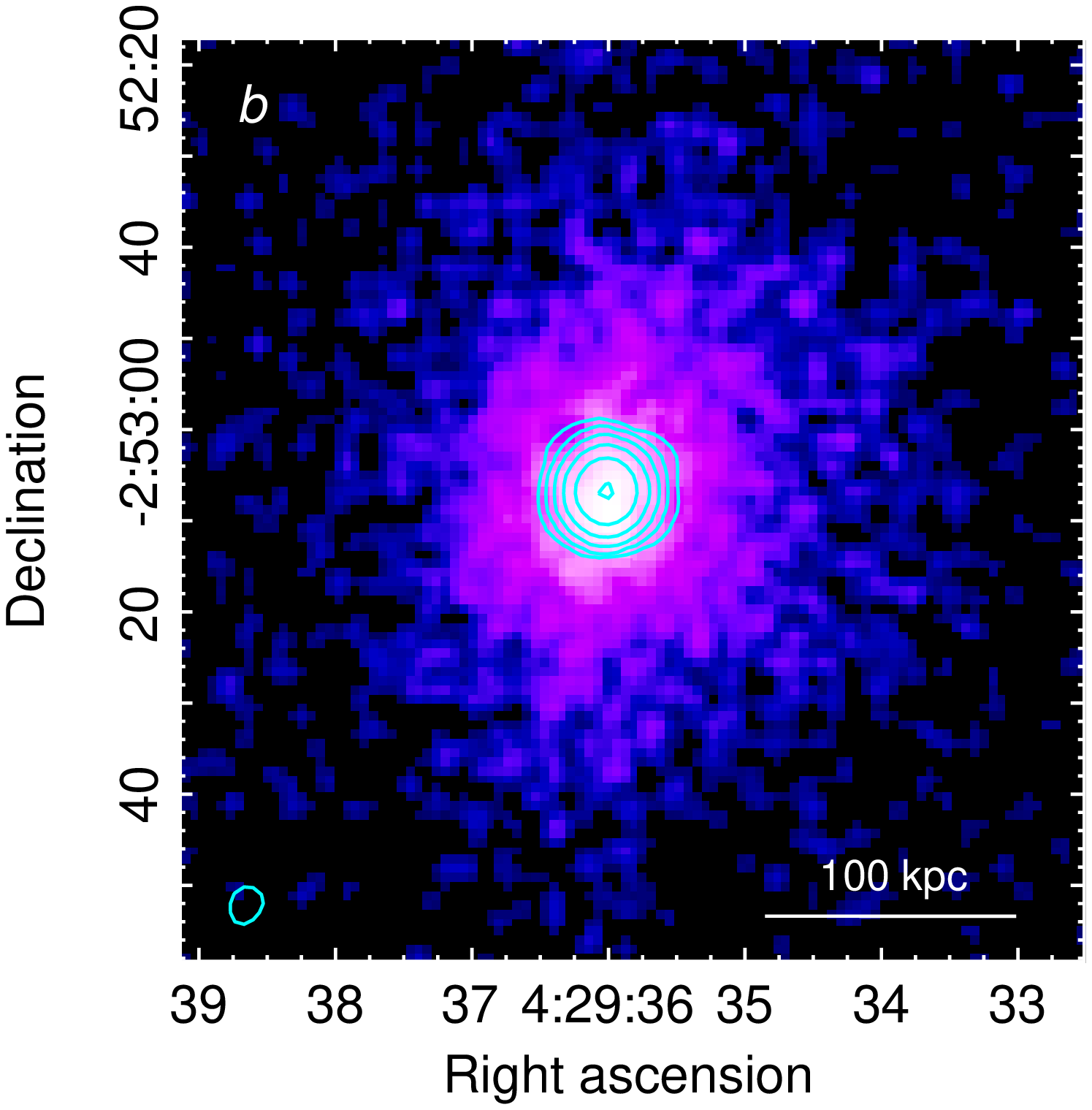}
\hspace{-0.5cm}
\includegraphics[width=6.2cm]{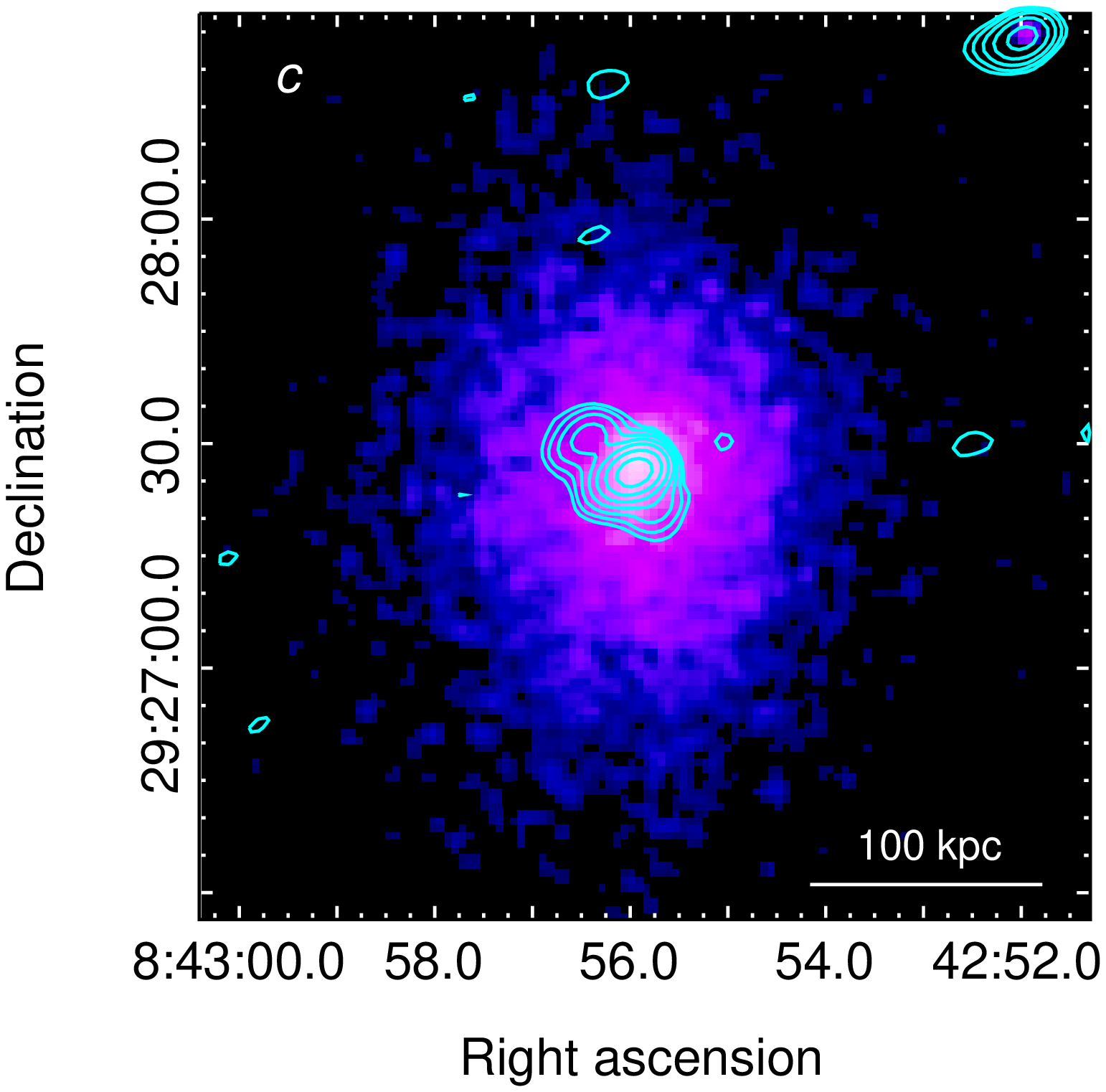}
\smallskip
\caption{{\em (a):} Radio minihalo in the cool core of A\,3444, 
  overlaid on the {\em Chandra} 0.5-4 keV image.  The radio contours are
  from a {\em VLA} BnA--configuration image at 1.4 GHz (from G17). The
  restoring beam is $5^{\prime\prime}$ and {\em rms} noise is 35 $\mu$Jy
  beam$^{-1}$. Contours are 0.09, 0.18, 0.36, 0.72, 1.44 mJy beam$^{-1}$.
  {\em (b,c):} examples of two cool-core clusters, MACS\,J0429.6-0253 and
  MS\,0839.8+2938, without a minihalo. For both clusters, the {\em VLA}
  B--configuration images at 1.4 GHz are overlaid as contours on the {\em
    Chandra} 0.5-4 keV image. For MACS\,J0429.6-0253, the restoring beam is
  $5^{\prime\prime}\times4^{\prime\prime}$ and {\em rms} noise is 40 $\mu$Jy
  beam$^{-1}$. Contours are 0.1, 0.4, 1.6, 6.4, 25.6, 104.4 mJy beam$^{-1}$.
  For MS\,0839.8+2938, the restoring beam is
  $6^{\prime\prime}\times4^{\prime\prime}$ and {\em rms} noise is 40 $\mu$Jy
  beam$^{-1}$. Contours start at $+3\sigma$ and then scale by a factor of
  2.}
\label{fig:a3444}
\end{figure*}

\begin{figure*}
\centering
\epsscale{1.1}
\includegraphics[width=8.5cm]{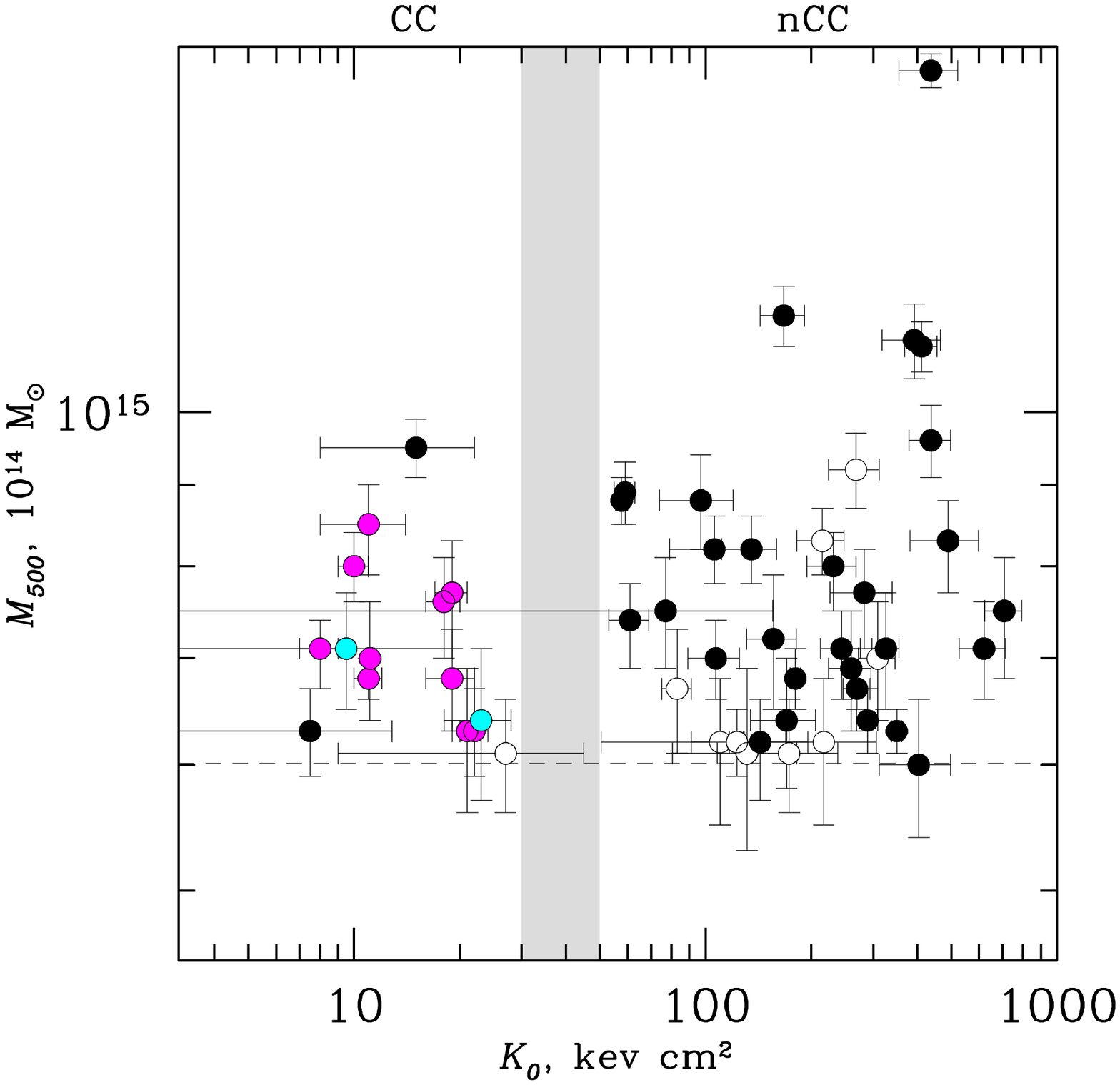}
\includegraphics[width=8.5cm]{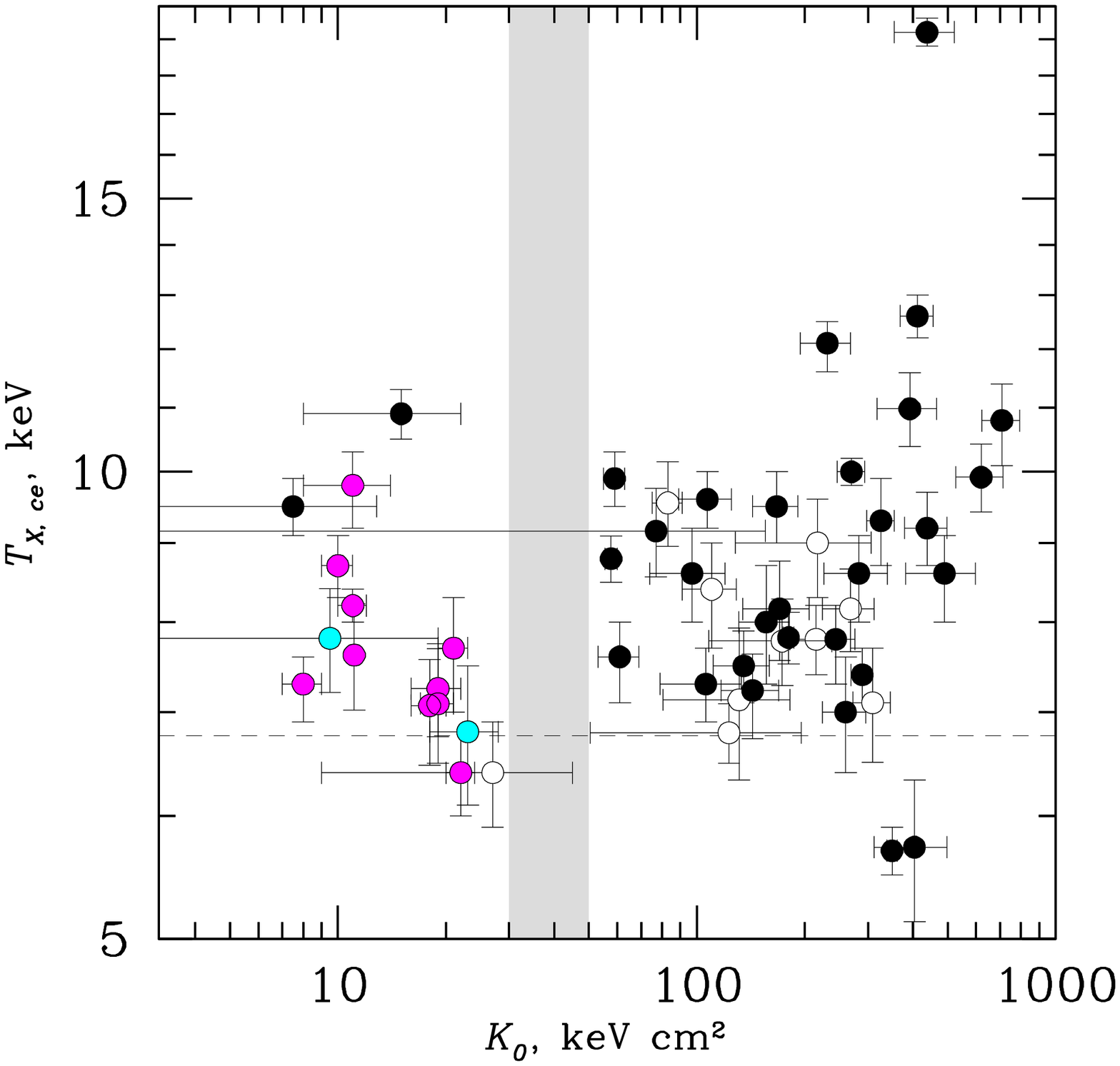}
\includegraphics[width=8.5cm]{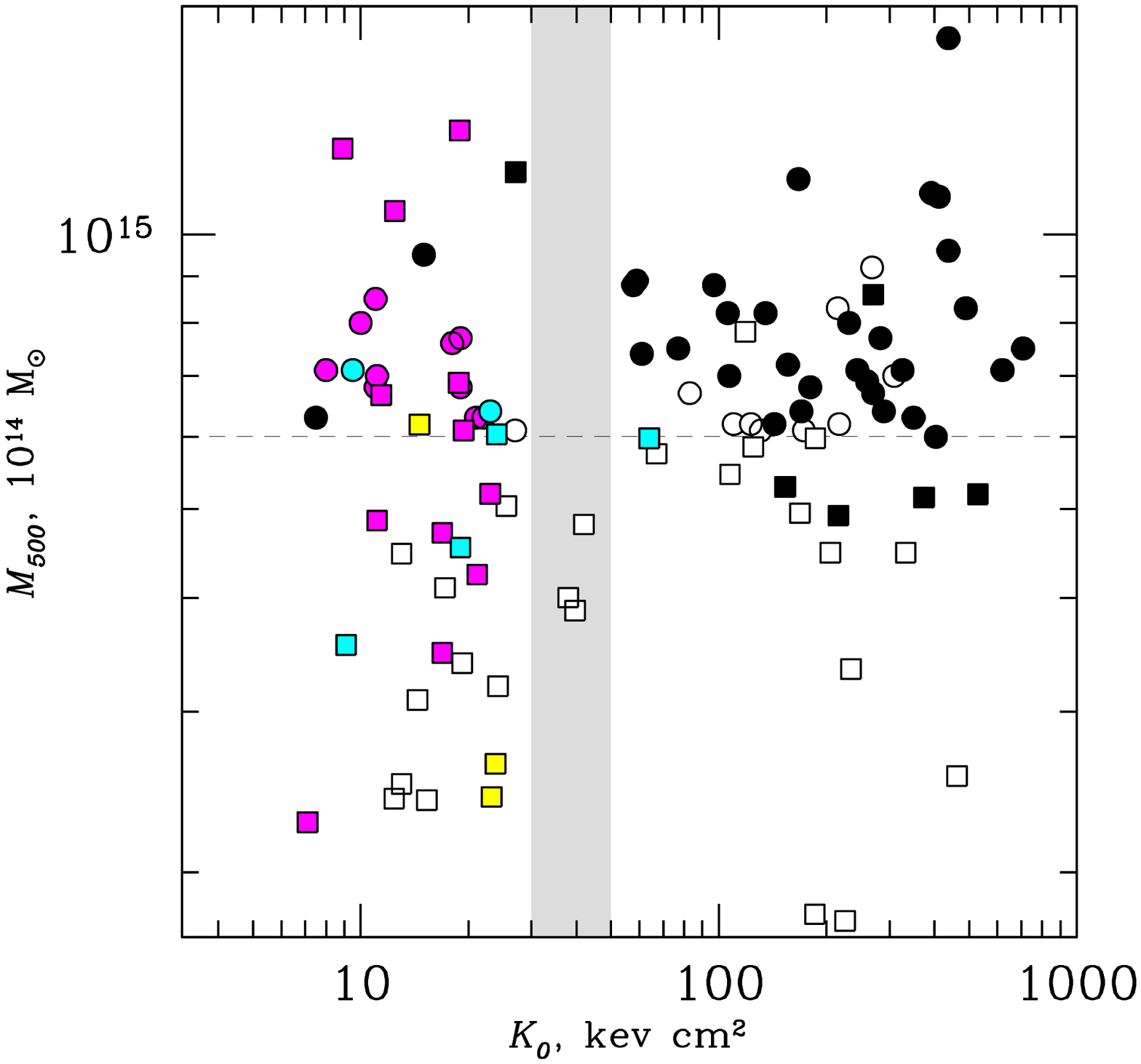}
\includegraphics[width=8.5cm]{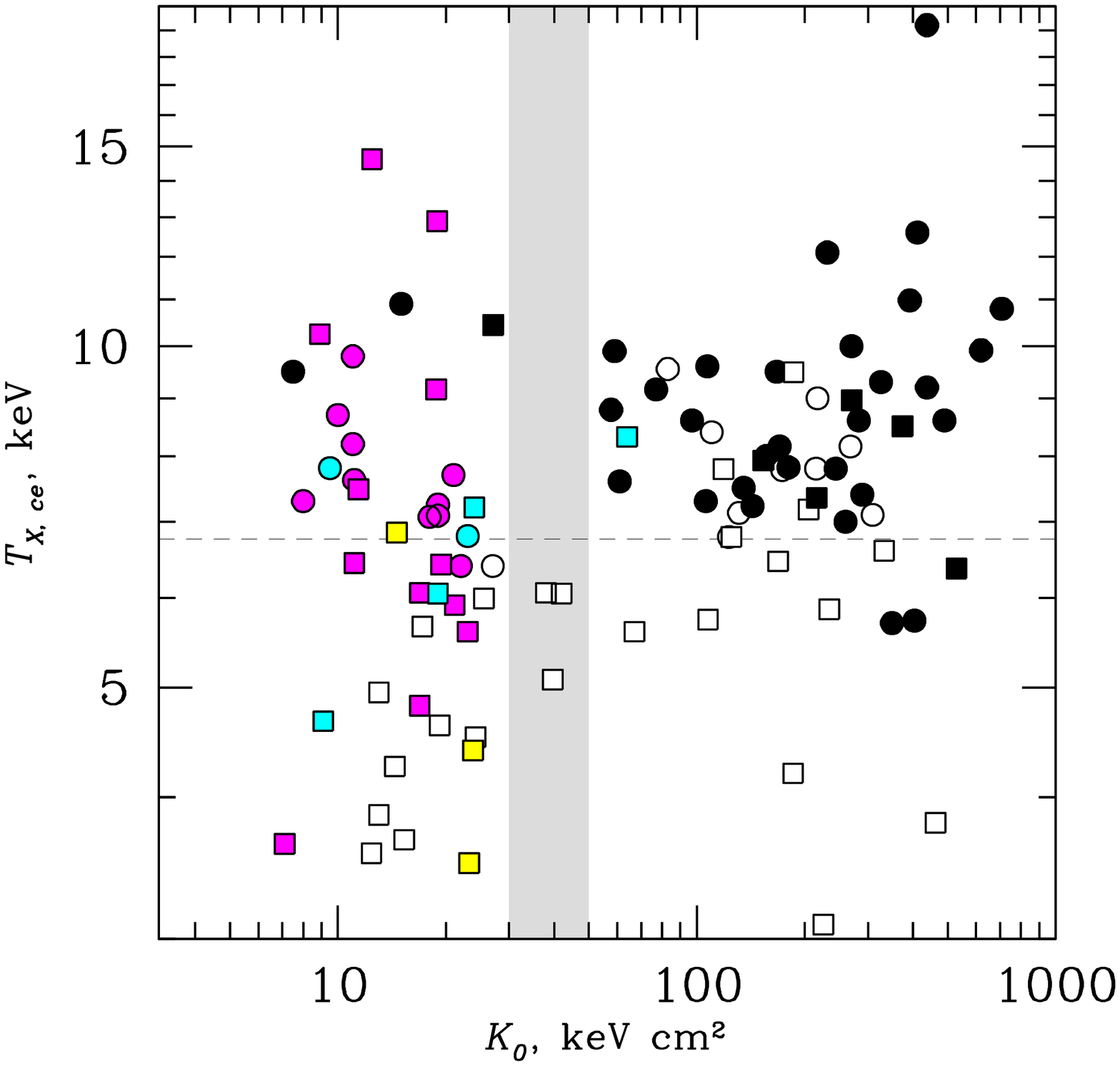}
\smallskip
\caption{
  {\em Upper panels}: $M_{\rm 500}-K_0$ and $T_{\rm X, \,ce}-K_0$ plots for
  the {\em Chandra} clusters in our statistical sample. Clusters with
  minihalos are shown as magenta (confirmed detections) and cyan
  (candidates), radio-halo clusters (including candidates) are shown in black
  and empty circles are clusters with no detected central diffuse radio
    emission.  Clusters without high-sensitivity radio observations
  (RXC\,J0510.7-0801, RXC\,J0520.7-1328, MACS\,J2135.2-0102, A\,2355 and
  A\,1733) are not included in the plots (they all have $K_0> 80$ kev
  cm$^2$; Table \ref{tab:k0}.) {\em Lower panels}: Same for the combined
  statistical sample (circles) and supplementary sample (squares). Error
  bars are omitted for clarity. Symbol colors are the same. Yellow shows
  clusters hosting a central diffuse radio source whose classification as a
  minihalo is uncertain (ZwCl 1742.1+3306, MACS\,J1931.8--2634, A2626;
  \S\ref{sec:supp}). The combined sample extends to lower masses, where
  minihalos appear to be less frequent.}
\label{fig:k0}
\end{figure*}

\section{Radio analysis}\label{sec:radio}

We analyzed {\em VLA} archival observations at 1.4 GHz of those clusters in
our supplementary sample with no high-sensitivity and multi-resolution radio
images on the minihalo angular scales in the literature (9 clusters). 
The observations
were chosen to be suitable to search for diffuse radio emission on the
cluster core scale, i.e., deep and with a good sampling of the $uv$--plane,
particularly at short spacings, that are crucial to properly image extended,
low surface brightness radio emission. Furthermore, we selected observations
with a range of angular resolutions that allow us to discriminate between
genuine diffuse emission and a blend of individual radio galaxies, as well
as to separate a possible minihalo from the radio source associated with the
cluster central galaxy.

Here we provide a brief description of the data reduction. The observation
details, radio images and notes on the individual clusters are presented in
Appendix A.  The data were calibrated and reduced in the standard fashion
using the NRAO Astronomical Image Processing System (AIPS) package.
Self-calibration was applied to reduce the effects of residual phase and
amplitude errors in the data and improve the quality of the final images.
For each cluster, the observations in different array configurations were
processed and imaged separately.
When possible, data from different dates and configurations were combined
together in the $uv$ or image plane. Sets of images were produced with
different weighting schemes, ranging from pure uniform weighting (Briggs
``robustness'' parameter (Briggs 1995) {\em ROBUST}$=-5$) to natural
weighting ({\em ROBUST}$=+5$), to enhance any possible extended emission.
The flux density scale was set using the Perley \& Butler (2013)
coefficients, and residual amplitude errors are within 3--5\%.

No new minihalos were detected among the 9 clusters analyzed here.  As an
illustrative example, in Fig.~\ref{fig:a3444} we show a comparison between a
known minihalo (in our statistical sample) and two non-detections based on
our radio analysis. Panel {\em a} shows the minihalo in the cool-core
cluster A\,3444 (from G17; see also Venturi et al. 2008, Kale et al. 2015).
Panels {\em b} and {\em c} show radio/X-ray overlays for MACS\,J0429.6-0253
and MS\,0839.8+2938, both cool cores in our supplementary sample. All three
radio images have similar angular resolution ($\sim 5^{\prime\prime}$) and
sensitivity ($1\sigma\sim 35-40$ $\mu$Jy beam$^{-1}$). The observations also
have similar sampling of short spacings in the $uv$--plane, that ensures the
possibility of detecting extended emission on a largest scale of $\sim
2^{\prime}$. This corresponds to a physical scale of $\sim 470$ kpc for
A\,3444, $\sim 640$ kpc for MACS\,J0429.6-0253 and $\sim 390$ kpc for
MS\,0839.8+2938, thus well beyond the region occupied by the central radio
galaxy. While a minihalo is clearly well detected in A\,3444, no indication
of diffuse radio emission is visible in the cores of the other clusters at a
similar sensitivity level.

\section{Discussion}\label{sec:disc}

The purpose of this study is to quantify how frequent are radio minihalos in
clusters.  Top panels in Fig.\ \ref{fig:k0} plot the clusters in our
mass-limited sample in the $M_{\rm 500}-K_0$ and $T_{\rm X, \, ce}-K_0$
planes (in the latter panel, the approximate temperature from the $M_{\rm
  500}-T_{\rm X}$ relation that corresponds to our mass cut is shown by a
dashed line).  As noted by C09, Rossetti et al.\ (2013) and others, the
cluster sample clearly separates into two populations, cool cores with
$K_0\lax 30$ keV~cm$^2$ and non-cool cores (most of which obvious mergers)
with $K_0\gax 50$ keV~cm$^2$. The fraction of cool cores in our {\em
  Planck}-selected sample is $26\%$ (15 out of 58), similar to the fraction
found in a much larger {\em Planck}-selected sample (Andrade-Santos et al.
2017) and lower than that seen in X-ray selected samples of nearby clusters
(e.g., Rossetti et al. 2016).

As shown before (e.g., Cassano et al. 2013, Rossetti et al. 2013, Cuciti et
al. 2015, Yuan et al.\ 2015), giant halos (black symbols) are found almost
exclusively in non-cool core clusters. We find that minihalos (magenta
symbols) are indeed found exclusively in cool cores, confirming previous
non-statistical findings (e.g., Giacintucci et al. 2014a, Yuan et al. 2015,
Kale et al. 2015). A new result that is evident in Fig.\ \ref{fig:k0} (top
panels) is that {\em almost all}\/ cool cores in a complete sample of
massive clusters --- 12 out of 15, or 80\% --- possess a minihalo.  Radio
minihalos are not that rare after all.

It is interesting to extend the mass range of the sample toward lower masses
and temperatures. According to Cuciti et al.\ (2015), the probability of
finding a giant radio halo in merging clusters increases with the total
mass.  This is instructive for the origin of giant halos, because it implies
a statistical link between the mechanical energy available during mergers
and the generation of these sources (e.g., Cassano et al.\ 2006, 2016). It
is interesting to check if minihalos exhibit a similar behavior.  Lower
panels in Fig.\ \ref{fig:k0} show our combined statistical + supplementary
sample, which includes all the known 22 confirmed minihalos (magenta points)
and 6 candidates (cyan points). This combined sample extends a factor 3
below the mass cut of the statistical one, as well as adding a number of
massive systems. We do see the lower frequency of the giant radio halos in
cooler clusters, observed in Cassano et al.\ (2013) and Cuciti et al.
(2015). Similarly, at lower masses, a large fraction of cool cores without
minihalos (not detected at a similar radio sensitivity; see \S5, Fig.~\ref{fig:a3444} 
and Appendix A) emerges, while almost all the additional {\em
  massive}\/ cool cores do have minihalos, consistently with our finding for
the statistical sample. Minihalos are still absent in the non-cool-core
clusters (with the exception of the candidate minihalo in A\,1413); a few
``warm cores'' that appear in the $K_0=30-50$ keV~cm$^2$ gap do not host
minihalos (or giant halos) either.  The apparent reduced frequency of
minihalos at lower masses needs a proper statistical investigation 
that accounts for selection effects  (e.g., what if the radio luminosity correlates 
with the cluster mass) and radio upper limits (for instance 
via injection of {\em fake} minihalos in the $uv$--data, e.g., Kale et al. 2015),  
which will be the subject of future work.

\section{Conclusions}

The two new results of our study are: (a) almost all (12 out of 15, i.e.,
80\%) massive clusters with cool cores ($M_{500}>6\times 10^{14}$\msun\ or
$T\gax 6$ keV) exhibit a radio minihalo, and (b) the fraction of minihalos
appears to drop in cool cores with lower cluster total masses (or global
temperatures). To make the former observation, we used a mass-limited
cluster sample. For the latter result, we extended the sample to include
more clusters with available \chandra\ and radio data, including
higher-redshift and lower-mass clusters and all the other known minihalos
(for a total of 28 minihalos, including 6 candidates). 

These findings may encode information on the origin of minihalos. In the
present study, we used only the presence or absence of a minihalo; we will
investigate the correlation of the radio power with cluster mass, as well as
with cool core thermodynamic parameters, in the follow-up works. A few
outliers may prove especially valuable, such as a minihalo in a low-mass
cool core 2A\,0335+096 \citep[]{1995ApJ...451..125S}, a giant halo in a
high-mass cool core CL\,1821+643
\citep{2014MNRAS.444L..44B,2016MNRAS.459.2940K}, a halo in A\,2390 (Sommer
et al. 2017) whose cool core has a similar, unusually large radius
(Vikhlinin et al. 2005), as well as the absence of minihalos in warm cores.
\\ \\ 

{\it Acknowledgements.}

We thank the referee for their critical and helpful comments. 
SG acknowledges the support by the National Aeronautics and Space
Administration, through {\em Chandra} Award Numbers G03-14140X, AR5-16013X
and G05-16136X. Basic research in radio astronomy at the Naval Research
Laboratory is supported by 6.1 Base funding. RC, TV and GB acknowledge 
partial support from PRIN INAF 2014. This research has made use of
the NASA/IPAC Extragalactic Database (NED) which is operated by the Jet
Propulsion Laboratory, California Institute of Technology, under contract
with the National Aeronautics and Space Administration. The National Radio
Astronomy Observatory is a facility of the National Science Foundation
operated under cooperative agreement by Associated Universities, Inc.

{}

\appendix

\section{RADIO OBSERVATIONS AND IMAGES}\label{sec:imagesr}

\begin{figure*}
\centering
\epsscale{1.1}
\includegraphics[width=6.5cm]{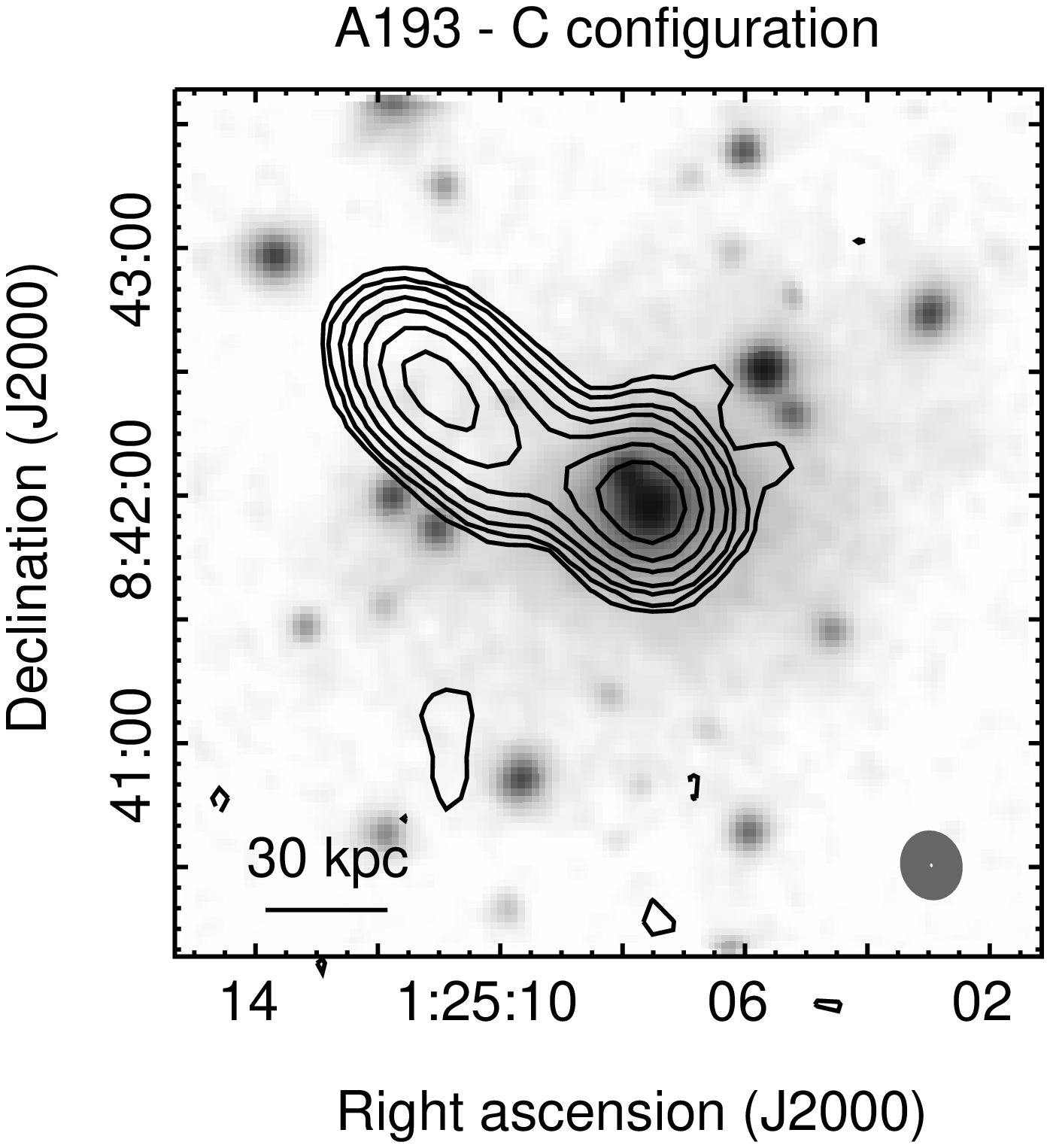}
\includegraphics[width=6.5cm]{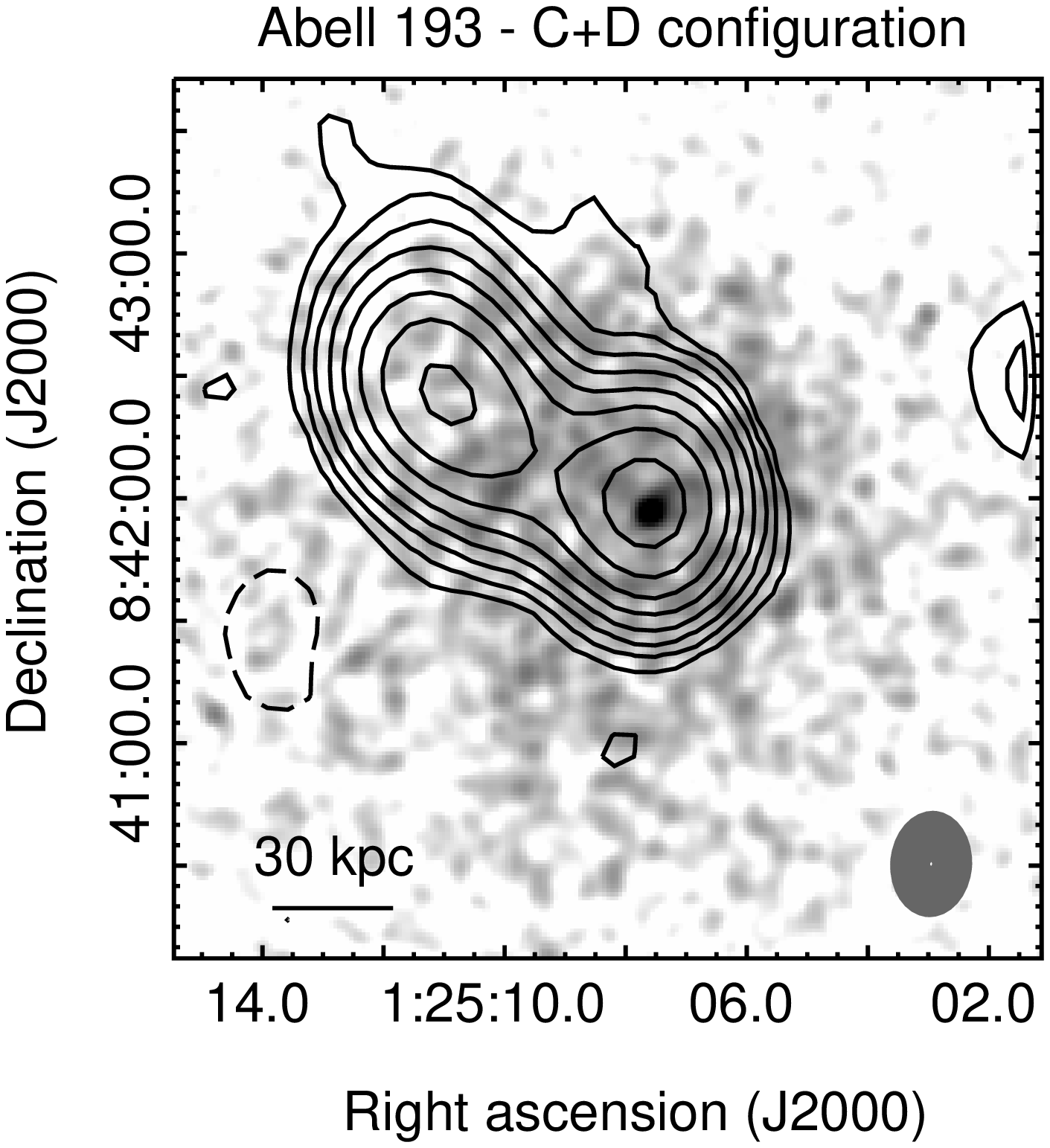}
\includegraphics[width=6.5cm]{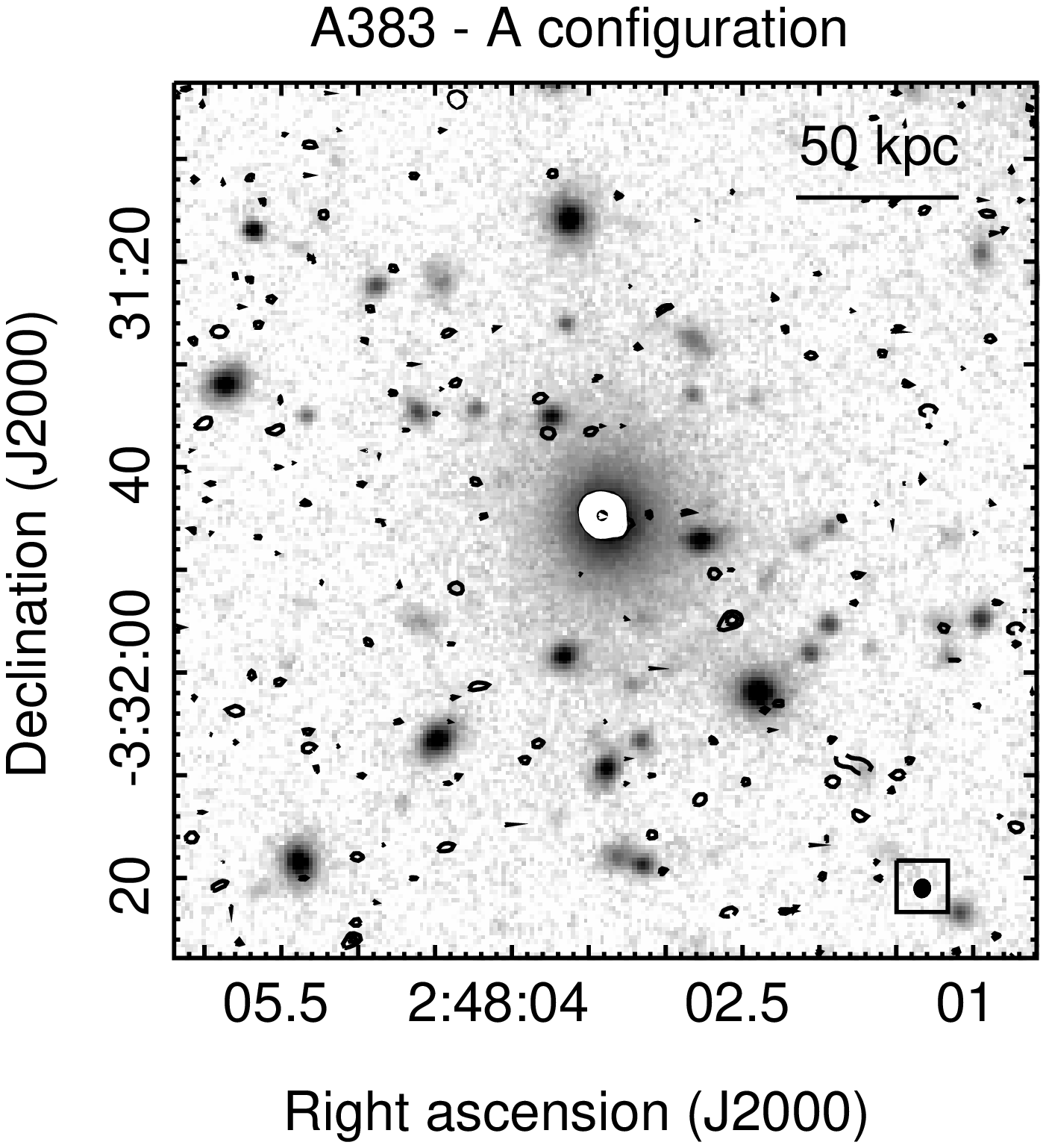}
\includegraphics[width=6.5cm]{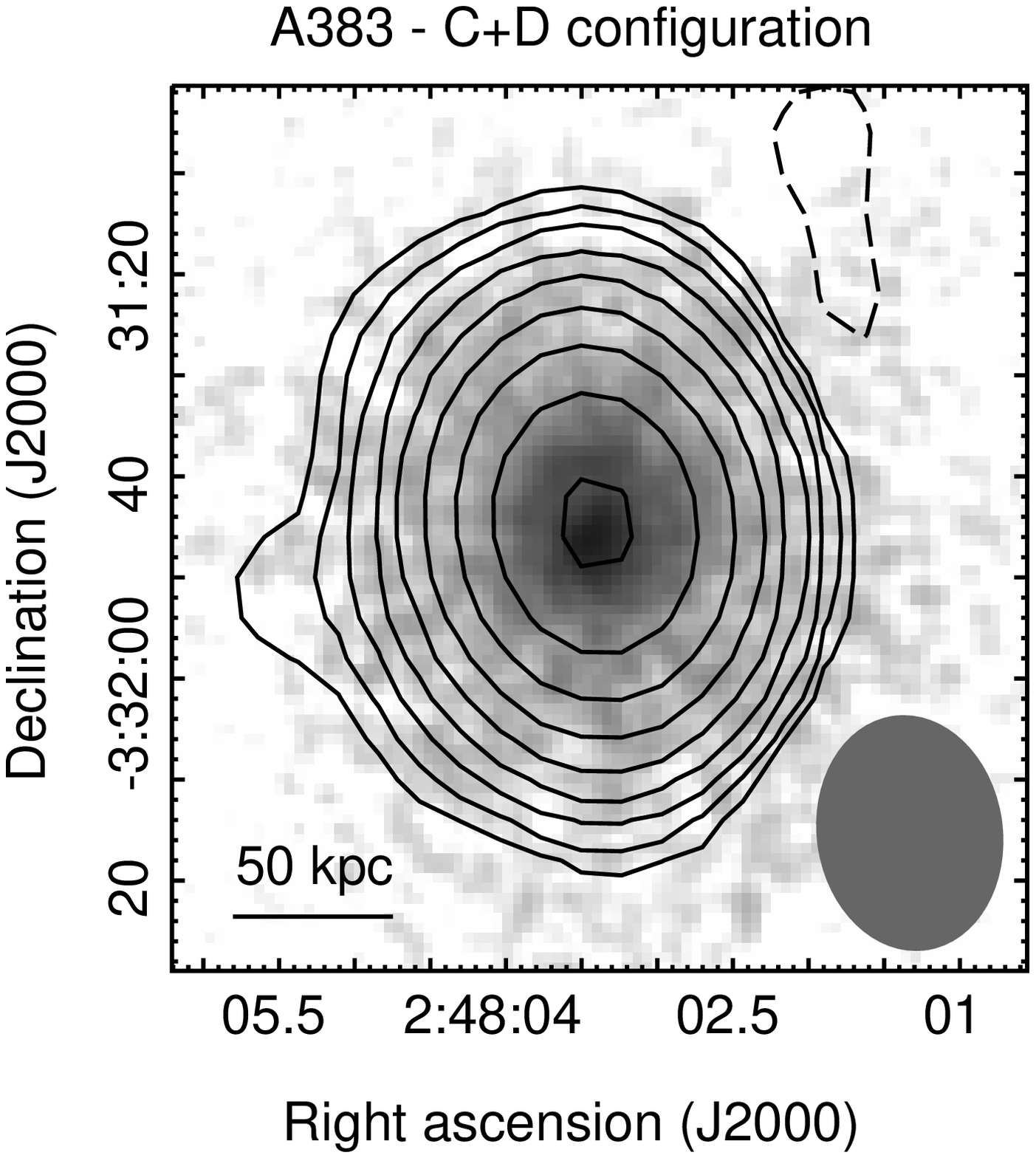}
\includegraphics[width=6.5cm]{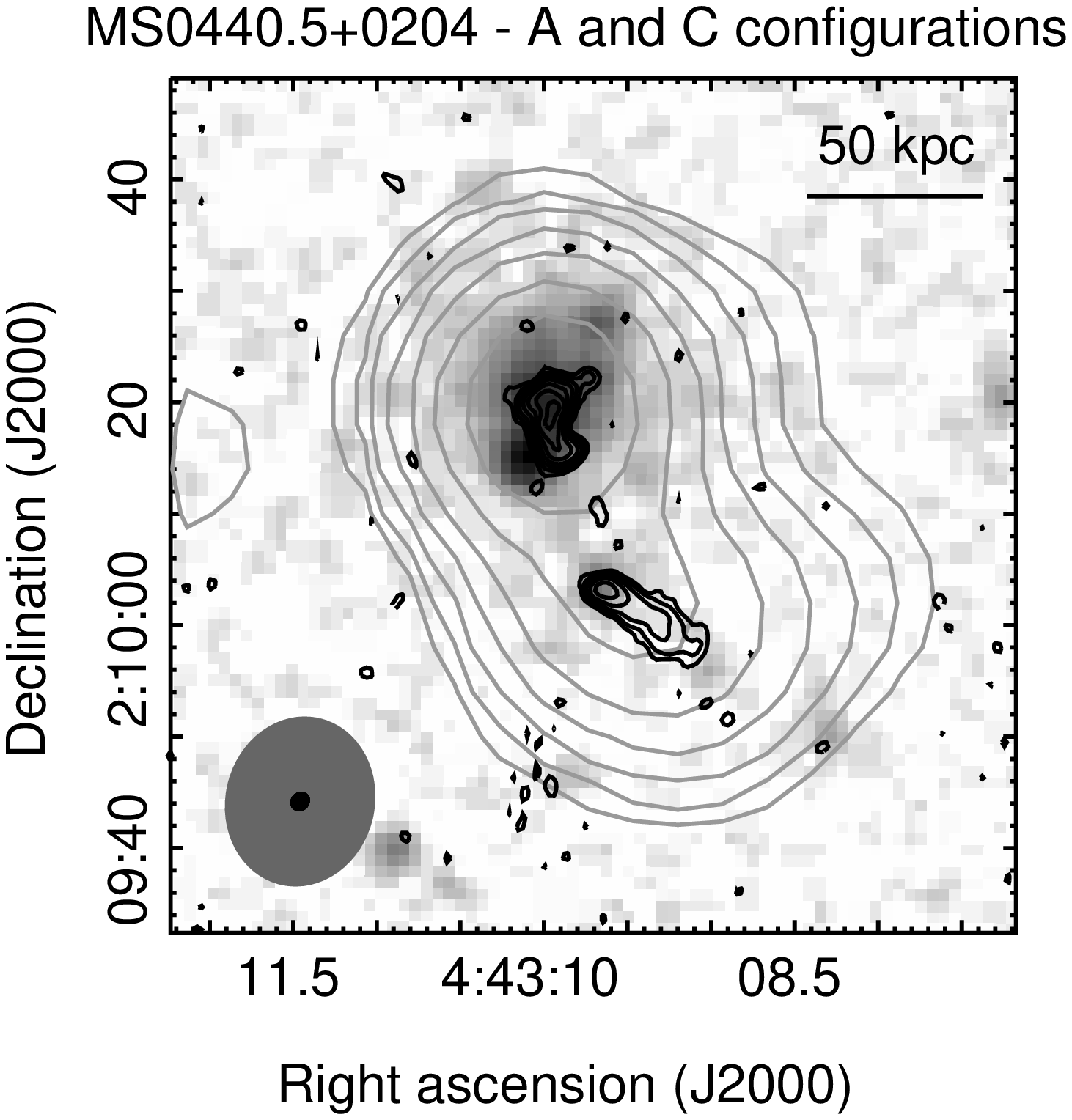}
\includegraphics[width=6.5cm]{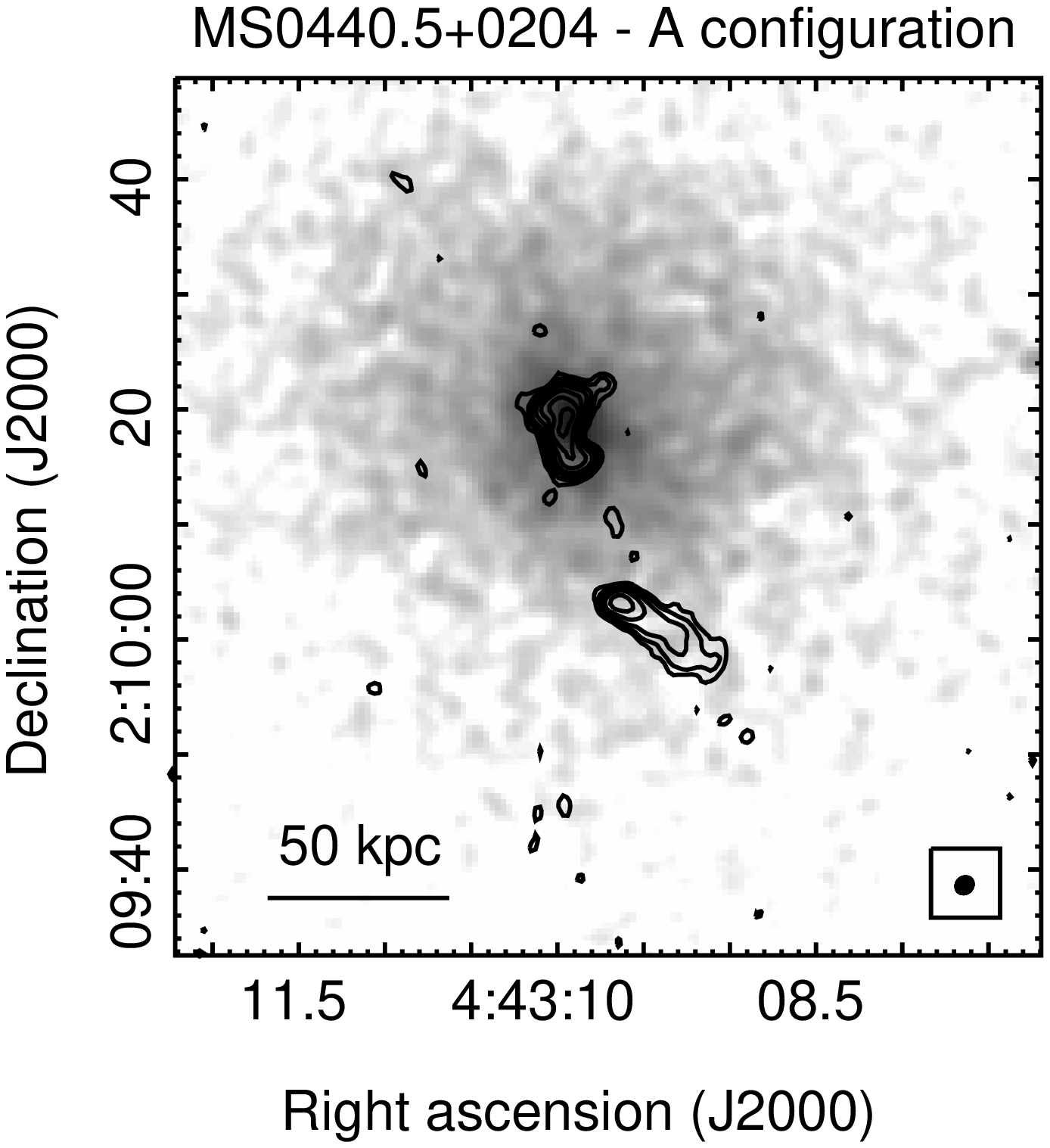}
\smallskip
\caption{Upper and middle panels: {\em VLA} 1.4 GHz combined radio images 
  of A\,193 and A\,383, overlaid on the optical $r$-band SDSS images (left)
  and smoothed X-ray {\em Chandra} images in the 0.5-4 keV band (right).
  Contours start at $+3\sigma$ and then scale by a factor of 2.  When
  present, contours at $-3\sigma$ are shown as dashed. Restoring beams (also
  shown as ellipses in the lower corner of each image) and {\em rms} noise values
  are as listed in Table \ref{tab:images}.  Bottom panels: {\em VLA}
  C--configuration (gray contours) and A--configuration (black contours)
  images of MS\,0440.5+020, overlaid on the POSS--2 red optical image
  (left). On the right, the A--configuration image is overlaid on the
  smoothed X-ray {\em Chandra} image in the 0.5-4 keV band. Contours start
  from $+3\sigma$ and then scale by a factor of 2. No levels at $-3\sigma$
  are present in the portion of the images shown. Restoring beams (also
  shown as ellipses in the lower right corner of each image) and {\em rms} noise
  values are as listed in Table \ref{tab:images}.}
\label{fig:images1}
\end{figure*}

\begin{figure*}
\centering
\epsscale{1.1}
\includegraphics[width=6.5cm]{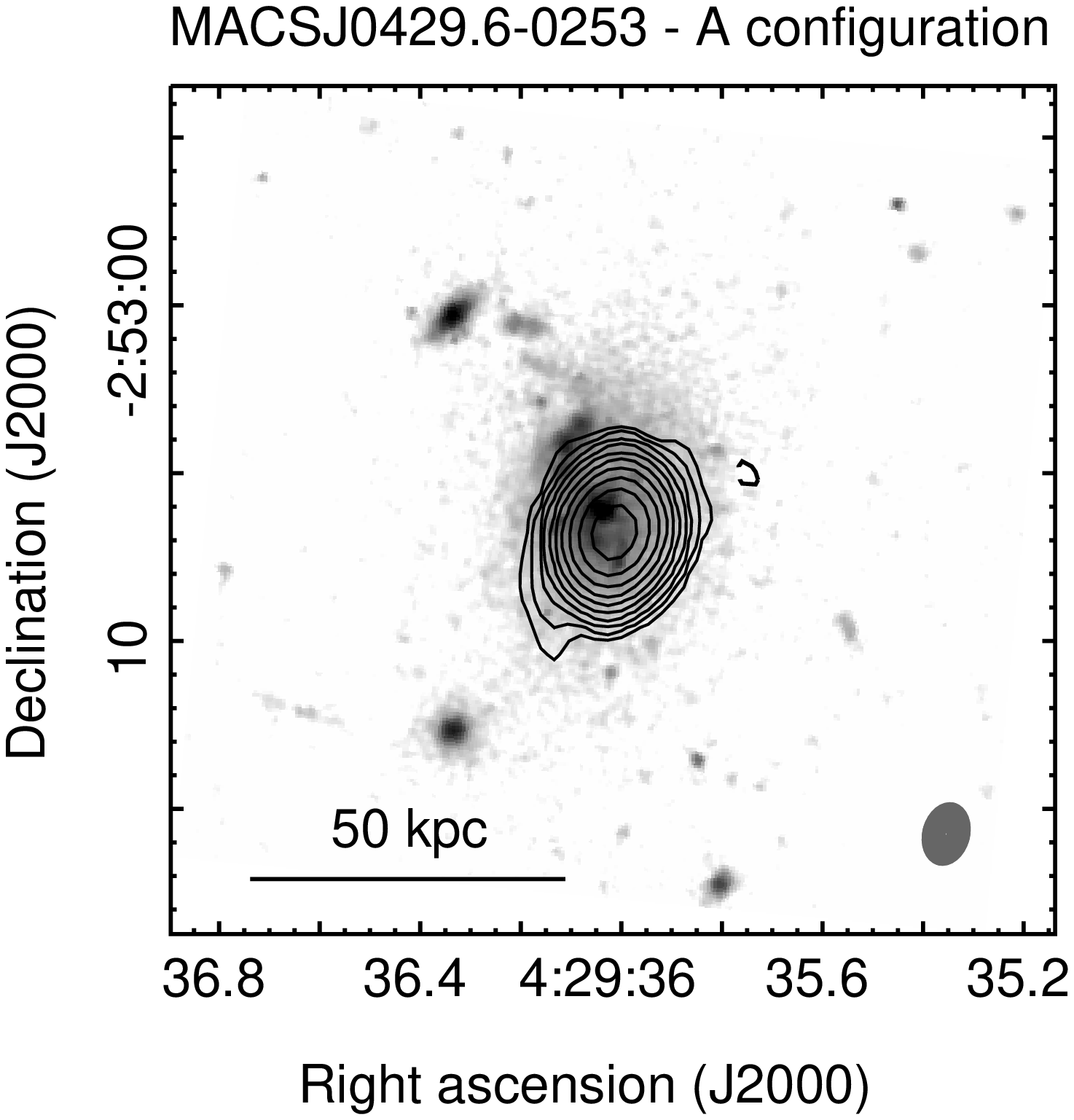}
\includegraphics[width=6.5cm]{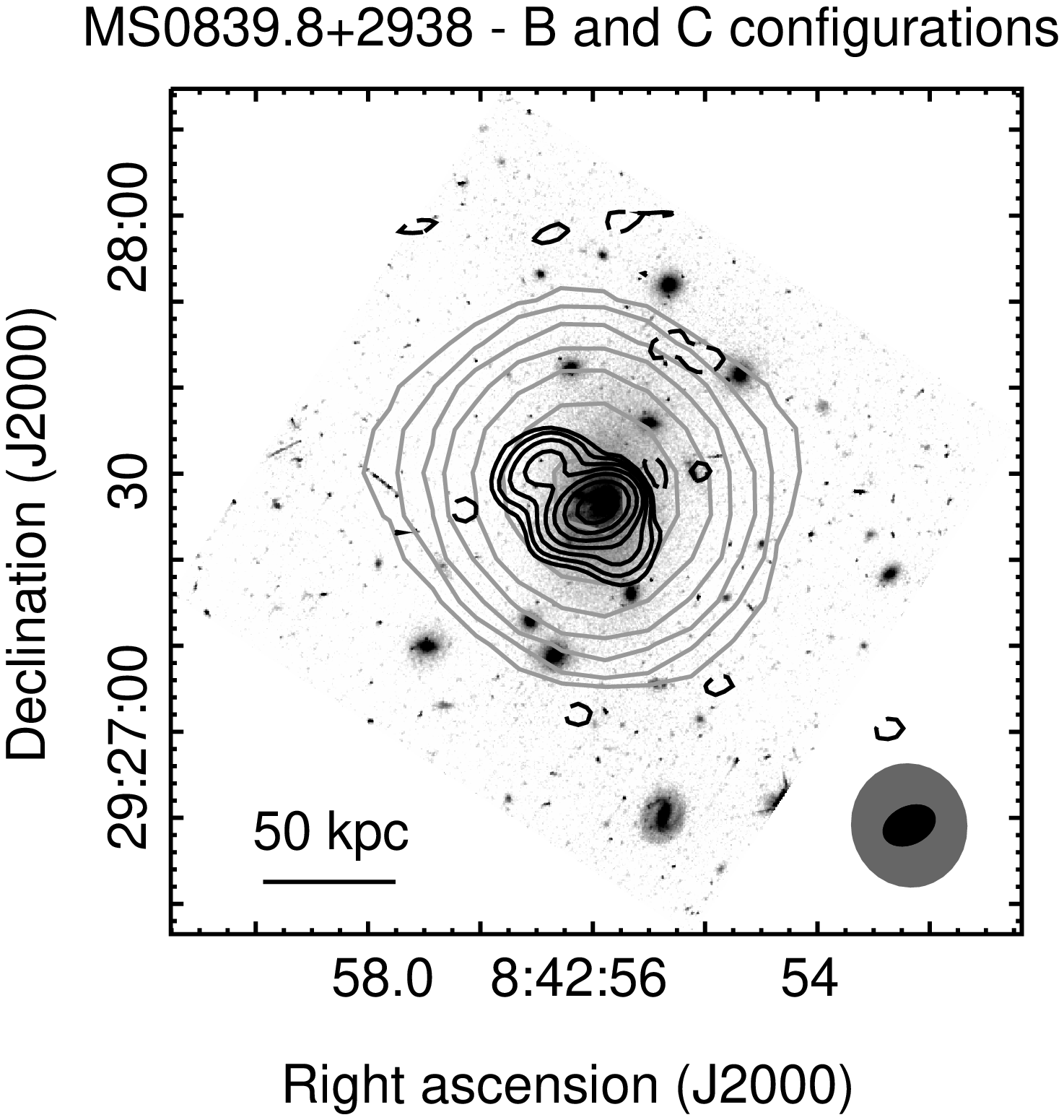}
\includegraphics[width=6.5cm]{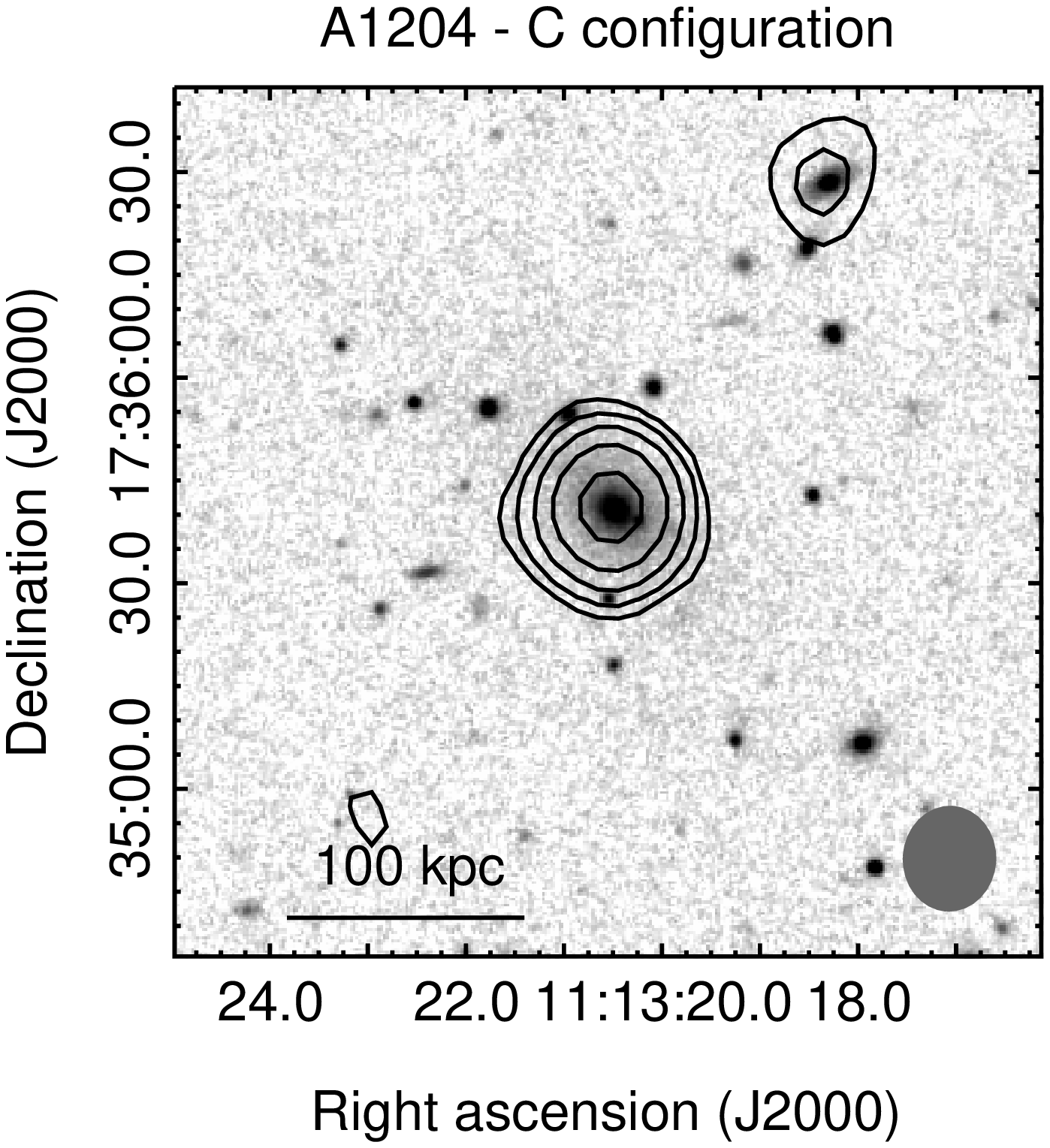}
\includegraphics[width=6.5cm]{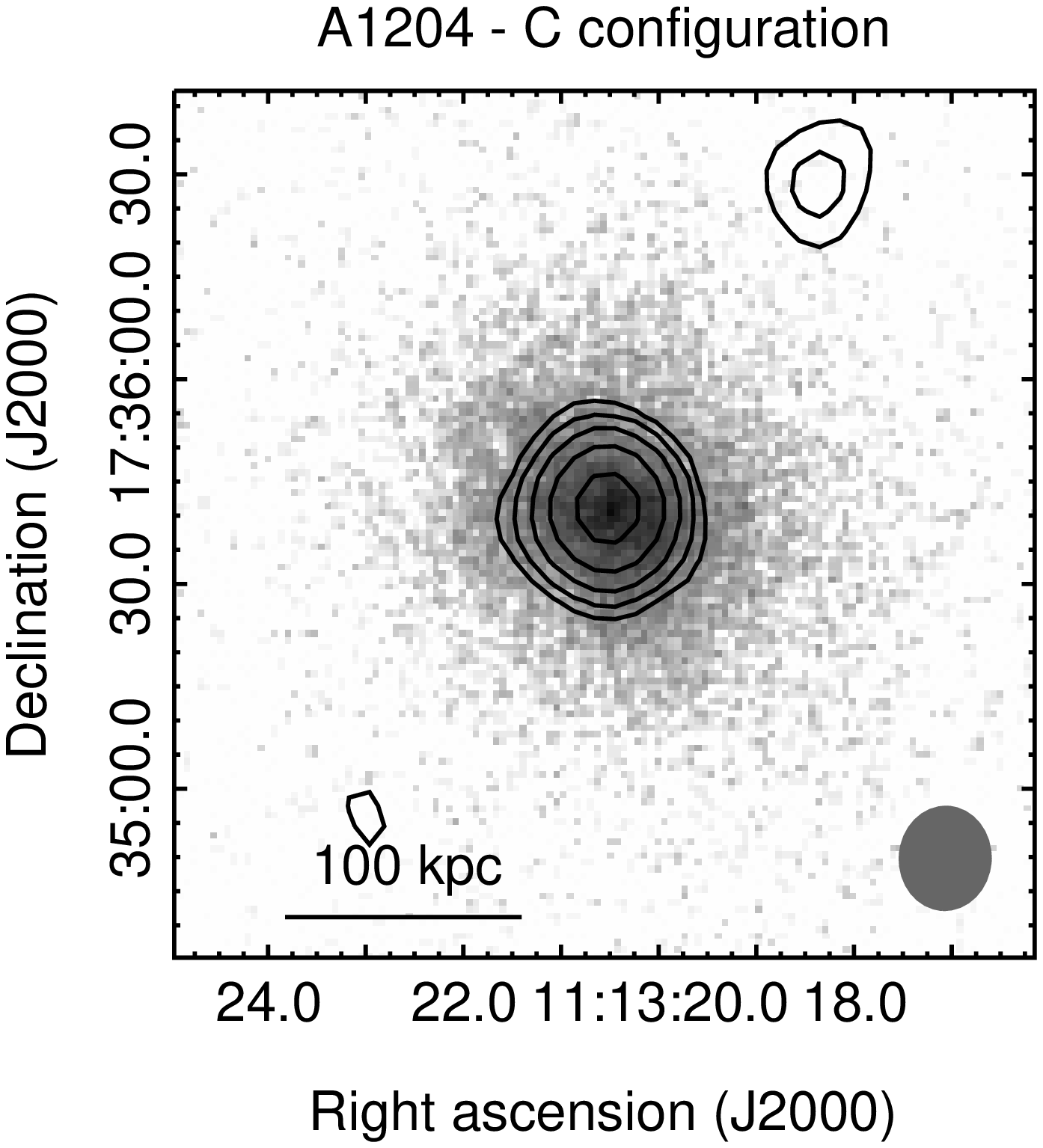}
\includegraphics[width=6.8cm]{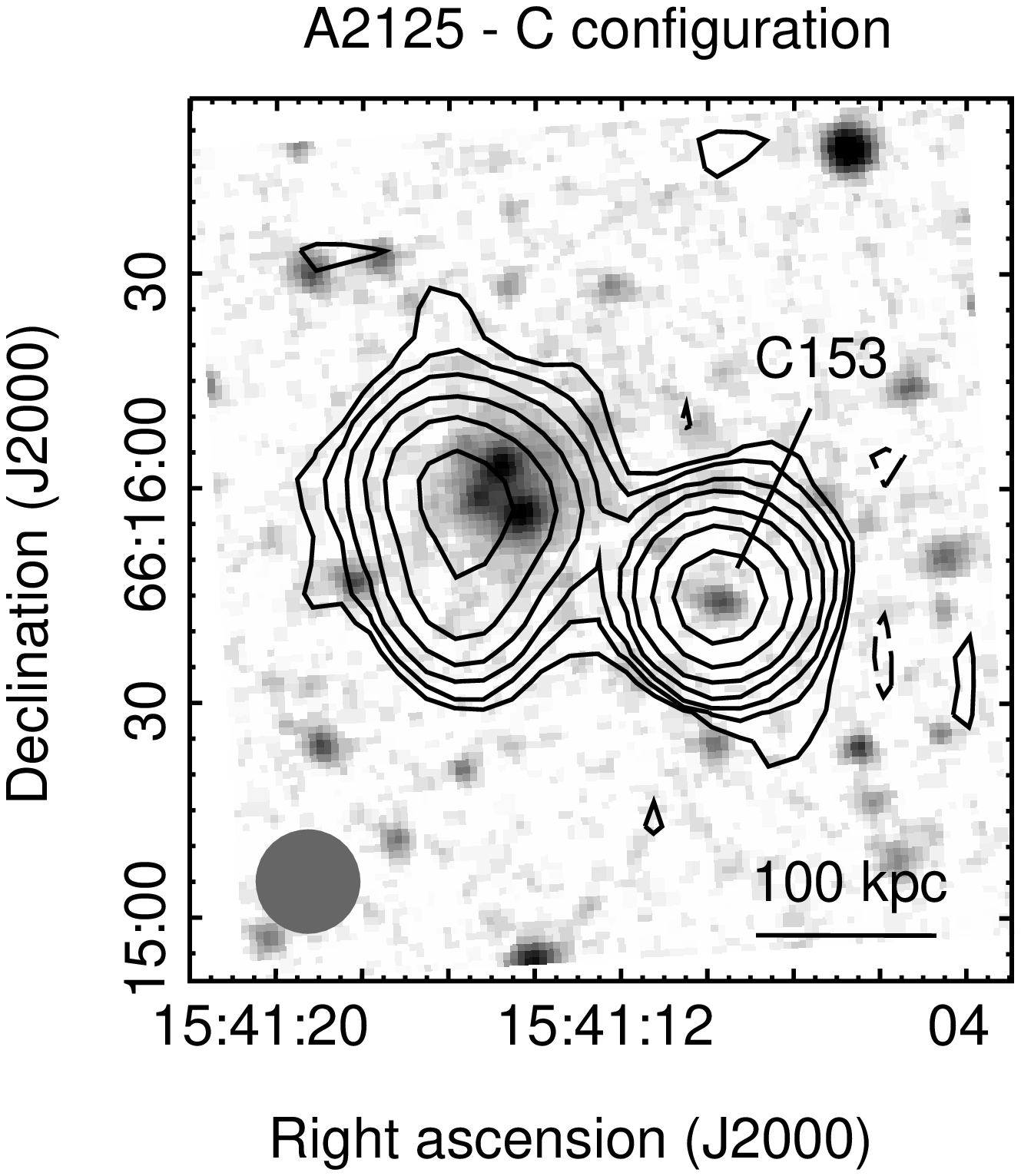}
\includegraphics[width=6.8cm]{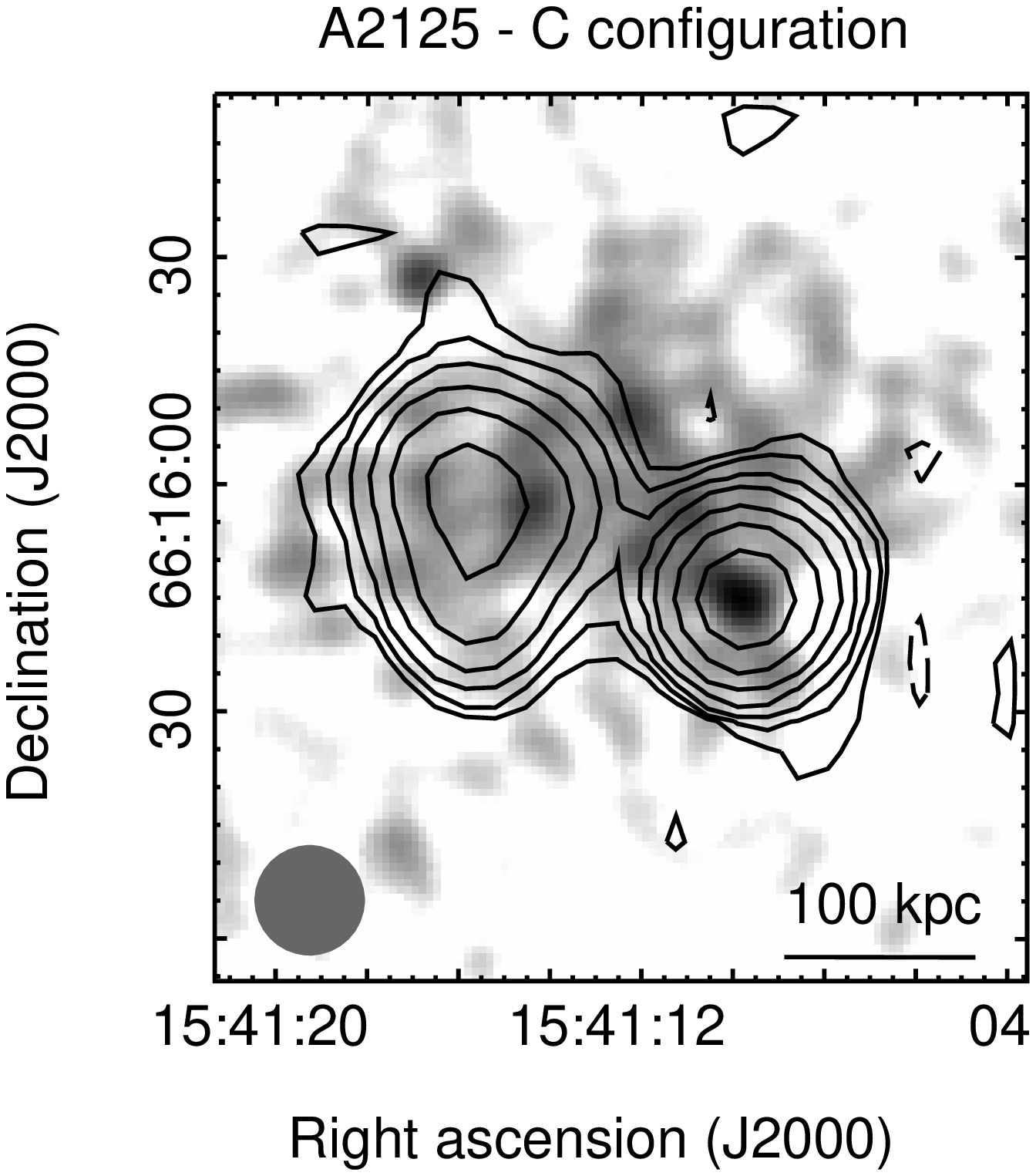}
\smallskip
\caption{Upper panels:  {\em VLA} 1.4 GHz radio images of 
  MACS J0429.6-0253 (left) and MS 0839.8+2938 (right), overlaid on the
  preview HST WFPC2 images (programs 11103 and 11312).  Middle panels: {\em
    VLA} 1.4 GHz radio image of A\,1204, overlaid on the optical $r$-band
  SDSS image (left) and smoothed X-ray {\em Chandra} image in the 0.5-4 keV
  band (right).  Bottom panels: {\em VLA} 1.4 GHz radio image of A\,2125,
  overlaid on the optical POSS--2 red optical images (left) and smoothed
  X-ray {\em Chandra} images in the 0.5-4 keV band (right). In all panels,
  contours start at $+3\sigma$ and then scale by a factor of 2. When
  present, contours at $-3\sigma$ are shown as dashed.  Restoring beams
  (also shown as ellipses in the lower corner of each image) and {\em rms} noise
  values are as listed in Table \ref{tab:images}.}
\label{fig:images2}
\end{figure*}

\begin{figure*}
\centering
\epsscale{1.1}
\includegraphics[width=6.5cm]{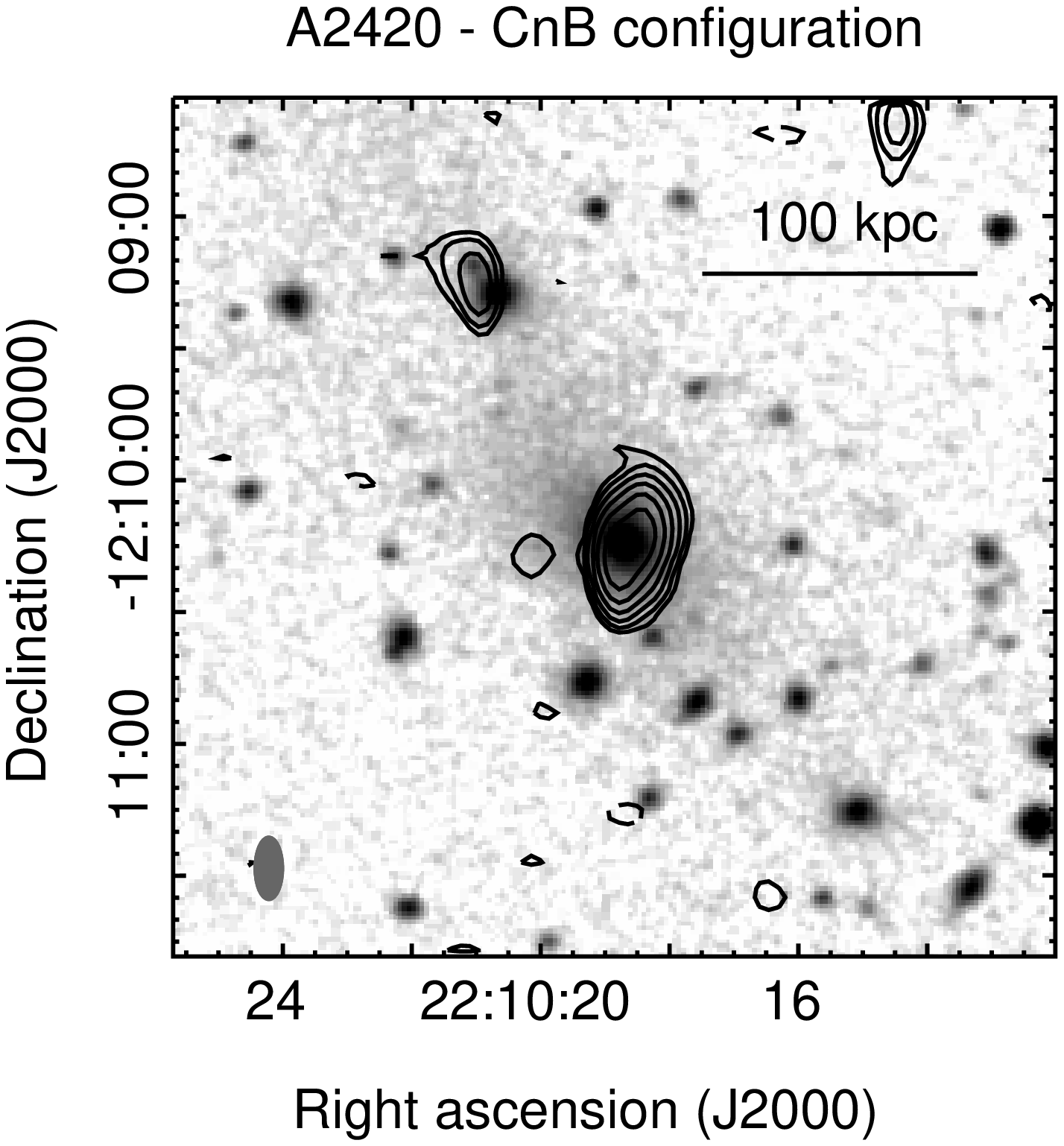}
\includegraphics[width=6.5cm]{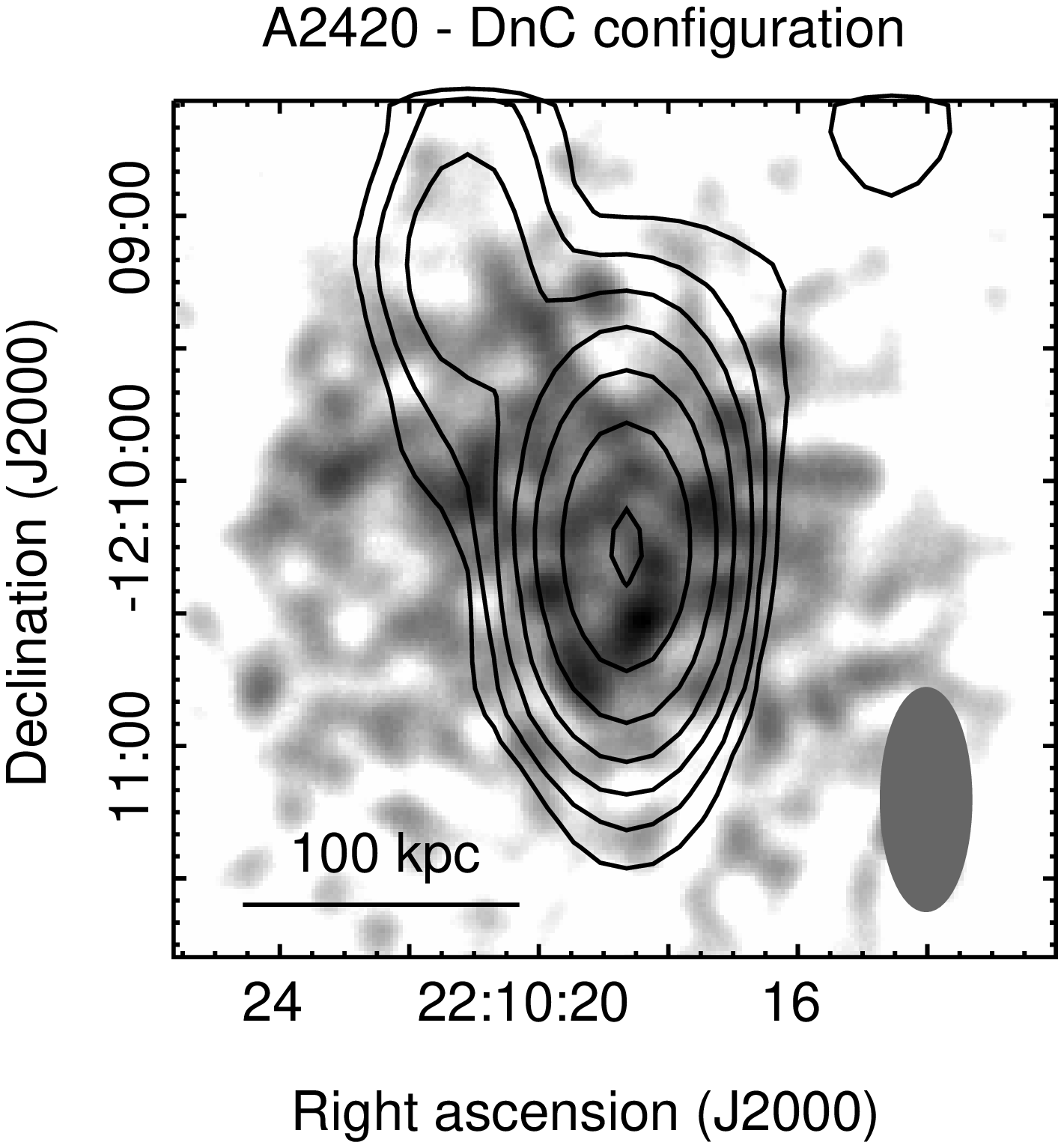}
\includegraphics[width=6.5cm]{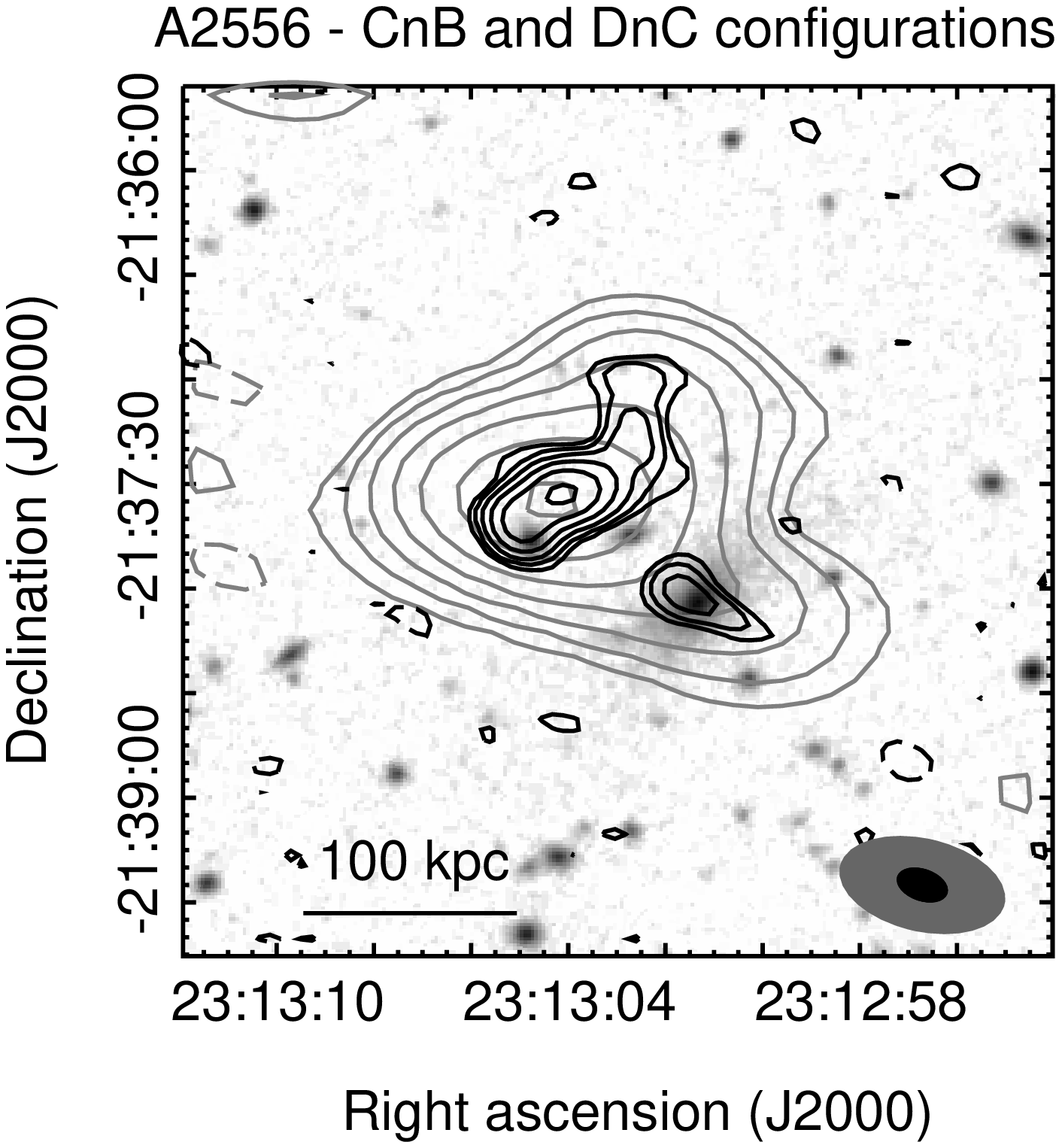}
\includegraphics[width=6.5cm]{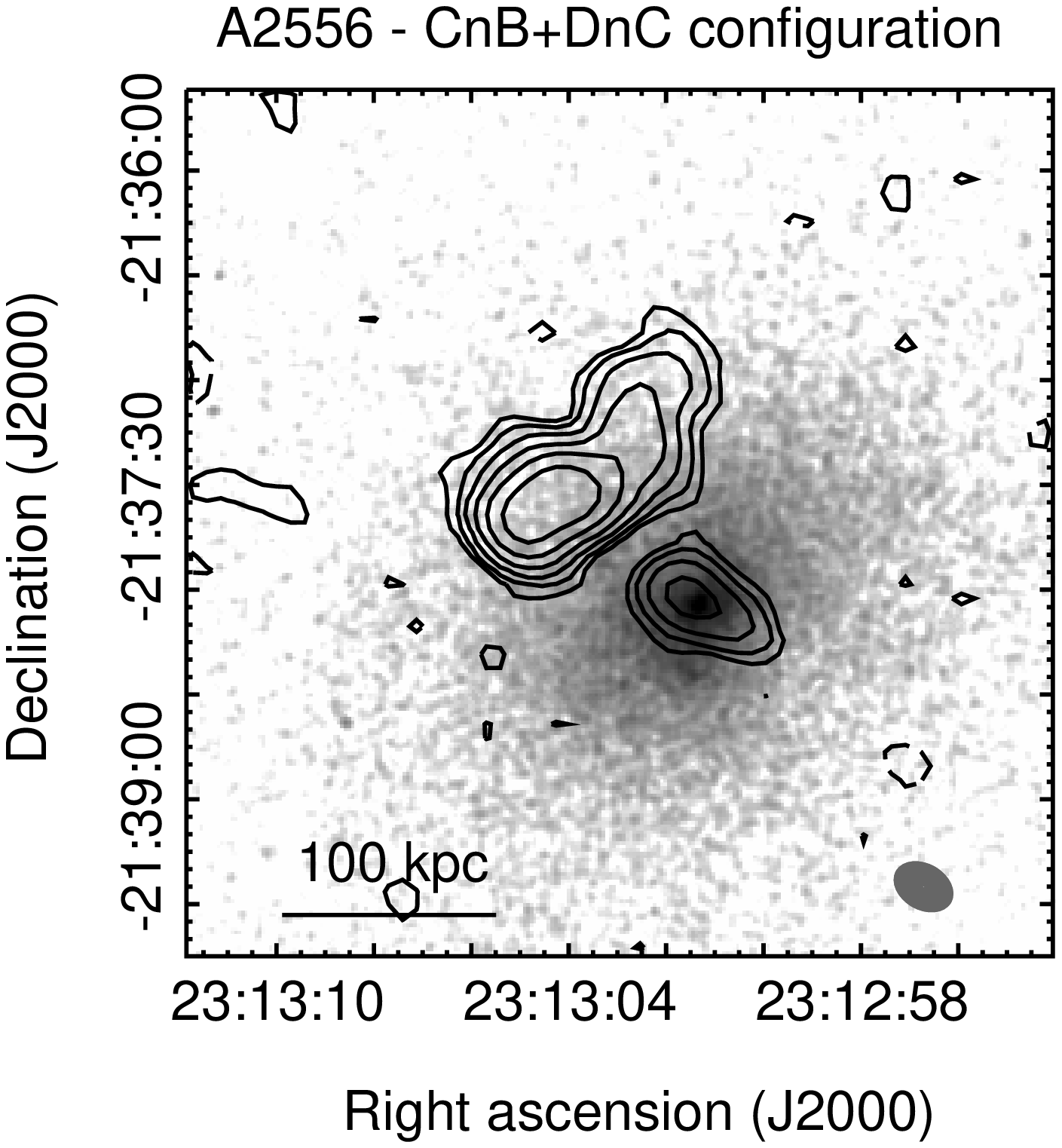}
\smallskip
\caption{{\em VLA} 1.4 GHz radio images of A\,2420 (upper panels) and 
  A\,2556 (bottom panels), overlaid on the optical POSS--2 red optical
  images (left) and smoothed X-ray {\em Chandra} images in the 0.5-4 keV
  band (right).  In all panels, contours start at $+3\sigma$ and then scale
  by a factor of 2.  When present, contours at $-3\sigma$ are shown as
  dashed.  Restoring beams (also shown as ellipses in the lower right corner
  of each image) and {\em rms} noise values are as listed in Table
  \ref{tab:images}.}
\label{fig:images3}
\end{figure*}

Table 8 lists the {\em VLA} archival observations at 1.4
GHz reanalyzed in this work, as described in \S\ref{sec:radio}.  Our final
radio images are presented and compared to the cluster
optical images (from SDSS%
\footnote{Sloan Digital Sky Survey.}, POSS-2%
\footnote{Second Palomar Observatory Sky Survey.}
or {\em HST} WFPC2%
\footnote{{\em Hubble Space Telescope} Wide-Field Planetary Camera 2.}
) and X-ray {\em Chandra} images in Figs.~\ref{fig:images1},
\ref{fig:images2} and Fig.~\ref{fig:images3}. The radio images of
MACS\,J0429.6-0253 and MS\,0839.8+2938 are overlaid on the {\em Chandra}
images in Fig.~\ref{fig:a3444} and on the optical {\em HST} 
images in Fig.~10 (upper panels). The image
properties are summarized in Table \ref{tab:images}. All the images shown
here have been obtained with the ROBUST parameter set to 0 in the AIPS task
IMAGR. The flux density of the unresolved radio galaxies was measured by
fitting the sources with a Gaussian model (task JMFIT). For extended radio
galaxies, the total flux density was determined using the task TVSTAT. All
fluxes were measured on images corrected for the primary-beam attenuation. A
brief description of the radio emission in each individual cluster is given
below.

\begin{deluxetable}{lcccc}
\tablecaption{Newly analyzed 1.4 GHz {\em VLA}\/ observations}
\label{tab:radioobs}
\tablehead{
\colhead{Cluster name}  & \colhead{Configuration}    & \colhead{Project} &  \colhead{Date} & \colhead{Time (min)}  \\ 
}
\startdata
A\,193             & C   & AM735  & 2002 Nov 30      & 118  \\
                   & C   & AM702  & 2002 Oct 28      & 37   \\
                   & D   & AM702  & 2001 Oct 15      & 55  \\
A\,383             & A   & AR369   & 1996 Nov 17    &  54   \\
                   & A   & AR369   & 1996 Nov 19    & 100   \\
                   & C   & AM072   & 2002 Oct 28    & 61    \\
                   & D   & AM072   & 2001 Oct 21    & 60    \\
MS\,0440.5+0204    & A   & AI0072  & 1998 Apr 19,20 & 360   \\
                   & C   & AS0873  & 2006 Dec 14    & 43    \\
MACS\,J0429.6-0253 & A   & AT0318  & 2006 Apr 03    & 52   \\
                   & B   & AT0318  & 2006 Sep 06    & 35   \\
MS\,0839.8+2938    & B   & AH0190  & 1985 Apr 25    & 30  \\
                   & C   & AH0491  & 1993 Jun 26    & 58  \\
A\,1204            & C   & AJ0242  & 1994 Dec 05    & 74 \\
A\,2125            & C   & AD0311  & 1993 Jul 29    & 343 \\
A\,2420            & CnB & AM0702  & 2002 Sep 24,29 & 60  \\
                   & DnC & AM0702  & 2001 Oct 6,7   & 60  \\
A\,2556            & CnB & AM0702  & 2002 Sep 29    & 60  \\
                   & DnC & AM0702  & 2001 Oct 07    & 60  \\
\enddata
\tablecomments{Column 1: cluster name. Column 2: {\em VLA} configuration. Columns 3 and 4: project code
and observation date. Column 5: total time on source.}
\end{deluxetable}

\begin{deluxetable}{lcrccccc}
\tablecaption{Properties of the {\em VLA} radio images}
\tablehead{
\colhead{Cluster}  & \colhead{Configuration}   & \colhead{FWHM, p.a.}   &   \colhead{{\em rms}}   & \colhead{$S_{\rm 1.4\, GHz}$} & LDS \\
    \colhead{name}     & \colhead{}    &  \colhead{($^{\prime \prime} \times^{\prime
  \prime}$, $^{\circ}$)\phantom{00}} & \colhead{($\mu$Jy beam$^{-1}$)} & \colhead{(mJy)} & (Mpc)\\
}
\startdata
A\,193             & C   & $16\times14$, 15       &  40 & $60\pm3$  & \phantom{0}0.8\tablenotemark{a}   \\
                   & C+D & $25\times19$, $-4$     &  35 & $61\pm3$  & 0.8    \\
A\,383             & A   & $1.5\times1.3$, $-10$  &  40 & $41\pm2$  & 0.12   \\
                   & C+D & $23\times18$, 8        &  45 & $41\pm2$  & 2.8    \\
MS\,0440.5+0204    & A   & $1.5\times1.3$, $-48$  &  30 & $43.6\pm2.2$\tablenotemark{b} & 0.12 \\
                   & C   & $15\times13$, $-13$    &  65 & $43\pm2$  & 2.8\tablenotemark{a}\\
MACS\,J0429.6-0253 & A   & $1.8\times1.3$, $-15$  &  38 & $133\pm7$ & 0.2 \\
                   & B   & $5\times4$, $-2$       &  40 & $138\pm7$ & 0.6 \\  
MS\,0839.8+2938    & B   & $6\times4$, $-65$      &  40 & $25\pm1$  & 0.4 \\
                   & C   & $14\times13$, 12       &  53 & $26\pm1$  & 2.9\tablenotemark{a}\\
A\,1204            & C   & $15\times13$, $-3$     &  28 & $1.8\pm0.1$ & 2.6\tablenotemark{a}\\ 
A\,2125            & C   & $14\times14$, $-3$     &  58 & $19\pm1$  & 3.5\tablenotemark{a}\\
A\,2420            & CnB & $14\times6$, 0         &  50 & $198\pm10$ & 1.4\tablenotemark{a} \\
                   & DnC & $50\times20$, 0        &  65 & $197\pm10$ & 1.4 \\
A\,2556            &  CnB &  $14\times8$, 71      &  45 & $20\pm1$\tablenotemark{c} & 1.5\tablenotemark{a} \\ 
                   &  DnC &  $47\times25$, 77     &  60 & $22\pm1$ & 1.5 \\
                   & CnB+DnC & $17\times12$, 60   &  35 & $22\pm1$\tablenotemark{c} & 1.5 \\
\enddata
\label{tab:images}
\tablenotetext{a}{Although the shortest baseline is the same as in D
    configuration, the surface brightness sensitivity to extended structure
    is significantly less that that of the D configuration, but still high
    enough to detect diffuse emission on the core scale at typical
    brightness of minihalos (Appendix \ref{sec:size}).}
\tablenotetext{b}{Sum of the flux densities of the double source
  ($33.0\pm1.7$ mJy) and head-tail radio galaxy ($10.6\pm0.5$ mJy).}
\tablenotetext{c}{Sum of the flux densities of the central radio galaxy
  ($1.8\pm0.1$ mJy in the CnB image and $1.9\pm0.1$ mJy in the CnB+DnC
  image) and head-tail radio galaxy ($18\pm1$ mJy and $20\pm1$ mJy).}
\tablecomments{Column 1: cluster name. Column 2: {\em VLA} configuration.
  Columns 3: Full width half maximum (FWHM) and position angle (p.a.)  of
  the radio beam. Column 4:image root mean square ({\em rms}) level ($1\sigma$)
  for $ROBUST=0$ in IMAGR. Column 5: total flux density of the central radio
  emission. Column 6: largest linear structure detectable by the
    observation (LDS), as derived from the maximum angular scale that can be
    imaged reasonably well by {\em VLA} long-synthesis observations.}
\end{deluxetable}

\subsection{Notes on individual clusters}

\smallskip\noindent{\bf A\,193.} The combined C and D--configuration image
(Fig. \ref{fig:images1}, right upper panel) shows a central double
radio source with a flux density of $61\pm3$ mJy. The combined
C--configuration image (Fig. \ref{fig:images1}, left upper panel)
reveals the source to be composed of a bright compact component, coincident
with the cluster central galaxy, and a $\sim 60$ kpc-long ``tail'' toward
North-East (see also \cite{1993ApJS...87..135O} and
\cite{1997ApJS..108...41O}). The total flux in this image is $60\pm3$ mJy,
of which $28\pm1$ mJy are in the compact component.

\smallskip\noindent{\bf A\,383.} A single point source with $41\pm2$ mJy is
detected at cluster central galaxy in all our images (Fig.
  \ref{fig:images1}, middle panels).  Its size is $<5$ kpc, based on a
Gaussian fit on our highest resolution image, obtained from the combination
of the A--configuration observations.

\smallskip\noindent{\bf MS\,0440.5+0204}. A central extended radio source is
detected in the C--configuration image (Fig. \ref{fig:images1}, bottom
  panels).  At the higher-resolution of the A--configuration observation,
the source separates into a $\sim 30$ kpc double, associated with the
cluster central galaxy, and a head-tail radio galaxy, identified as a member
galaxy ($z=0.193$, Gioia et al. 1998).  A total flux of $43.6\pm2.2$ mJy is
measured on the C--configuration image.  This value is consistent with the
sum of the flux densities in the double ($33.0\pm1.7$ mJy) and head tail
($10.6\pm0.5$) measured at higher resolution.

\smallskip\noindent{\bf MACS\,J0429.6-0253}. An unresolved source is
detected at the cluster center in both A-- and B--configuration images 
(Fig. \ref{fig:images2}, left upper panel; see also Fig.~7), with a flux density
of $133\pm7$ mJy and $138\pm7$ mJy, respectively. Based on the
A--configuration image, the source is $<10$ kpc in size.

\smallskip\noindent{\bf MS\,0839.8+2938}. The C--configuration observation
detects a central compact radio source, which the B--configuration reveals
to be a $80$ kpc-long, core-dominated double radio source (Fig.
  \ref{fig:images2}, right upper panel; see also Fig.~7). Its flux density is $26\pm1$ mJy
(C configuration) and $25\pm1$ mJy (B configuration).

\smallskip\noindent{\bf A\,1204}. An unresolved source (size $<40$ kpc) is
detected at the cluster center in the C-configuration image (Fig. \ref{fig:images2}, middle panels).  
Its flux density is $1.8\pm0.1$ mJy.
An unresolved source is also detected by a slightly higher resolution
observation at 5 GHz with the {\em VLA} in C configuration (Hogan et al.
2015), with a similar flux of 1.8 mJy, implying a flat spectral index
between 1.4 GHz and 5 GHz.

\smallskip\noindent{\bf A\,2125}. A deep radio survey of the cluster region
has been carried out by Owen et al. (2005) using the {\em VLA} at 1.4 GHz in
its A and B configurations. An image of the central cluster region is
presented in Owen et al. (2006). The region is occupied by a triple system
of optical galaxies, each hosting an extended radio source, and a nearby
bright compact radio galaxy, named C153. Here we present a lower resolution
image of the cluster center, obtained from an archival {\em VLA} observation
in C configuration (Fig. \ref{fig:images2}, bottom panels). The three
central sources, individually detected by Owen et al., appear here blended
into an single source with a total flux density of $19\pm1$ mJy.  This value
is in agreement with the the sum of the flux densities measured by Owen et
al. (2005) for these three sources (19.3 mJy).  C153 is unresolved in our
image; its flux density of $22.9\pm0.7$ mJy is in good agreement with that
reported by Owen et al. (22.9 mJy).

\smallskip\noindent{\bf A\,2420}. A slightly extended radio source, oriented
north-south and with a flux density of $198\pm10$ mJy, is detected at the
position of the central galaxy in the CnB--configuration image (Fig.~\ref{fig:images3}, left upper panel).  
In the DnC--configuration image (Fig.~\ref{fig:images3}, right upper panel), 
the source has a flux of
$197\pm10$ mJy. The compact radio galaxy visible to the north-east has a
flux of $1.5\pm0.1$ mJy (CnB configuration) and $2.1\pm0.1$ mJy (DnC
configuration).

\smallskip\noindent{\bf A\,2556}. The CnB--configuration and combined
CnB+DnC-configuration images (Fig.~\ref{fig:images3}, bottom panels),
show an extended radio source associated with the dominant cluster galaxy,
with a size of $\sim 65$ kpc along the NE-SW axis. A radio galaxy with
head-tail morphology is located at a projected distance of about $90$ kpc
from the central galaxy. The flux densities of these sources, measured on
the combined image are $1.9\pm0.1$ mJy and $20\pm1$ mJy, respectively.  The
two radio galaxies are blended together in the lower-resolution image from
the DnC configuration (Fig.~\ref{fig:images3}, left bottom panel))
with a total flux density of $22\pm1$ mJy.

\section{Radio size and average surface brightness of minihalos and halos}
\label{sec:size}

In Table 10 we summarize the properties --- frequency, angular
resolution, sensitivity and the radius of the largest detectable
structure (based on full-synthesis nominal values for {\em GMRT}, {\em
  VLA}\/ and {\em WSRT}) --- of the radio images used to estimate the
extent of the radio sources in our statistical and
supplementary samples (candidates are not included).  For each source,
the table also lists the measured radius (the average radius of a
$3\sigma$ isocontour, \S~\ref{sec:def}) and the ratio of the source
average surface brightness to the {\em rms}\/ noise level of the
corresponding radio image.


\startlongtable
\begin{deluxetable*}{lccccccc}
\tablecaption{Properties of the radio images and central diffuse radio sources}
\tabletypesize{\scriptsize}  
 \tablehead{
    \colhead{Cluster name} &
    \colhead{$\nu$} &
    \colhead{FWHM} &
    \colhead{{\em rms} ($1\sigma$)} &
    \colhead{$\frac{LDS}{2}$} &
    \colhead{$R_{\rm radio}$} &
    \colhead{$\frac{<SB_{\rm radio}>}{1\sigma}$} &
    \colhead{Reference} 
    \\[0.5mm]
    \colhead{} &
    \colhead{(GHz)} & 
    \colhead{($^{\prime\prime}\times^{\prime \prime}$)} &
    \colhead{(mJy/beam)} &
    \colhead{(Mpc)} &
    \colhead{(Mpc)} &
    \colhead{} &
    \colhead{} 
}
\startdata
\multicolumn{8}{c}{Clusters with minihalos in statistical sample} \\
\hline\noalign{\smallskip}
A\,478              & 1.4 & $30\times30$  & 0.05   & 0.7  & 0.16 &   7.9   &  4 \\ 
RBS\,797            & 1.4 & $19\times12$  & 0.04   & 2.2  & 0.12 &  12.2   &  2 \\ 
A\,3444             & 0.6 & $7\times7$    & 0.04   & 2.0  & 0.12 &  10.4   &  3 \\  
A\,1835             & 1.4 & $51\times45$  & 0.04   & 1.8  & 0.24 &  23.7   &  2 \\ 
AS\,780             & 0.6 & $6\times4$    & 0.07   & 1.9  & 0.05 &  17.0   &  5 \\ 
RXC\,J1504.1$-$0248 & 0.3 & $11\times10$  & 0.10   & 3.3  & 0.14 &  22.7   &  6 \\ 
A\,2029             & 1.4 & $57\times45$  & 0.04   & 0.6  & 0.25 &  10.8   &  2 \\ 
A\,2204             & 1.4 & $6\times5$    & 0.03   & 0.16 & 0.05 &   7.1   &  4 \\ 
RXC\,J1720.1+2638   & 0.6 & $8\times6$    & 0.03   & 1.4  & 0.14 &  28.1   &  7 \\ 
A\,2667             & 1.2 & $5\times5$    & 0.04   & 1.5  & 0.06 &   5.0   &  2 \\ 
\hline\noalign{\smallskip}
\multicolumn{8}{c}{Clusters with minihalos in supplementary sample} \\
\hline\noalign{\smallskip}
MACS\,J0159.8$-$0849& 1.4 & $5\times5$    & 0.015  & 0.3  & 0.09 &   3.6    &  4 \\ 
Perseus             & 0.3 & $51\times77$  & 1.4    & 0.9  & 0.13 &  98.7    &  8 \\ 
MACS\,J0329.6$-$0214& 1.3 & $5\times5$    & 0.03   & 1.2  & 0.07 &   6.3    &  2 \\ 
2A\,0335+096        & 1.4 & $30\times27$  & 0.04   & 0.3  & 0.07 &  10.2    &  2 \\ 
A\,907              & 0.6 & $5\times5$    & 0.05   & 1.5  & 0.06 &  11.0    &  2 \\ 
ZwCl\,3146          & 0.6 & $9\times7$    & 0.09   & 2.2  & 0.08 &   6.1    &  5 \\ 
RX\,J1347.5$-$1145  & 1.4 & $17\times17$  & 0.04   & 2.6  & 0.26 &  30.4    &  2 \\ 
MS\,1455.0+2232     & 0.6 & $6\times5$    & 0.07   & 2.0  & 0.12 &  4.1     &  9 \\
RX\,J1532.9+3021    & 1.4 & $3\times3$    & 0.01   & 0.3  & 0.10 &  4.3     &  4 \\ 
Ophiuchus           & 1.4 & $110\times60$ & 0.12   & 0.25 & 0.25 &  5.8     &  2 \\ 
RXC\,J2129.6+0005   & 0.6 & $11\times11$  & 0.11   & 1.9  & 0.08 &  4.8     &  5 \\ 
Phoenix             & 0.6 & $14\times6$   & 0.04   & 3.4  & 0.23 &  6.7     &  10 \\ 
\hline\noalign{\smallskip}
\multicolumn{8}{c}{Clusters with radio halos in statistical sample} \\
\hline\noalign{\smallskip}
A\,2744             & 1.4  & $50\times50$  & 0.09  & 2.0  & 0.8  &   12.7    &  34 \\   
A\,209              & 1.4  & $60\times40$  & 0.05  & 1.5  & 0.5  &    9.3    &  12 \\   
A\,401              & 1.4  & $60\times60$  & 0.08  & 0.6  & 0.4  &    3.8    &  2  \\   
PSZ1\,G171.96$-$40.64&1.4  & $50\times50$  & 0.12  & 1.9  & 0.6  &    4.5    &  13 \\   
A\,521              & 0.2  & $35\times35$  & 0.22  & 5.1  & 0.7  &    6.5    &  14 \\    
A\,520              & 1.4  & $26\times26$  & 0.03  & 1.5  & 0.5  &    4.2    &  15 \\   
A\,665              & 1.4  & $52\times42$  & 0.07  & 1.4  & 0.7  &    7.0    &  16 \\  
A\,697              & 0.3  & $47\times41$  & 0.15  & 4.1  & 0.6  &    9.6    &  17 \\ 
A\,754              & 0.3  & $109\times74$ & 1.0   & 1.0  & 0.6  &    5.1    &  18 \\   
A\,773              & 1.4  & $30\times30$  & 0.03  & 1.6  & 0.5  &    4.7    &  34 \\
A\,1300             & 0.3  & $28\times28$  & 0.50  & 4.3  & 0.5  &    4.2    &  11 \\ 
A\,1351             & 1.4  & $20\times18$  & 0.05  & 2.1  & 0.5  &    3.2    &  19 \\   
A\,1443             & 0.3  & $27\times27$  & 0.36  & 3.9  & 0.6  &    1.8    &  20 \\ 
A\,1689             & 1.4  & $67\times52$  & 0.10  & 1.3  & 0.7  &    5.0    &  2 \\
A\,1758a            & 0.3  & $35\times35$  & 0.40  & 4.1  & 0.8  &     3.3   &  11 \\ 
A\,1914             & 1.4  & $50\times45$  & 0.05  & 1.3  & 0.6  &    16.7   &  21 \\   
RXC\,J1514.9$-$1523 & 0.3  & $53\times41$  & 0.27  & 3.4  & 0.7  &    5.5    &  22 \\   
A\,2142             & 1.4  & $60\times60$  & 0.18  & 0.7  & 0.4  &    2.0    &  23 \\    
A\,2163             & 1.4  & $45\times60$  & 0.05  & 1.5  & 1.0  &    23.4   &  24 \\   
A\,2218             & 0.6  & $25\times25$  & 0.07  & 1.5  & 0.4  &    1.7    &  2  \\
A\,2219             & 1.4  & $50\times45$  & 0.07  & 1.6  & 0.7  &    18.1   &  21 \\   
A\,2256             & 0.3  & $67\times67$  & 0.60  & 3.5  & 0.5  &    7.2    &  25 \\   
A\,2261             & 1.4  & $50\times50$  & 0.05  & 1.6  & 0.5  &    2.8    &  26 \\   
CL\,1821+643        & 0.3  & $30\times26$  & 0.20  & 4.5  & 0.4  &    8.6    &  27 \\    
RXC\,J2003.5$-$2323 & 1.4  & $35\times35$  & 0.03  & 2.1  & 0.9  &    9.4    &  28 \\   
A\,2390             & 1.4  & $30\times30$  & 0.04  & 1.7  & 0.4  &    8.1    &  26  \\
RXC\,J1314.4$-$2515 & 0.6  & $25\times22$  & 0.06  & 1.9  & 0.3  &    3.7    &  12 \\
\hline\noalign{\smallskip}
\multicolumn{8}{c}{Clusters with radio halos in supplementary sample} \\
\hline\noalign{\smallskip}
A\,399              & 1.4 & $45\times45$   & 0.04   & 0.6  & 0.3  &     4.2   &  33 \\
MACS\,J0417.5$-$1154& 0.6 & $20\times20$   & 0.15   & 2.9  & 0.5  &     4.7   &  29 \\    
A\,1995             & 1.4 & $30\times30$   & 0.05   & 2.1  & 0.3  &     4.4   &  30 \\ 
A\,2034             & 1.4 & $44\times40$   & 0.04   & 0.9  & 0.3  &     7.0   &  30  \\
A\,2255             & 0.3 & $54\times64$   & 0.10   & 3.7  & 0.8  &    15.7   &  31 \\   
A\,2319             & 1.4 & $119\times110$ & 0.40   & 0.9\tablenotemark{a}  & 0.6 &  6.4 &  32 \\ 
\hline\noalign{\smallskip}                         
\enddata
\tablenotetext{a}{combination of two VLA pointings offset by
  $12.2^{\prime}$.}
\tablecomments{Column 1: cluster name. Columns 2--4: frequency, beam FWHM
  and {\em rms}\/ noise ($1\sigma$, per beam) of the images used to measure the source radius.
  Images at 1.4 GHz are from {\em VLA} observations, all other frequencies
  are {\em GMRT} observations; for Perseus and A\,2255 we used {\em WSRT} images. Column 5:
  radius of the largest linear structure detectable by the observations, 
as derived from the maximum angular scale that can be imaged reasonably well by 
long-synthesis observations with the {\em VLA}, {\em GMRT} and {\em WSRT}. 
Column 6: radius of the central diffuse radio
  source, defined as in \S\ref{sec:def}, derived from the $3\sigma$ surface
  brightness isocontour. Column 7: ratio of the source average surface
  brightness ($SB_{\rm radio}$) and $1\sigma$ noise level of the image. Column 8: reference for
  the radio images. Reference code:
(1) this work, 
(2) G17a,
(3) G17b, 
(4) Giacintucci et al. (2014a),
(5) Kale et al. (2015),
(6) Giacintucci et al. (2011a),
(7) Giacintucci et al. (2014b),
(8) Sijbring (1993),
(9) Venturi et al. (2008),
(10) van Weeren et al. (2014),
(11) Venturi et al. (2013), 
(12) Venturi et al. (2007),
(13) Giacintucci et al. (2013),
(14) Brunetti et al. (2008),
(15) Vacca et al. (2014),
(16) Giovannini \& Feretti (2000),
(17) Macario et al. (2010),
(18) Macario et al. (2011),
(19) Giacintucci et al. (2009a),
(20) Bonafede et al. (2015),
(21) Bacchi et al. (2003),
(22) Giacintucci et al. (2011b),
(23) Venturi et al. (2017),
(24) Feretti et al. (2001),
(25) Brentjens (2008),
(26) Sommer et al. (2017),
(27) Bonafede et al. (2014),
(28) Giacintucci et al. (2009b),
(29) Parekh et al. (2017),
(30) Giovannini et al. (2009),
(31) Pizzo \& de Bruyn (2009),
(32) Storm et al. (2015),
(33) Murgia et al. (2010),
(34) Govoni et al. (2001),
(35) Vacca et al. (2015).}
\label{tab:radii}
\end{deluxetable*}

\begin{figure*}
\centering
\includegraphics[width=17cm]{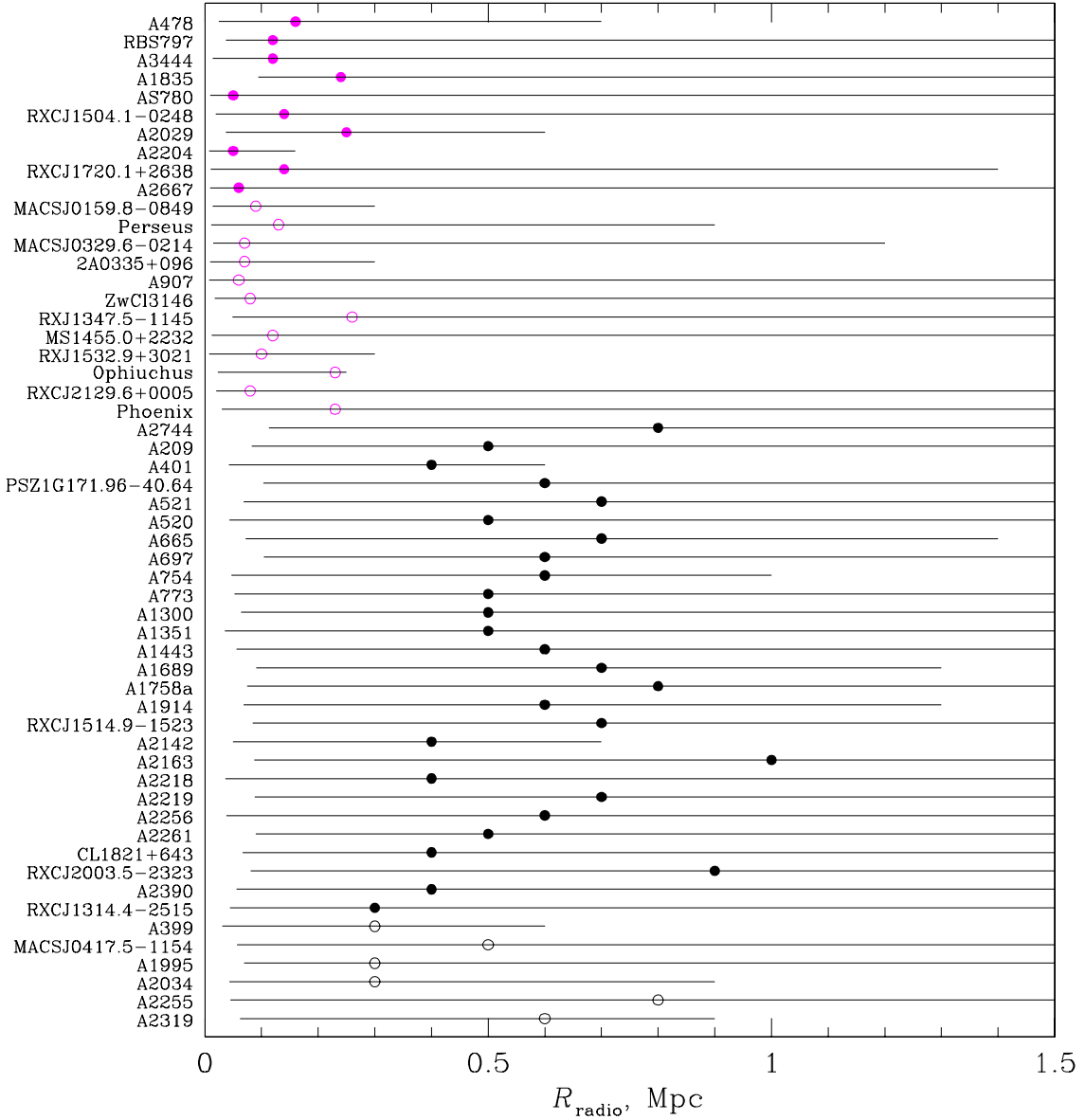}
\caption{Radius of minihalos (magenta) and halos (black) in the 
  statistical sample (filled symbols) and supplementary sample (empty
  symbols).  Horizontal lines show the range of linear scales that can be
  detected in the corresponding radio images: the minimum scale is the
  angular resolution of the image, the maximum scale is the radius of the
  largest structure that can be imaged by the observations (Table 10).}
\label{fig:radii}
\end{figure*}

\begin{figure*}
\centering
\includegraphics[width=7.5cm]{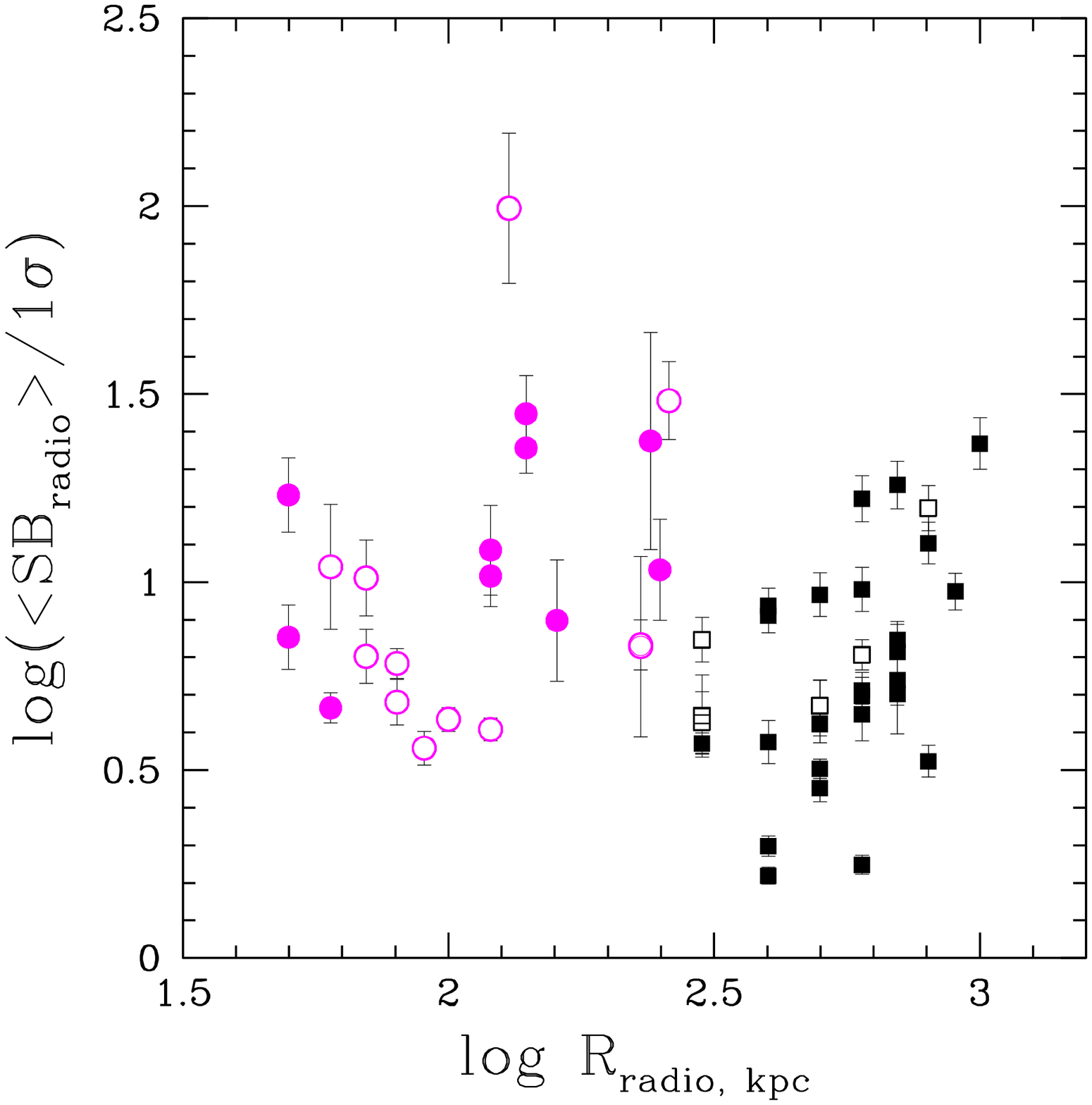}
\includegraphics[width=7.5cm]{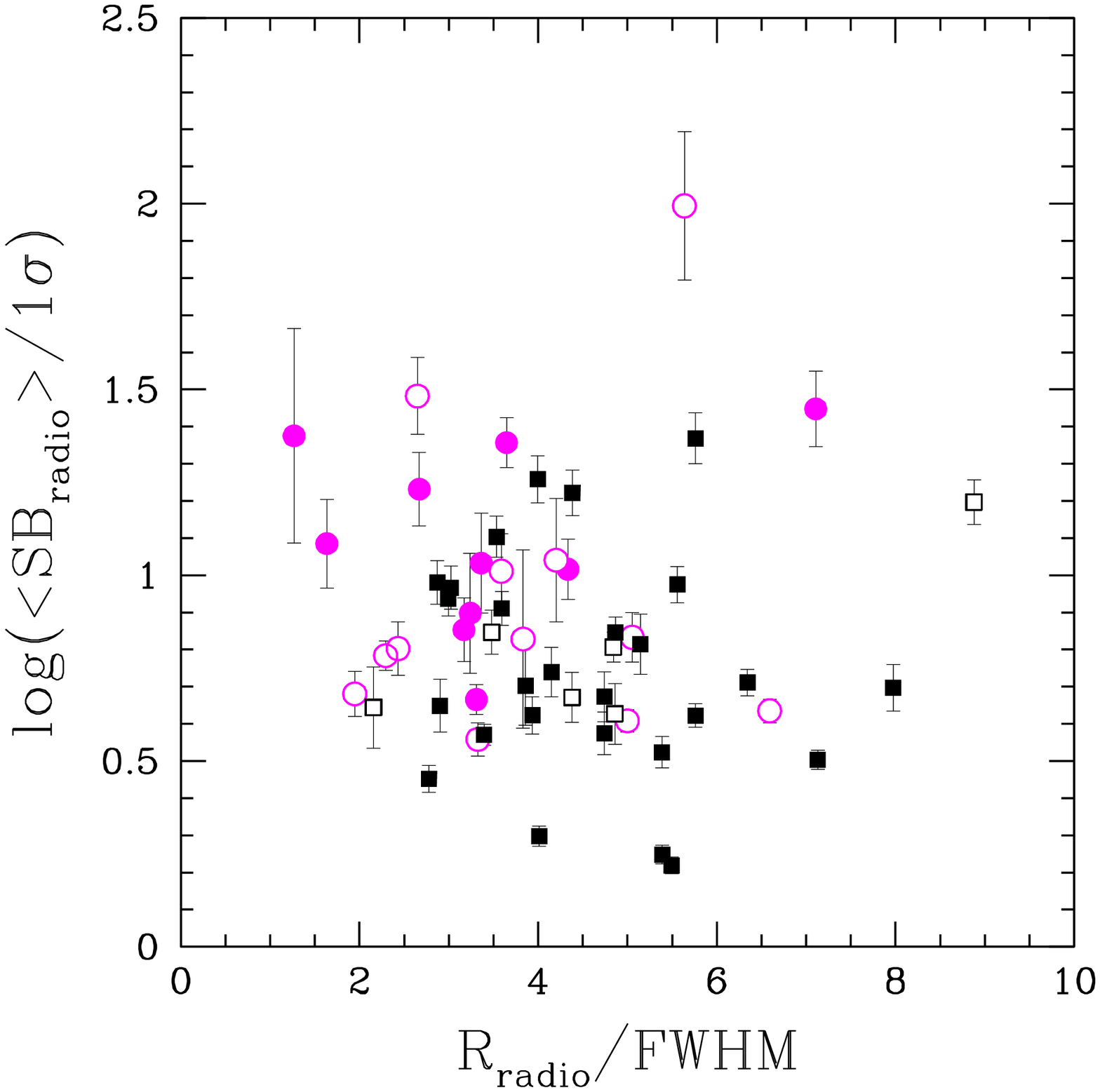}
\caption{Ratios of the average surface brightness of minihalos
  (magenta) and large halos (black) to the {\em rms}\/ noise level
  (per beam) of the
  corresponding radio images (Table 10). The ratios are plotted as a
  function of the radius in kpc (left) and beam sizes (right).
  Filled symbols are clusters in the statistical sample and empty
  symbols are those in the supplementary sample.}
\label{fig:sb}
\end{figure*}

In Fig~\ref{fig:radii}, we compare the measured source radii to the
interval of linear scales that can be detected in the respective radio
images.  The minimum scale is set by the angular resolution of the
image (average FWHM, Table 10). We note here that almost all 
clusters have complementary observations at higher 
resolution. The maximum scale is the radius of the largest structure 
that can be detected in this particular observation (Table 10). 
Even though the actual scale that can be reliably imaged by the observation 
can be less than the nominal value (due, for instance, to a short exposure 
or the exclusion of bad data at short baselines), the minihalo and halo radii 
are typically much less than $LDS/2$ (one exception is Ophiuchus). This 
indicates that the observations are not limited by the interferometric 
coverage; if there was emission outside the minihalo, provided
sufficiently low noise, we should be able to detect it well beyond the
cool core and, in most cases, on scales comparable to those of large
halos. The other obvious instrumental limitation is the noise level
--- if minihalos and large halos were similar sources with a smoothly
declining radial brightness profile, a more sensitive observation
would yield a larger radius of the $3\sigma$ isophote. In
Fig.~\ref{fig:sb}, we show the ratio of the source average surface
brightness to the image {\em rms}\/ noise (Table 10) as a function of
the radius in kpc (left panel) or in beam sizes (right panel). There
is no correlation between the signal-to-noise ratio and the radius of
the halo; the signal-to-noise ratios for the minihalos are similar
to those for the large halos and in several cases even slightly
higher. This is in agreement with a similar plot for giant halos in
Cassano et al.\ (2007). Based on Figs.~\ref{fig:radii} and \ref{fig:sb}, we
conclude that most of the minihalo datasets used in this work
possess the requisite sensitivity to detect diffuse emission on
angular scales greater than the minihalo at a surface brightness level
typical of large halos, thus the measured difference between the sizes
of minihalos and large halos is physically meaningful. This answers a 
frequently asked question of whether minihalos are simply the peaks of 
larger halos in a cluster with the central density peak (and where the 
radio emission roughly follows the ICM density) --- in such a picture, 
we would be able to detect the rest of the halo emission outside the 
cool core, and we do not.


\begin{thebibliography}{}



\bibitem[Andrade-Santos et al.(2017)]{Santos2017apjinprep} Andrade-Santos, F., Jones, C., Forman, W., et al., 2017, ApJ, submitted, arXiv:1703.08690

\bibitem[Ascasibar \& Markevitch(2006)]{2006ApJ...650..102A} Ascasibar, Y., \& Markevitch, M.\ 2006, \apj, 650, 102 

\bibitem[Bacchi et al.(2003)]{2003A&A...400..465B} Bacchi, M., Feretti, L., Giovannini, G., \& Govoni, F.\ 2003, \aap, 400, 465 

\bibitem[Barrena et al.(2007)]{2007A&A...469..861B} Barrena, R., Boschin, W., Girardi, M., \& Spolaor, M.\ 2007, \aap, 469, 861 

\bibitem[Bonafede et al.(2009)]{2009A&A...494..429B} Bonafede, A., Giovannini, G., Feretti, L., Govoni, F., \& Murgia, M.\ 2009, \aap, 494, 429 

\bibitem[Bonafede et al.(2014)]{2014MNRAS.444L..44B} Bonafede, A., Intema, H.~T., Br{\"u}ggen, M., et al.\ 2014, \mnras, 444, L44 

\bibitem[Bonafede et al.(2015)]{2015MNRAS.454.3391B} Bonafede, A., Intema, H., Br{\"u}ggen, M., et al.\ 2015, \mnras, 454, 3391 

\bibitem[Briggs(1995)]{1995AAS...18711202B} Briggs, D.~S.\ 1995, Bulletin of the American Astronomical Society, 27, 112.02 

\bibitem[Bravi et al.(2016)]{2016MNRAS.455L..41B} Bravi, L., Gitti, M., \& Brunetti, G.\ 2016, \mnras, 455, L41 

\bibitem[Brentjens(2008)]{2008A&A...489...69B} Brentjens, M.~A.\ 2008, \aap, 489, 69 

\bibitem[Brunetti et al.(2008)]{2008Natur.455..944B} Brunetti, G., Giacintucci, S., Cassano, R., et al.\ 2008, Nature, 455, 944 

\bibitem[Brunetti \& Jones(2014)]{2014IJMPD..2330007B} Brunetti, G., \& Jones, T.~W.\ 2014, International Journal of Modern Physics D, 23, 1430007 

\bibitem[Burns et al.(1992)]{1992ApJ...388L..49B} Burns, J.~O., Sulkanen, M.~E., Gisler, G.~R., \& Perley, R.~A.\ 1992, \apjl, 388, L49 


\bibitem[Cassano \& Brunetti(2005)]{2005MNRAS.357.1313C} Cassano, R., \& Brunetti, G.\ 2005, \mnras, 357, 1313 

\bibitem[Cassano et al.(2006)]{2006MNRAS.369.1577C} Cassano, R., Brunetti, G., \& Setti, G.\ 2006, \mnras, 369, 1577 

\bibitem[Cassano et al.(2007)]{2007MNRAS.378.1565C} Cassano, R., Brunetti, G., Setti, G., Govoni, F., \& Dolag, K.\ 2007, \mnras, 378, 1565 

\bibitem[Cassano et al.(2008)]{2008A&A...486L..31C} Cassano, R., Gitti, M., \& Brunetti, G.\ 2008, \aap, 486, L31 

\bibitem[Cassano et al.(2010)]{2010ApJ...721L..82C} Cassano, R., Ettori, S., Giacintucci, S., et al.\ 2010, \apjl, 721, L82 

\bibitem[Cassano et al.(2013)]{2013ApJ...777..141C} Cassano, R., Ettori, S., Brunetti, G., et al.\ 2013, \apj, 777, 141 

\bibitem[Cassano et al.(2016)]{2016A&A...593A..81C} Cassano, R., Brunetti, G., Giocoli, C., \& Ettori, S.\ 2016, \aap, 593, A81 

\bibitem[Cavagnolo et al.(2008)]{2008ApJ...682..821C} Cavagnolo, K.~W., Donahue, M., Voit, G.~M., \& Sun, M.\ 2008, \apj, 682, 821 (C08)

\bibitem[Cavagnolo et al.(2009)]{2009ApJS..182...12C} Cavagnolo, K.~W., Donahue, M., Voit, G.~M., \& Sun, M.\ 2009, \apjs, 182, 12 (C09)

\bibitem[Cavaliere \& Fusco-Femiano(1978)]{1978A&A....70..677C} Cavaliere, A., \& Fusco-Femiano, R.\ 1978, \aap, 70, 677 

\bibitem[Clarke \& Ensslin(2006)]{2006AJ....131.2900C} Clarke, T.~E., \& Ensslin, T.~A.\ 2006, \aj, 131, 2900 

\bibitem[Crawford et al.(1999)]{1999MNRAS.306..857C} Crawford, C.~S., Allen, S.~W., Ebeling, H., Edge, A.~C., \& Fabian, A.~C.\ 1999, \mnras, 306, 857 

\bibitem[Cuciti et al.(2015)]{2015A&A...580A..97C} Cuciti, V., Cassano, R., Brunetti, G., et al.\ 2015, \aap, 580, A97 

\bibitem[Dallacasa et al.(2009)]{2009ApJ...699.1288D} Dallacasa, D., Brunetti, G., Giacintucci, S., et al.\ 2009, \apj, 699, 1288 

\bibitem[David et al.(1993)]{1993ApJ...412..479D} David, L.~P., Slyz, A., Jones, C., et al.\ 1993, \apj, 412, 479 

\bibitem[Doria et al.(2012)]{2012ApJ...753...47D} Doria, A., Gitti, M., Ettori, S., et al.\ 2012, \apj, 753, 47 

\bibitem[Ehlert et al.(2011)]{2011MNRAS.411.1641E} Ehlert, S., Allen, S.~W., von der Linden, A., et al.\ 2011, \mnras, 411, 1641 

\bibitem[Farnsworth et al.(2013)]{2013ApJ...779..189F} Farnsworth, D., Rudnick, L., Brown, S., \& Brunetti, G.\ 2013, \apj, 779, 189 

\bibitem[Feretti et al.(1997a)]{1997A&A...317..432F} Feretti, L., Boehringer, H., Giovannini, G., \& Neumann, D.\ 1997, \aap, 317, 432 

\bibitem[Feretti et al.(1997b)]{1997NewA....2..501F} Feretti, L., Giovannini, G., Boehringer, H.\ 1997, New Astronomy, 2, 501 

\bibitem[Feretti et al.(2001)]{2001A&A...373..106F} Feretti, L., Fusco-Femiano, R., Giovannini, G., \& Govoni, F.\ 2001, \aap, 373, 106 

\bibitem[Feretti et al.(2005)]{2005A&A...444..157F} Feretti, L., Schuecker, P., B{\"o}hringer, H., Govoni, F., \& Giovannini, G.\ 2005, \aap, 444, 157 

\bibitem[Feretti et al.(2012)]{2012A&ARv..20...54F} Feretti, L., Giovannini, G., Govoni, F., \& Murgia, M.\ 2012, \aapr, 20, 54 

\bibitem[Ferrari et al.(2011)]{2011A&A...534L..12F} Ferrari, C., Intema, H.~T., Orr{\`u}, E., et al.\ 2011, \aap, 534, L12 

\bibitem[Fujita et al.(2007)]{2007ApJ...663L..61F} Fujita, Y., Kohri, K., Yamazaki, R., \& Kino, M.\ 2007, \apjl, 663, L61 

\bibitem[Fujita \& Ohira(2011)]{2011ApJ...738..182F} Fujita, Y., \& Ohira, Y.\ 2011, \apj, 738, 182 

\bibitem[Fujita \& Ohira(2013)]{2013MNRAS.428..599F} Fujita, Y., \& Ohira, Y.\ 2013, \mnras, 428, 599 

\bibitem[Gendron-Marsolais et al.(2017)]{2017arXiv170103791G} Gendron-Marsolais, M., Hlavacek-Larrondo, J., van Weeren, R.~J., et al.\ 2017, arXiv:1701.03791 

\bibitem[Giacintucci et al.(2009a)]{2009ApJ...704L..54G} Giacintucci, S., Venturi, T., Cassano, R., Dallacasa, D., \& Brunetti, G.\ 2009, \apjl, 704, L54 

\bibitem[Giacintucci et al.(2009b)]{2009A&A...505...45G} Giacintucci, S., Venturi, T., Brunetti, G., et al.\ 2009, \aap, 505, 45 

\bibitem[Giacintucci et al.(2011a)]{2011A&A...525L..10G} Giacintucci, S., Markevitch, M., Brunetti, G., Cassano, R., \& Venturi, T.\ 2011, \aap, 525, L10 

\bibitem[Giacintucci et al.(2011b)]{2011A&A...534A..57G} Giacintucci, S., Dallacasa, D., Venturi, T., et al.\ 2011, \aap, 534, A57 

\bibitem[Giacintucci et al.(2012)]{2012ApJ...755..172G} Giacintucci, S., O'Sullivan, E., Clarke, T.~E., et al.\ 2012, \apj, 755, 172 

\bibitem[Giacintucci et al.(2013)]{2013ApJ...770..161G} Giacintucci, S., Kale, R., Wik, D.~R., Venturi, T., \& Markevitch, M.\ 2013, \apj, 770, 161 

\bibitem[Giacintucci et al.(2014a)]{2014ApJ...781....9G} Giacintucci, S., Markevitch, M., Venturi, T., et al.\ 2014a, \apj, 781, 9 

\bibitem[Giacintucci et al.(2014b)]{2014ApJ...795...73G} Giacintucci, S., Markevitch, M., Brunetti, G., et al.\ 2014b, \apj, 795, 73 

\bibitem[Giacintucci et al.(2017)]{} Giacintucci, S., et al., 2017, in preparation (G17)

\bibitem[Gioia et al.(1998)]{1998ApJ...497..573G} Gioia, I.~M., Shaya, E.~J., Le F{\`e}vre, O., et al.\ 1998, \apj, 497, 573 

\bibitem[Giovannini \& Feretti(2000)]{2000NewA....5..335G} Giovannini, G., \& Feretti, L.\ 2000, New A, 5, 335 

\bibitem[Giovannini et al.(2009)]{2009A&A...507.1257G} Giovannini, G., Bonafede, A., Feretti, L., et al.\ 2009, \aap, 507, 1257 

\bibitem[Gitti et al.(2015)]{2015aska.confE..76G} Gitti, M., Tozzi, P., Brunetti, G., et al.\ 2015, Advancing Astrophysics with the Square Kilometre Array (AASKA14), 76 

\bibitem[Gitti(2013)]{2013MNRAS.436L..84G} Gitti, M.\ 2013, \mnras, 436, L84 

\bibitem[Gitti et al.(2002)]{2002A&A...386..456G} Gitti, M., Brunetti, G., \& Setti, G.\ 2002, \aap, 386, 456 

\bibitem[Gitti et al.(2004)]{2004A&A...417....1G} Gitti, M., Brunetti, G., Feretti, L., \& Setti, G.\ 2004, \aap, 417, 1 

\bibitem[Gitti et al.(2006)]{2006A&A...448..853G} Gitti, M., Feretti, L., \& Schindler, S.\ 2006, \aap, 448, 853 

\bibitem[Gitti et al.(2007)]{2007A&A...470L..25G} Gitti, M., Ferrari, C., Domainko, W., Feretti, L., \& Schindler, S.\ 2007, \aap, 470, L25 

\bibitem[Govoni et al.(2001)]{2001A&A...376..803G} Govoni, F., Feretti, L., Giovannini, G., et al.\ 2001, \aap, 376, 803 

\bibitem[Govoni et al.(2005)]{2005A&A...430L...5G} Govoni, F., Murgia, M., Feretti, L., et al.\ 2005, \aap, 430, L5 

\bibitem[Govoni et al.(2009)]{2009A&A...499..371G} Govoni, F., Murgia, M., Markevitch, M., et al.\ 2009, \aap, 499, 371 

\bibitem[Govoni et al.(2011)]{2011A&A...529A..69G} Govoni, F., Murgia, M., Giovannini, G., Vacca, V., \& Bonafede, A.\ 2011, \aap, 529, A69 

\bibitem[Guo \& Oh(2008)]{2008MNRAS.384..251G} Guo, F., \& Oh, S.~P.\ 2008, \mnras, 384, 251 

\bibitem[Hitomi Collaboration et al.(2016)]{2016Natur.535..117H} Hitomi Collaboration, Aharonian, F., Akamatsu, H., et al.\ 2016, \nat, 535, 117 

\bibitem[Hlavacek-Larrondo et al.(2013)]{2013ApJ...777..163H} Hlavacek-Larrondo, J., Allen, S.~W., Taylor, G.~B., et al.\ 2013, \apj, 777, 163 

\bibitem[Hogan et al.(2015)]{2015MNRAS.453.1201H} Hogan, M.~T., Edge, A.~C., Hlavacek-Larrondo, J., et al.\ 2015, \mnras, 453, 1201 

\bibitem[Ikebe et al.(2002)]{2002A&A...383..773I} Ikebe, Y., Reiprich, T.~H., B{\"o}hringer, H., Tanaka, Y., \& Kitayama, T.\ 2002, \aap, 383, 773 

\bibitem[Jacob \& Pfrommer(2017a)]{2016arXiv160906321J} Jacob, S., \& Pfrommer, C.\ 2017a, \mnras, arXiv:1609.06321 

\bibitem[Jacob \& Pfrommer(2017b)]{2016arXiv160906322J} Jacob, S., \& Pfrommer, C.\ 2017b, \mnras, arXiv:1609.06322 

\bibitem[Jones \& Forman(1984)]{1984ApJ...276...38J} Jones, C., \& Forman, W.\ 1984, \apj, 276, 38 

\bibitem[Kalberla et al.(2005)]{2005A&A...440..775K} Kalberla, P.~M.~W., Burton, W.~B., Hartmann, D., et al.\ 2005, \aap, 440, 775 

\bibitem[Kale et al.(2013)]{2013A&A...557A..99K} Kale, R., Venturi, T., Giacintucci, S., et al.\ 2013, \aap, 557, A99 

\bibitem[Kale et al.(2015)]{2015A&A...579A..92K} Kale, R., Venturi, T., Giacintucci, S., et al.\ 2015, \aap, 579, A92 

\bibitem[Kale \& Parekh(2016)]{2016MNRAS.459.2940K} Kale, R., \& Parekh, V.\ 2016, \mnras, 459, 2940 

\bibitem[Kassim et al.(2001)]{2001ApJ...559..785K} Kassim, N.~E., Clarke, T.~E., En{\ss}lin, T.~A., Cohen, A.~S., 
\& Neumann, D.~M.\ 2001, \apj, 559, 785 

\bibitem[Macario et al.(2010)]{2010A&A...517A..43M} Macario, G., Venturi, T., Brunetti, G., et al.\ 2010, \aap, 517, A43 

\bibitem[Macario et al.(2011)]{2011ApJ...728...82M} Macario, G., Markevitch, M., Giacintucci, S., et al.\ 2011, \apj, 728, 82 

\bibitem[Macario et al.(2013)]{2013A&A...551A.141M} Macario, G., Venturi, T., Intema, H.~T., et al.\ 2013, \aap, 551, A141 

\bibitem[Macario et al.(2014)]{2014A&A...565A..13M} Macario, G., Intema, H.~T., Ferrari, C., et al.\ 2014, \aap, 565, A13 

\bibitem[Markevitch et al.(2000)]{2000ApJ...541..542M} Markevitch, M., Ponman, T.~J., Nulsen, P.~E.~J., et al.\ 2000, \apj, 541, 542 

\bibitem[Markevitch et al.(2003)]{2003ApJ...586L..19M} Markevitch, M., Mazzotta, P., Vikhlinin, A., et al.\ 2003, \apjl, 586, L19 

\bibitem[Mazzotta et al.(2004)]{2004MNRAS.354...10M} Mazzotta, P., Rasia, E., Moscardini, L., \& Tormen, G.\ 2004, \mnras, 354, 10 

\bibitem[Mazzotta \& Giacintucci(2008)]{2008ApJ...675L...9M} Mazzotta, P., \& Giacintucci, S.\ 2008, \apjl, 675, L9 

\bibitem[McDonald et al.(2015)]{2015ApJ...811..111M} McDonald, M., McNamara, B.~R., van Weeren, R.~J., et al.\ 2015, \apj, 811, 111 

\bibitem[McDonald et al.(2017)]{2017arXiv170205094M} McDonald, M., Allen, S.~W., Bayliss, M., et al.\ 2017, arXiv:1702.05094 

\bibitem[McNamara \& Nulsen(2007)]{2007ARA&A..45..117M} McNamara, B.~R., \& Nulsen, P.~E.~J.\ 2007, \araa, 45, 117

\bibitem[Mittal et al.(2009)]{2009A&A...501..835M} Mittal, R., Hudson, D.~S., Reiprich, T.~H., \& Clarke, T.\ 2009, \aap, 501, 835

\bibitem[Murgia et al.(2009)]{2009A&A...499..679M} Murgia, M., Govoni, F., Markevitch, M., et al.\ 2009, \aap, 499, 679 

\bibitem[Murgia et al.(2010)]{2010A&A...509A..86M} Murgia, M., Govoni, F., Feretti, L., \& Giovannini, G.\ 2010, \aap, 509, A86 

\bibitem[Murgia et al.(2010b)]{2010A&A...514A..76M} Murgia, M., Eckert, D., Govoni, F., et al.\ 2010, \aap, 514, A76 

\bibitem[Murgia et al.(2011)]{2011A&A...526A.148M} Murgia, M., Parma, P., Mack, K.-H., et al.\ 2011, \aap, 526, A148 

\bibitem[Murgia et al.(2012)]{2012A&A...548A..75M} Murgia, M., Markevitch, M., Govoni, F., et al.\ 2012, \aap, 548, A75 

\bibitem[Nagai(2006)]{2006ApJ...650..538N} Nagai, D.\ 2006, \apj, 650, 538 

\bibitem[Owen et al.(2006)]{2006AJ....131.1974O} Owen, F.~N., Keel, W.~C., Wang, Q.~D., Ledlow, M.~J., \& Morrison, G.~E.\ 2006, \aj, 131, 1974 

\bibitem[Owen et al.(2005)]{2005AJ....129...31O} Owen, F.~N., Ledlow, M.~J., Keel, W.~C., Wang, Q.~D., \& Morrison, G.~E.\ 2005, \aj, 129, 31 

\bibitem[Owen et al.(1993)]{1993ApJS...87..135O} Owen, F.~N., White, R.~A., \& Ge, J.\ 1993, \apjs, 87, 135 

\bibitem[Owen \& Ledlow(1997)]{1997ApJS..108...41O} Owen, F.~N., \& Ledlow, M.~J.\ 1997, \apjs, 108, 41 

\bibitem[Pandey-Pommier et al.(2016)]{2016arXiv161200225P} Pandey-Pommier, M., Richard, J., Combes, F., et al.\ 2016, arXiv:1612.00225 

\bibitem[Parekh et al.(2015)]{2015A&A...575A.127P} Parekh, V., van der Heyden, K., Ferrari, C., Angus, G., \& Holwerda, B.\ 2015, \aap, 575, A127 

\bibitem[Parekh et al.(2017)]{2017MNRAS.464.2752P} Parekh, V., Dwarakanath, K.~S., Kale, R., \& Intema, H.\ 2017, \mnras, 464, 2752 

\bibitem[Perley \& Butler(2013)]{2013ApJS..204...19P} Perley, R.~A., \& Butler, B.~J.\ 2013, \apjs, 204, 19 

\bibitem[Pizzo \& de Bruyn(2009)]{2009A&A...507..639P} Pizzo, R.~F., \& de Bruyn, A.~G.\ 2009, \aap, 507, 639 

\bibitem[Planck Collaboration et al.(2014)]{2014A&A...571A..29P} Planck Collaboration, Ade, P.~A.~R., Aghanim, N., et al.\ 2014, A\&A, 571, 29

\bibitem[Pfrommer \& En{\ss}lin(2004)]{2004A&A...413...17P} Pfrommer, C., \& En{\ss}lin, T.~A.\ 2004, \aap, 413, 17 

\bibitem[Pizzo \& de Bruyn(2009)]{2009A&A...507..639P} Pizzo, R.~F., \& de Bruyn, A.~G.\ 2009, \aap, 507, 639 

\bibitem[Reid et al.(1999)]{1999MNRAS.302..571R} Reid, A.~D., Hunstead, R.~W., Lemonon, L., \& Pierre, M.~M.\ 1999, \mnras, 302, 571 

\bibitem[Rossetti et al.(2013)]{2013A&A...556A..44R} Rossetti, M., Eckert, D., De Grandi, S., et al.\ 2013, \aap, 556, A44 

\bibitem[Rossetti et al.(2016)]{2016MNRAS.457.4515R} Rossetti, M., Gastaldello, F., Ferioli, G., et al.\ 2016, \mnras, 457, 4515 

\bibitem[Saikia \& Jamrozy(2009)]{2009BASI...37...63S} Saikia, D.~J., \& Jamrozy, M.\ 2009, Bulletin of the Astronomical Society of India, 37, 63 

\bibitem[Sarazin et al.(1995)]{1995ApJ...451..125S} Sarazin, C.~L., Baum, S.~A., \& O'Dea, C.~P.\ 1995, \apj, 451, 125 

\bibitem[Sijbring (1993)]{} Sijbring L.G., PhD Thesis, University of Groningen, A radio continuum and HI Line study of the Perseus cluster (1993)

\bibitem[Sommer et al.(2017)]{2017MNRAS.466..996S} Sommer, M.~W., Basu, K., Intema, H., et al.\ 2017, \mnras, 466, 996 

\bibitem[Vacca et al.(2011)]{2011A&A...535A..82V} Vacca, V., Govoni, F., Murgia, M., et al.\ 2011, \aap, 535, A82 

\bibitem[Vacca et al.(2014)]{2014A&A...561A..52V} Vacca, V., Feretti, L., Giovannini, G., et al.\ 2014, \aap, 561, A52 

\bibitem[van Weeren et al.(2010)]{2010Sci...330..347V} van Weeren, R.~J., R{\"o}ttgering, H.~J.~A., Br{\"u}ggen, M., \& Hoeft, M.\ 2010, Science, 330, 347 

\bibitem[van Weeren et al.(2014)]{2014ApJ...786L..17V} van Weeren, R.~J., Intema, H.~T., Lal, D.~V., et al.\ 2014, \apjl, 786, L17 

\bibitem[Venturi et al.(2007)]{2007A&A...463..937V} Venturi, T., Giacintucci, S., Brunetti, G., et al.\ 2007, \aap, 463, 937 

\bibitem[Venturi et al.(2008)]{2008A&A...484..327V} Venturi, T., Giacintucci, S., Dallacasa, D., et al.\ 2008, \aap, 484, 327 

\bibitem[Venturi et al.(2011a)]{2011MNRAS.414L..65V} Venturi, T., Giacintucci, G., Dallacasa, D., et al.\ 2011, \mnras, 414, L65 

\bibitem[Venturi et al.(2011b)]{2011JApA...32..501V} Venturi, T., Giacintucci, S., \& Dallacasa, D.\ 2011, Journal of Astrophysics and Astronomy, 32, 501 

\bibitem[Venturi et al.(2013)]{2013A&A...551A..24V} Venturi, T., Giacintucci, S., Dallacasa, D., et al.\ 2013, \aap, 551, A24 

\bibitem[Venturi et al.(2017)]{} Venturi T., et al., 2017, A\&A, in press (arXiv:1703.06802)

\bibitem[Vikhlinin et al.(2005)]{2005ApJ...628..655V} Vikhlinin, A., Markevitch, M., Murray, S.~S., et al.\ 2005, \apj, 628, 655 

\bibitem[Vikhlinin et al.(2006)]{2006ApJ...640..691V} Vikhlinin, A., Kravtsov, A., Forman, W., et al.\ 2006, \apj, 640, 691 

\bibitem[Vikhlinin et al.(2009)]{2009ApJ...692.1033V} Vikhlinin, A., Burenin, R.~A., Ebeling, H., et al.\ 2009, \apj, 692, 1033 

\bibitem[White et al.(1997)]{1997MNRAS.292..419W} White, D.~A., Jones, C., \& Forman, W.\ 1997, \mnras, 292, 419 

\bibitem[Yuan et al.(2015)]{2015ApJ...813...77Y} Yuan, Z.~S., Han, J.~L., \& Wen, Z.~L.\ 2015, \apj, 813, 77

\bibitem[Zandanel et al.(2014)]{2014MNRAS.438..124Z} Zandanel, F., Pfrommer, C., \& Prada, F.\ 2014, \mnras, 438, 124 

\bibitem[Zhuravleva et al.(2014)]{2014Natur.515...85Z} Zhuravleva, I., Churazov, E., Schekochihin, A.~A., et al.\ 2014, \nat, 515, 85 

\bibitem[ZuHone et al.(2013)]{2013ApJ...762...78Z} ZuHone, J.~A., Markevitch, M., Brunetti, G., \& Giacintucci, S.\ 2013, \apj, 762, 78 

\end{thebibliography}
\end{document}